\documentclass[prx,reprint,preprintnumbers,nofootinbib,superscriptaddress,amsmath,amssymb,aps, floatfix]{revtex4-2}

\usepackage[colorlinks=true,linkcolor=blue, citecolor=blue,
urlcolor=blue]{hyperref}

\usepackage{multirow,graphics}
\usepackage{enumerate}
\usepackage{amstext}
\usepackage{amssymb}
\usepackage{amsmath}
\usepackage{graphicx}
\usepackage{color}
\usepackage{tikz}
\usepackage{psfrag}
\usepackage{soul}

\usepackage{pgfplots}
\pgfplotsset{compat=1.3}

\usepackage{thumbpdf}
\usepackage{esint,shuffle,psfrag}
\usepackage[utf8]{inputenc}

\allowdisplaybreaks
\usepackage{varioref}
\usepackage{amsmath,amsfonts,amsthm,mathrsfs}
\usepackage{enumerate}
\usepackage{verbatim}
\usepackage{wrapfig}
\usepackage{appendix}
\usepackage{amstext}
\usepackage{amssymb}
\usepackage{graphicx}
\usepackage{color}
\usepackage{capt-of}
\usepackage[caption=false]{subfig}
\usepackage{varioref}
\usepackage{multirow,graphics}
\usepackage{epstopdf}
\usepackage{bbm}
\usepackage{slashed}
\usepackage{psfrag}
\usepackage{relsize}
\usepackage{bbm}
\usepackage{mathrsfs}
\usepackage{epsfig}
\usepackage{upgreek}
\usepackage{longtable}

\usepackage{booktabs}

\usepackage{braket}

\usepackage{mathtools}
       \usepackage{dsfont}

\setcitestyle{square,numbers}

\begin{document}

\newcommand{\IM}{{\rm Im}\,}
\newcommand{\card}{\#}
\newcommand{\la}[1]{\label{#1}}
\newcommand{\eq}[1]{(\ref{#1}),} 
\newcommand{\figref}[1]{Fig.~\ref{#1}}
\newcommand{\abs}[1]{\left|#1\right|}
\newcommand{\comD}[1]{{\color{red}#1\color{black}}}
\newcommand{\p}{{\partial}}
\newcommand{\Tr}{{\text{Tr}}}
\newcommand{\tr}{{\text{tr}}}
\newcommand{\como}[1]{{\color[rgb]{0.0,0.1,0.9} {\bf \"O:} #1} }

\makeatletter
\newcommand{\subalign}[1]{%
  \vcenter{%
    \Let@ \restore@math@cr \default@tag
    \baselineskip\fontdimen10 \scriptfont\tw@
    \advance\baselineskip\fontdimen12 \scriptfont\tw@
    \lineskip\thr@@\fontdimen8 \scriptfont\thr@@
    \lineskiplimit\lineskip
    \ialign{\hfil$\m@th\scriptstyle##$&$\m@th\scriptstyle{}##$\crcr
      #1\crcr
    }%
  }
}
\makeatother

\newcommand{\mzvv}[2]{
  \zeta_{
    \subalign{
      &#1,\\
      &#2
    }
}
}

\newcommand{\mzvvv}[3]{
  \zeta_{
    \subalign{
      &#1,\\
      &#2,\\
      &#3
    }
}
  }

\makeatletter
     \@ifundefined{usebibtex}{\newcommand{\ifbibtexelse}[2]{#2}} {\newcommand{\ifbibtexelse}[2]{#1}}
\makeatother

\preprint{}

\def\corrlen{\x,{0.7*exp(-((\x-3.5)^2)*4)}}
\def\wavepacket{\x,{1.3*exp(-((\x-3.5)^2)/4)}}

\definecolor{gradientblue}{RGB}{110,195,240}
\definecolor{gradienteta2}{RGB}{211,105,120}
\definecolor{gradientFF}{RGB}{253,70,70}

\usetikzlibrary{decorations.pathmorphing}
\usetikzlibrary{decorations.markings}
\usetikzlibrary{intersections}
\usetikzlibrary{calc}

\tikzset{
photon/.style={decorate, decoration={snake}},
particle/.style={postaction={decorate},
    decoration={markings,mark=at position .5 with {\arrow{>}}}},
antiparticle/.style={postaction={decorate},
    decoration={markings,mark=at position .5 with {\arrow{<}}}},
gluon/.style={decorate, decoration={coil,amplitude=2pt, segment length=4pt},color=purple},
wilson/.style={color=blue, thick},
scalarZ/.style={postaction={decorate},decoration={markings, mark=at position .5 with{\arrow[scale=1]{stealth}}}},
scalarX/.style={postaction={decorate}, dashed, dash pattern = on 4pt off 2pt, dash phase = 2pt, decoration={markings, mark=at position .53 with{\arrow[scale=1]{stealth}}}},
scalarZw/.style={postaction={decorate},decoration={markings, mark=at position .75 with{\arrow[scale=1]{stealth}}}},
scalarXw/.style={postaction={decorate}, dashed, dash pattern = on 4pt off 2pt, dash phase = 2pt, decoration={markings, mark=at position .60 with{\arrow[scale=1]{stealth}}}}
}

 \newcommand{\doublewheelsmall}{
   \begin{minipage}[c]{1cm}
     
     \begin{center}
       \begin{tikzpicture}[scale=0.3]
         \foreach \m in {1,2} {
           \draw (0.75*\m,0) arc[radius = 0.75*\m,start angle = 0, end angle = 300] ;
           \draw[black, densely dotted] (300:0.78*\m) arc[radius = 0.75*\m,start angle = -60, end angle = 0];
         }
         \foreach \t in {1,2,...,5} {
           \draw (0,0) -- (60*\t:1.5);
         }
         \draw[black,densely dotted] (0,0) -- (0:1.5);
       \end{tikzpicture}
     \end{center}
   \end{minipage}
 }


\newcommand{\footnoteab}[2]{\ifbibtexelse{%
\footnotetext{#1}%
\footnotetext{#2}%
\cite{Note1,Note2}%
}{%
\newcommand{\textfootnotea}{#1}%
\newcommand{\textfootnoteab}{#2}%
\cite{thefootnotea,thefootnoteab}}}

\definecolor{green(pigment)}{rgb}{0.0, 0.65, 0.31}
\def\e{\epsilon}
     \def\bT{{\bf T}}
    \def\bQ{{\bf Q}}
    \def\wT{{\mathbb{T}}}
    \def\wQ{{\mathbb{Q}}}
    \def\ttQ{{\bar Q}}
    \def\tQ{{\tilde \bP}}
        \def\bP{{\bf P}}
        \def\dq{{\dot q}}
    \def\CF{{\cal F}}
    \def\cC{\CF}
    
     \def\l{\lambda}
\def\hbZ{{\widehat{ Z}}}
\def\bZ{{\resizebox{0.28cm}{0.33cm}{$\hspace{0.03cm}\check {\hspace{-0.03cm}\resizebox{0.14cm}{0.18cm}{$Z$}}$}}}

\definecolor{forestgreen(traditional)}{rgb}{0.0, 0.27, 0.13}

\title{Real-Time Scattering in Ising Field Theory using Matrix Product States}

\author{Raghav G.~Jha}
\affiliation{Thomas Jefferson National Accelerator Facility, Newport News, VA 23606, USA}
\author{Ashley Milsted}
\affiliation{AWS Center for Quantum Computing, Pasadena, CA 91125, USA}
\author{Dominik Neuenfeld} 
\affiliation{Institute for Theoretical Physics and Astrophysics, Julius-Maximilians-Universität Würzburg, Am Hubland, 97074 Würzburg, Germany}
\author{John Preskill}
\affiliation{AWS Center for Quantum Computing, Pasadena, CA 91125, USA}
\affiliation{Institute for Quantum Information and Matter, California Institute of Technology, Pasadena, CA 91125, USA}
\author{Pedro Vieira}
\affiliation{Perimeter Institute for Theoretical Physics, Waterloo, Ontario N2L 2Y5, Canada}
\affiliation{ICTP South American Institute for Fundamental Research, IFT-UNESP, S\~ao Paulo, SP Brazil 01440-070}

\begin{abstract}
\vspace{5mm}
We study scattering in Ising Field Theory (IFT) using matrix product states and the time-dependent variational principle. IFT is a one-parameter family of strongly coupled non-integrable quantum field theories in 1+1 dimensions, interpolating between massive free fermion theory and Zamolodchikov's integrable massive $E_8$ theory. Particles in IFT may scatter either elastically or inelastically. In the post-collision wavefunction, particle tracks from all final-state channels occur in superposition; processes of interest can be isolated by projecting the wavefunction onto definite particle sectors, or by evaluating energy density correlation functions. Using numerical simulations we determine the time delay of elastic scattering and the probability of inelastic particle production as a function of collision energy. We also study the mass and width of the lightest resonance near the $E_8$ point in detail. Close to both the free fermion and $E_8$ theories, our results for both elastic and inelastic scattering are in good agreement with expectations from form-factor perturbation theory. Using numerical computations to go beyond the regime accessible by perturbation theory, we find that the high energy behavior of the two-to-two particle scattering probability in IFT is consistent with a conjecture of Zamolodchikov. Our results demonstrate the efficacy of tensor-network methods for simulating the real-time dynamics of strongly coupled quantum field theories in 1+1 dimensions.
\\
\\ 
\rule{0.8\textwidth}{0.4pt}
 \begin{center}
\includegraphics[scale=0.7]{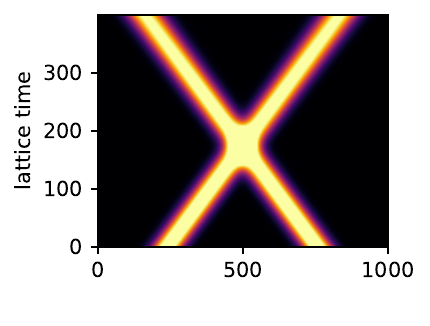}
\includegraphics[scale=0.7]{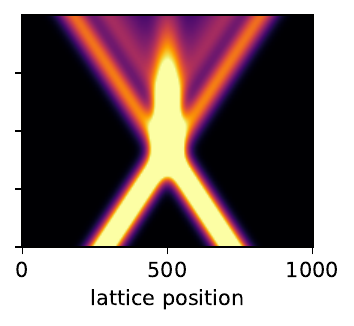}
\includegraphics[scale=0.7]{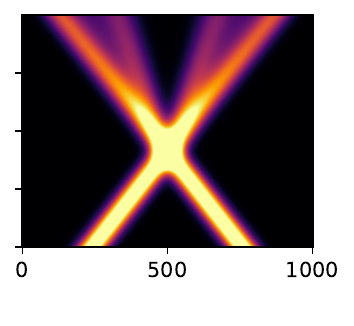}
    \includegraphics[scale=0.7]{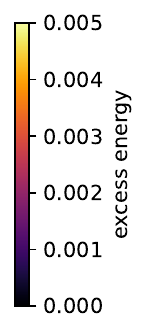}
    \hspace*{0.15\linewidth}\parbox{0.235\linewidth}{(a)}\parbox{0.235\linewidth}{{(b)}}\parbox{0.235\linewidth}{{(c)}}
    {\normalfont
    \captionof{figure}{Scattering processes can be visualized by computing energy expectation values at every time step. There are three characteristic scattering processes. (a) Elastic scattering at low energies, (b) scattering near a resonance, (c) inelastic scattering above threshold. Those processes allow us to extract the momentum dependence of the S-matrix via time delays,  the masses and widths of resonances, and inelastic scattering probabilities, respectively. 
    }
    \label{fig:two_tracks}
    }
\end{center}
\rule{0.8\textwidth}{0.4pt}
 {}\vspace{2em}
\end{abstract}    
\maketitle

\section{Introduction}
Much of what we know about fundamental particle physics has been inferred from collider experiments in which high-energy collisions produce many outgoing particles. However, our ability to analyze these processes from fundamental principles is currently restricted due to the lack of theoretical tools. Feynman diagram methods can be effective in the weak-coupling regime of a quantum field theory, and in some special cases scattering data can be extracted from Euclidean lattice QCD computations \cite{Luscher:1990ux,He:2005ey}, but real-time analysis of particle production at strong coupling and high energy is largely beyond the reach of existing methods. For example, there is no known technique for computing the high energy behavior of glueball-glueball scattering in pure Yang-Mills theory. 

The situation is less dire for strongly coupled quantum field theory in 1+1 spacetime dimensions. We can succinctly represent the state of a quantum spin chain as a matrix product state (MPS), and if the state does not become too highly entangled, we can efficiently simulate its evolution under a local Hamiltonian. Furthermore, relatively simple spin chains close to criticality can approximate a variety of nonintegrable continuum quantum field theories. By studying elastic and inelastic scattering processes in such field theories using the MPS method, we can hope to address some longstanding open questions about these processes.

In this work, we demonstrate the feasibility of this strategy by using MPS methods to study Ising Field Theory (IFT), a parameterized family of massive continuum theories obtained by deforming the critical two-dimensional Ising model. The 
action of IFT is
\begin{equation}\label{eq:eta-definition}
I_{\text{IFT}} = I_{\text{ICFT}} - \tau \int d^{2}x \, \epsilon(x) - h \int d^{2}x \, \sigma(x), 
\end{equation}
where $I_{\text{ICFT}}$ is the action of the Ising conformal field theory (ICFT) and $\epsilon$ and $\sigma$ are the relevant energy and spin operators with scaling dimensions $\Delta_{\epsilon} = 1$ and $\Delta_{\sigma} = \frac 1 8$ respectively. 
The term in Eq.~\eqref{eq:eta-definition} that contains the energy operator $\epsilon$ is called the thermal deformation of ICFT because it arises from perturbing the temperature in the Ising model away from the critical temperature. The term that contains the spin operator $\sigma$ is called the magnetic deformation of ICFT because it arises from a $\mathbb{Z}_{2}$-symmetry-breaking external magnetic field in the Ising model. Because $\tau$ has scale dimension 1 (in units of mass) and $h$ has scale dimension $15/8$, we characterize the relative weight of the two perturbations using the dimensionless ratio 
\begin{equation}
\label{eq:cont_eta}
    \eta \equiv \frac{\tau}{|h|^{8/15}} 
\end{equation}
where $\tau \propto T-T_{c}$. Our conventions are the same as in Ref.~\cite{Delfino:2005bh}.

For $\eta=\infty$ (purely thermal deformation, $h=0$), IFT is the theory of a massive free fermion for which scattering is trivial. For $\eta=0$ (purely magnetic deformation, $\tau =0$), Zamolodchikov \cite{Zamolodchikov:1989hfa, zamolodchikov1989integrals} found that IFT is an integrable theory with 
eight stable particles called the $E_8$ theory.\footnote{This integrable limit is commonly referred to as the $E_{8}$ limit since the masses of the eight stable particles found by Zamolodchikov match the components of the Perron-Frobenius vector (the eigenvector corresponding to the largest eigenvalue) of the Cartan matrix of the $E_{8}$ Lie algebra; the structure of the S-matrices reflects this underlying symmetry. } The five heaviest particles lie above the two-particle threshold and are stable only due to integrability. The peculiar excitation spectrum of IFT close to the $E_8$ limit has been observed experimentally in quasi-one-dimensional Ising ferromagnets probed by neutron scattering~\cite{Coldea2010}. 

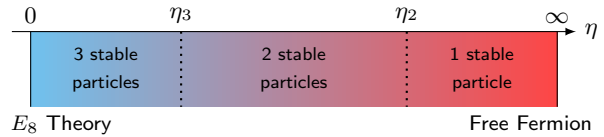
\begin{figure}[h!]
    \centering
    \hspace*{-3em}
    \begin{tikzpicture}[font=\sffamily]
    \fill [left color=gradientblue, right color=gradientFF] (0,0) rectangle (7,-1);
    \draw[-latex] (-0.25,0) -- (7.25,0) node[right] {$\eta$};
    \draw (0,0) node [above]{$0$} -- (0,-1) node[below] {\hspace{2.5em}\footnotesize $E_8$ Theory};
    \draw (7,0) node [above]{$\infty$} -- (7,-1) node[below] {\hspace{-2.5em} \footnotesize Free Fermion};
    \draw[black] (1,-0.5) node[align=center] {\scriptsize 3 stable\\\scriptsize particles};
    \draw[black] (3.5,-0.5) node[align=center] {\scriptsize 2 stable\\\scriptsize particles};
    \draw[black] (6,-0.5) node[align=center] {\scriptsize 1 stable\\\scriptsize particle};
    \draw[dotted, thick] (2,0) node [above] {$\eta_3$} -- (2,-1);
    \draw[dotted, thick] (5,0) node [above] {$\eta_2$} -- (5,-1);
    \end{tikzpicture}
    \caption{IFT interpolates between the integrable $E_8$ theory ($\eta=0$) and massive free field theory ($\eta=\infty$). As $\eta$ increases, the number of stable particles changes from 3 to 2 at $\eta_3 \approx 0.022$ and changes from 2 to 1 at $\eta_2 \approx 0.333$~\cite{Delfino:2005bh}.}
    \label{fig:eta}
\end{figure}

For intermediate values of $\eta$, IFT is non-integrable and strongly coupled. The number of stable particles in IFT is three for small $\eta$ (close to $E_8$ theory), one for large $\eta$ (close to the free fermion theory), and two for intermediate values of $\eta$, as indicated in Fig.~\ref{fig:eta}.
IFT has been widely explored in previous works \cite{PhysRevD.18.1259, Delfino:1995zk, Fonseca:2001dc, Delfino:2007qe, Gabai:2019ryw, Fitzpatrick:2023aqm}. 

\begin{table}[h!]
    \centering
    \includegraphics[scale=0.65]{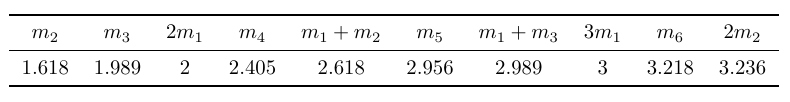}
    \caption{Mass spectrum at $E_{8}$ and multi-particle thresholds in units where $m_{1} = 1$.}
    \label{fig:E8_spec_det}
\end{table}

IFT arises in the scaling limit of a simple spin-chain model with Hamiltonian
\begin{equation}
\label{spinChain}
H = -\sum_{j=0}^{N}\Big(\sigma^{z}_{j} \sigma^{z}_{j+1}  + g_x \sigma^{x}_{j} + g_z \sigma^{z}_{j}\Big),  
\end{equation} 
where $g_x = T/T_{c}$ and $g_z$ is the $\mathbb{Z}_{2}$-symmetry-breaking magnetic field. $H$ generates time translations in lattice time $t_\text{latt}$. The low-lying excitations are described by IFT as the couplings approach the critical point, $g_x \to 1$ and $ g_z \to 0$; the resulting continuum theory may be parametrized by  
\begin{equation}
    \label{eq:lattice_eta}
    \eta_{\text{latt}}= \frac{g_x{-}1}{~|g_z|^{8/15}},
\end{equation} 
which is to be held fixed as the critical point is approached.
The relationship between the continuum $\eta$ defined in Eq.~\eqref{eq:cont_eta} and $\eta_{\text{latt}}$ is not obvious but can be determined in principle by comparing continuum and lattice computations. See App.~\ref{app:Ising-chain_and_conversion} for further details. Lattice energies, momenta, times and positions need to be converted to their dimensionful ``physical'' continuum counterparts as explained in App.~\ref{app:conversion}.

An object of fundamental interest in any gapped quantum field theory is the S-matrix, which encodes the particle spectrum and couplings. Therefore, following the previous work of some of the authors \cite{Milsted:2020jmf}, we use MPS methods to study the scattering of wave packets in this spin chain. 

We can describe a scattering experiment involving two particles of ``type 1'' (the lightest particle in the theory) at center-of-mass energy $E$ as
\begin{align}
    \ket{p_1, p_2}_\text{in} ={ } &S_{11\to 11}(E; \eta) \ket{p_1, p_2}_\text{out} + \sum_{n=2}^\infty \left(\prod_{i=1}^n \int \frac{dq_i}{4\pi E_i} \right)  \nonumber \\
    & \times \frac{(2\pi)^2}{\mathfrak{s}_n} \delta^{(2)}(P_\text{in} - P_\text{out}) \\ 
    & \times S_{11 \to X_n}(p_1, p_2, q_1, \dots, q_n ; \eta) \ket{q_1, \dots, q_n}_\text{out}, \nonumber
\end{align}
where the subscripts of the states denote that they are defined in the asymptotic in- and out-regions where particles do not interact.
The first term on the right-hand side describes elastic scattering; in two dimensions, energy-momentum conservation requires that for elastic two-particle scattering the incoming and outgoing momenta agree. The remaining terms describe inelastic scattering into states with two or more particles. 
$\mathfrak s_n$ is a symmetry factor that accounts for the indistinguishability of outgoing particles. The delta-function imposes energy-momentum conservation. We will refer to $S_{11 \to \dots}$ as S-matrix elements.\footnote{In higher dimensions the quantities $S_{11 \to X}$ are often called matrix elements.} They are analytic functions of the involved momenta. We have also explicitly indicated a dependence on the couplings through $\eta$.
 
Using our MPS simulation methods we are not able to obtain the full S-matrix. We can, however, obtain two observables that are sensitive to the magnitude and phase of the S-matrix: scattering probabilities and time delays. For incoming wave packets that are well-localized in momentum space with a center-of-mass energy around $E$, the probability for an elastic collision is given by integrating the absolute value squared of the S-matrix element $S_{11\to 11}(E,\eta)$ against the incoming wave packets,
\begin{align}
\begin{split}
    \label{eq:s-matrix_to_prob}
    P_{11\to 11}(E,\eta) & = \int dE' \rho(E'; E) |S_{11\to 11}(E',\eta)|^2 \\ & \approx |S_{11\to 11}(E,\eta)|^2.
\end{split}
\end{align}
The last equality holds if the $S$-matrix element is approximately constant on scales set by the variance of the incoming energy probability density $\rho(E';E)$. To compute the probability $P_{11 \to X}(E, \eta)$ of an inelastic process with final state $X$, the square of the S-matrix element in Eq.~\eqref{eq:s-matrix_to_prob} needs to be replaced by the total cross-section $\sigma(E)_{11 \to X}$ which can be computed from the S-matrix element $S_{11 \to X}(p_1, p_2, q_1, \dots, q_n ; \eta)$ using standard textbook methods. Following the same logic as above, for a slowly varying inelastic cross-section one can show that
\begin{align}
\begin{split}
    P_{11\to X}(E,\eta) \approx \sigma(E, \eta)_{11 \to X}.
\end{split}
\end{align}

Through our simulations we also have access to the time delay,
\begin{align}
    \label{eq:def_time_delay}
    \Delta t = - i \partial_E \log S_{11\to 11}(E,\eta),
\end{align}
which roughly captures how trajectories are displaced when particles scatter elastically and was first studied by Wigner \cite{PhysRev.98.145}; see App.~\ref{app:time_delay} for details. By extracting the time delay from the scattering process, we can reconstruct the derivative of the phase of $S_{11\to 11}$ with respect to the incoming momenta. In Sec.~\ref{sec:methods} we describe in detail how both these quantities can be extracted from our simulations. Depending on the scattering energy, a variety of effects are visible and our numerical results can then be compared to theoretical predictions. 

MPS methods can access features of the S-matrix of strongly coupled ($1{+}1$)-dimensional quantum field theories which are beyond the reach of other computational methods. In addition, aside from capturing properties of asymptotic scattering states encoded in the S-matrix, these methods provide an engaging visualization of scattering processes, with high resolution in both space and time, such as Fig.~\ref{fig:two_tracks}. By unveiling aspects of real-time evolution that are not easily extracted from more traditional quantum field theory calculations or from typical scattering experiments, tensor-network simulations can strengthen our intuition about field-theory dynamics and stimulate us to raise and probe new questions concerning strong-coupling phenomena. However, the correct interpretation of the images produced in the simulations is not always straight-forward and in Sec.~\ref{sec:multiverses} we explain the correct reading of our figures.

Whenever we have theoretical predictions to compare to, our simulations achieve a good approximation to scattering in IFT if the correlation length is large compared to the lattice spacing and our wave packets are broad compared to the correlation length (see Fig.~\ref{fig:MPS_diagrams}). In practice, the correlation length for all our simulations is in the range of $4 - 10$ lattice sites, the wave packets typically extend over $70-120$ sites, and the simulated window spans $1000 - 2000$ sites.

Consider the image in Fig.~\ref{fig:two_tracks}a, which depicts a simulated scattering event at low energies, in the kinematic regime where only elastic scattering is allowed. In this regime, $S_{11\to11}$ is merely an energy-dependent phase. We can extract the energy dependence of the time delay in our simulations. In particular, when IFT is close to an integrable theory we find that it agrees with form-factor perturbation theory; see Secs.~\ref{sec:nff} and \ref{sec:ne8}.

As noted, the integrable $E_8$ theory contains stable particles with masses above the two-particle threshold; the masses of three of these particles are shown in Table \ref{fig:E8_spec_det}. In the regime of IFT where $\eta$ is nonzero and small, these particles become narrow resonances which leave an imprint on the time delay. When the total energy is close to the resonant energy, real-time evolution produces a lump of energy in the scattering region that dissipates slowly and produces a set of secondary tracks as shown in Fig.~\ref{fig:two_tracks}b. As discussed in Sec.~\ref{sec:resonances} the lifetime of the resonance can be inferred from the rate of decay of this lump. We used a variant of this method to determine the mass $m_4$ and the width $\Gamma_4$ of the lightest resonance of IFT as a function of $\eta$. We find good agreement with perturbative calculations to leading order in $\eta$ when $\eta$ is small, and also obtain results for values of $\eta$ well beyond the domain of validity of those perturbative estimates, in good agreement with existing results \cite{Gabai:2019ryw}.

At sufficiently high energy, particle production can occur. Consider for example a collision of two particles of type 1 at $\eta$ roughly midway between the two integrable endpoints with center-of-mass energy greater than $m_1 + m_2$, the sum of the masses of the stable particles of types 1 and 2. Fig.~\ref{fig:two_tracks}c depicts the post-collision energy density as a function of space and time. The tracks shown provide clear evidence of particle production. From our simulations, we can extract the scattering probabilities and near the integrable points compare them to predictions from form-factor perturbation theory. We do so in Secs.~\ref{sec:nff} and \ref{sec:ne8} and again find good agreement.

However, due to the strongly coupled nature of IFT, for a large swath of the $\eta$ parameter space no theoretical predictions are available for comparison. Thus using MPS methods to address questions about IFT in this regime is particularly informative. For energies above the threshold for inelastic particle production $P_{11\to11}(E,\eta)$ can be less than one. A natural question concerns the value of
\begin{equation}
    P^\infty(\eta)= \lim_{E\to\infty}P_{11\to11}(E,\eta). 
\end{equation}

Is this quantity zero (inelastic scattering dominates at high energy), one (elastic scattering dominates), or some other value? Since the
free fermion theory and the $E_8$ theory are both integrable, we know that $P^\infty(\infty)= P^\infty(0) = 1$, but the answer for intermediate values of $\eta$ is not obvious. We have aimed to address this question by computing $P_{11\to11}(E,\eta)$ for values of $E$ up to several times the inelastic scattering threshold in Sec.~\ref{sec:high}. Our results are consistent with a proposal by Zamolodchikov~\cite{Zam_2013} which implies that $P^\infty(\eta) = 0$ for $\eta < \eta_c$ and $P^\infty(\eta) = 1$ for $\eta > \eta_c$,  where $\eta_c$ is a critical value of $\eta$.


\section{Methods}
\label{sec:methods}
\subsection{Matrix Product States}
This paper concerns the scattering of low-lying excitations of gapped, one-dimensional spin chains in the thermodynamic limit. At every site $i$, the chain hosts a spin $s_i$ which lives in a $d$-dimensional Hilbert space $\mathcal H_i \cong \mathbb C^d$. Since gapped systems have finite correlation length, these excitations can be efficiently captured by matrix product states (MPS) \cite{Perez-Garcia2006-um,Klumper1993-sy,Fannes1992-nz}, which take the form
\begin{align}
    \label{eq:mps}
        \ket \psi = \sum_{\{\mathbf s\}} \left( \vec v_L^\dagger \dots A_{-1}^{(s_{-1})} A_{0}^{(s_{0})} A_{1}^{(s_{1})} A_{2}^{(s_{2})} \dots \vec v_R \right)  \ket{ \mathbf s },
\end{align}
where
\begin{align}
    \ket{\mathbf s} = \ket{\dots s_{-1}s_0s_1s_2\dots}
\end{align}
denotes a computational basis state of the spin chain.
Here, $A^{(s_n)}_n$ are complex $D_{n-1} \times D_{n}$ matrices. The parameters $D_{n}$ are called \emph{bond dimensions} and equal the maximal number of non-zero eigenvalues of the reduced density matrix of a bipartition between sites $n$ and $n-1$. Thus, $\log D_n$ gives an upper bound on the von Neumann entropy between spins $s_l$ with $l\leq n$ and spins $s_r$ with $r >n$. While this limits the class of states that can be represented as an MPS, in practice we can choose all $D_n$ large enough such that for our scattering ``experiments'' the resulting states are well-approximated by the ansatz~\eqref{eq:mps}. The $D_{\pm \infty}$-dimensional vectors $\vec v_L$ and $\vec v_R$ live at spatial infinity. In the generic case, due to exponentially decaying correlations, their values do not affect expectation values or dynamics and are thus immaterial. As is not hard to see from~\eqref{eq:mps}, a given state has many different MPS representations related by $A^{(s_n)}_n \mapsto g_{n-1} \;A_n^{(s_n)} \; g_n^{-1}$. This arbitrariness is fixed by imposing gauge conditions on $A_n$~\cite{PhysRevLett.100.167202,https://doi.org/10.48550/arxiv.quant-ph/0608197}, leaving $D^2 (d-1)$ degrees of freedom. Moreover, these gauge transformations can be used to bring the $A_n$ into form which makes numerical evaluation of operator expectation values in MPS particularly efficient. For a pedagogical introduction see for example~\cite{Vanderstraeten:2019voi}.

\subsection{Construction of Initial States}
To study scattering we follow a strategy inspired by real-life experiments. We prepare an initial state of a pair of well-separated and localized particles with opposite momenta on top of the MPS approximation of the ground state. Those excitations approach each other and scatter in a finite region. We let excitations emanating from the scattering region propagate freely until they are sufficiently well separated and study the properties of the resulting outgoing state, such as particle content and time delays. Due to the mass gap the initial and final states are well approximated by a superposition of products of one-particle states. Therefore, representations of single-particle states suffice for constructing incoming states and analyzing outgoing states, including correlations among outgoing particles.

\paragraph{Ground State.} To find the ground state approximation $\ket \Omega$ of the spin chain we consider a translationally-symmetric ansatz of the general form given in Eq.~\eqref{eq:mps} where all tensors are chosen to be identical, i.e., $A_n^{(s_n)} = A^{(s_n)}$ with bond dimension $D_\text{vac}$. In practice, we choose $D_\text{vac}$ such that the smallest Schmidt coefficients are about $10^{-8}$; $D_\text{vac} = 24$ typically suffices. We then minimize the energy in this state with respect to $A$.\footnote{From here on, we will suppress the upper index of $A^{(s_n)}$ and simply write this site-independent tensor as $A$.} The resulting state is close to the ground state. For the numerical computations in this paper we use the evoMPS software \cite{evomps} which implements the conjugate-gradient method \cite{Milsted:2013rxa} for optimizing $A$.

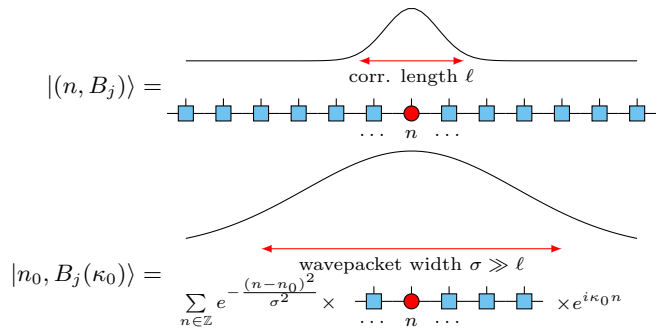
\begin{figure}
    \centering
    \begin{tikzpicture}
    
    \begin{scope}[yshift=2.5cm]
    \draw (0.3,0.35) node [left] {$\ket{(n,B_j)} = $};
    \foreach \i in {1,2,3,4,5,6,8,9,10,11,12,13}
    {
        \draw[fill=gradientblue] (0.5*\i-0.1, -0.1) rectangle (0.5*\i+0.1, 0.1);
       
        \draw (0.5*\i, 0.1) --  (0.5*\i, 0.2);
        \draw (0.5*\i+0.1, 0) --  (0.5*\i+0.25, 0);
         \draw (0.5*\i-0.1, 0) --  (0.5*\i-0.25, 0);
    }
    \draw[fill=red] (0.5*7, 0) circle (0.1);
    \draw (0.5*7,-0.1) node [below] {\scriptsize $n$};
    \draw (0.5*7, 0.1) --  (0.5*7, 0.2);
    \draw (0.5*7+0.1, 0) --  (0.5*7+0.25, 0);
    \draw (0.5*7-0.1, 0) --  (0.5*7-0.25, 0);
    
    \draw (0.5*6,-0.15) node [below] {\scriptsize $\dots$};
    \draw (0.5*8,-0.15) node [below] {\scriptsize $\dots$};
    \begin{scope}[yshift = 0.7cm]
    \draw[latex-latex, red] (2.8,0) -- (4.2,0);
    \draw (3.5,0) node [below] {\scriptsize corr. length $\ell$};
    \draw[domain=0.5:6.5, samples=70] plot (\corrlen) node[right] {};
    \end{scope}
    \end{scope}

    \begin{scope}
    \draw (0.3,0.35) node [left] {$\ket{n_0,B_j(\kappa_0)} = $};
    
    \draw (0.3,0) node [right] {\scriptsize $\sum\limits_{n \in \mathbb Z} e^{-\frac{(n - n_0)^2}{\sigma^2}} \times$};
    \draw (5.3,0) node [right] {\scriptsize $\times e^{i \kappa_0 n}$};
    \foreach \i in {6,8,9,10}
    {
        \draw[fill=gradientblue] (0.5*\i-0.1, -0.1) rectangle (0.5*\i+0.1, 0.1);
       
        \draw (0.5*\i, 0.1) --  (0.5*\i, 0.2);
        \draw (0.5*\i+0.1, 0) --  (0.5*\i+0.25, 0);
         \draw (0.5*\i-0.1, 0) --  (0.5*\i-0.25, 0);
    }
    \draw[fill=red] (0.5*7, 0) circle (0.1);
    \draw (0.5*7,-0.1) node [below] {\scriptsize $n$};
    \draw (0.5*7, 0.1) --  (0.5*7, 0.2);
    \draw (0.5*7+0.1, 0) --  (0.5*7+0.25, 0);
    \draw (0.5*7-0.1, 0) --  (0.5*7-0.25, 0);
    
    \draw (0.5*6,-0.15) node [below] {\scriptsize $\dots$};
    \draw (0.5*8,-0.15) node [below] {\scriptsize $\dots$};
    \begin{scope}[yshift = 0.7cm]
    \draw[latex-latex, red] (1.5,0) -- (5.5,0);
    \draw (3.5,0) node [below] {\scriptsize wavepacket width $\sigma \gg \ell$};
    \draw[domain=0.5:6.5, samples=70] plot (\wavepacket) node[right] {};
    \end{scope}
    \end{scope}
    
    \end{tikzpicture}
    \caption{Pictorial representation of MPS states. MPS tensors are represented by blue squares (vacuum $A$ tensor) or red circles (excitation $B$ tensor). The horizontal lines represent the contraction of the bond dimension indices. The vertical black lines correspond to spin degrees of freedom. Upper figure: A single excitation tensor affects expectation values about a correlation length away from its insertion locus. Lower figure: Our scattering states are wave packets built from superpositions of localized basis states with $\sigma \gg \ell$.}
    \label{fig:MPS_diagrams}
\end{figure}

\paragraph{1-Particle Excited States.} To construct one-particle states, we start from the vacuum state and replace the vacuum tensor $A$ at a single site $n$ by an independent \emph{excitation tensor} $B$, following \cite{Haegeman:2011lcd} as shown in Fig.~\ref{fig:MPS_diagrams}. Such an ansatz represents a quasi-local excitation around $n$. Its effect on correlation functions is not confined to the site $n$, but dies off exponentially away from that site with a rate set by the correlation length $\ell$.

We are interested in identifying $(D^2_\text{vac} \times d)$-dimensional tensors $B$ that yield excitations corresponding to the stable particles of the theory. To achieve this we construct a state of fixed momentum $\kappa \in (-\pi, \pi]$ by performing a Fourier transformation with respect to the location of $B$,
\begin{align}
\label{eq:momentum_eigenstate}
    \ket {(\kappa, j)} = {}&\sum_{n \in \mathbb Z} \sum_{s_n} e^{i \kappa n} \vec v_L^\dagger (\dots A^{(s_{n-1})} B_j^{(s_n)} A^{(s_{n+1})} \dots) \vec v_R \ket{ \mathbf s }\nonumber \\
    =: {}&\sum_{n \in \mathbb Z} e^{i \kappa n} \ket{ (n, B_j) },
\end{align}
where the last step defines $\ket{ (n,B_j) }$ as the state obtained from the vacuum state by replacing the vacuum tensor at site $n$ by $B_j$. The additional index $j$ enumerates different particle species with mass below the two-particle threshold. 
Note that states of the form Eq.~\eqref{eq:momentum_eigenstate} do not generally live in the MPS manifold defined by Eq.~\eqref{eq:mps} but in its tangent space at the vacuum \cite{Haegeman2013-tg}. Analogous to the gauge freedom in~\eqref{eq:mps}, $B$ has the gauge freedom
\begin{align}
    \label{eq:gauge_freedom_B}
    B_j^{(s_n)} \mapsto B_j^{(s_n)} + A^{(s_n)} X - e^{- i \kappa }X A^{(s_n)},
\end{align}
where $X$ is an arbitrary $D_\text{vac} \times D_\text{vac}$ matrix whose value does not affect the momentum eigenstates. The gauge freedom can be fixed such that $\langle (n,B_j) | (m,B_k) \rangle = \delta_{jk} \delta_{nm}$, with $\ket{ (n,B_j) }$ forming an orthonormal ``position basis'' for excitations of species $j$.

In order to construct $B_j^{(s_n)}$ we can now demand that the momentum eigenstates \eqref{eq:momentum_eigenstate} are orthogonal and delta-function normalized, while at the same time requiring that the effective Hamiltonian $\bra{(\kappa',j')} H - H_\text{vac} \ket{(\kappa, j)} \propto H_{\kappa,\kappa'}^{j,j'}$ for states of the form Eq.~\eqref{eq:momentum_eigenstate} is diagonal, thus obtaining a basis of approximate eigenstates of the Hamiltonian; here $H_\text{vac}$ denotes the vacuum energy. The diagonal entries of $H_{\kappa,\kappa'}^{j,j'}$ then give the energy $E_j(\kappa)$ of excitation $j$ at momentum $\kappa$ \cite{Haegeman:2011lcd}. The corresponding tensors $B_j(\kappa)$ generally depend not only on the particle species $j$, but also on the momentum $\kappa$. 

\paragraph{2-Particle Scattering States.} To obtain particles that are well localized before and after scattering, we can construct superpositions of energy eigenstates to obtain finite-width wave packets centered at site $n_0$ and with central momentum $\kappa_0$,\footnote{Here the parameter $\sigma$ characterizing the width of the wave packet is defined following the conventions of the evoMPS code~\cite{evomps}. The probability distribution governing the particle's momentum is a Gaussian with variance $\sigma^{-2}$. In position space the wave packet has variance $\sigma^2/ 4$.}
\begin{align}\label{eq:finite-width}
    \ket{n_0, \kappa_0, j} = \sum_{n \in \mathbb Z} e^{-\frac{(n-n_0)^2}{\sigma^2}} e^{i \kappa_0 n} \ket{(n, B_j(\kappa_0))};
\end{align}
see Fig.~\ref{fig:MPS_diagrams}

Note that here we have fixed the argument of the excitation tensor to be $\kappa_0$. However, since we use this excitation tensor to construct a localized excitation, the resulting state is necessarily a superposition of momentum eigenstates with a finite spread around the central value $\kappa_0$. Furthermore,
the localized wave packets are not invariant under \eqref{eq:gauge_freedom_B}. Fortunately, the momentum dependence of the excitation tensors can be safely neglected if the wave packets are suitably broad in position space (compared to the correlation length $\ell$) and correspondingly narrow in momentum space. 
Spatially broad wave packets are desirable in any case, both to reduce lattice artifacts and to characterize the momentum dependence of the S-matrix. 
Our wave packets have a width of $\sigma\approx 70-120$ lattice sites and are constructed over around $400-800$ sites, wide enough for Eq.~\eqref{eq:finite-width} to be an adequate approximation given that in our simulations $\ell$ is $4 - 10$ lattice sites.

Importantly, we can also approximate multi-particle states by inserting copies of the excitation tensor at several sites and convoluting them with a Gaussian. For example, if we want to scatter two incoming particles of mass $m_1$ with opposite momenta $\pm \kappa_0$, the in-state is constructed as 
\begin{align}
    \label{eq:in_state}
\begin{split}
    \ket{\text{in}} = \sum_{n < n'} e^{-\frac{(n-n_0)^2}{\sigma^2}} e^{-\frac{(n'-n'_0)^2}{\sigma'^2}} \times \\ e^{i \kappa_0( n - n')}\ket{(n, B_1(\kappa_0)),(n', B_1(-\kappa_0))},
\end{split}
\end{align}
Here, $\ket{(n,B_j),(n',B_j')}$ is defined analogously to $\ket{(n,B_j)}$ as the state obtained from the vacuum state after replacing tensors at locations $n$ and $n'$ by tensors $B_j^{(s_n)}$ and $B^{(s_{n'})}_{j'}$, respectively. This will be a good approximation to a two-particle state as long as the wave packets are sufficiently wide and sufficiently far separated. In practice, the states are constructed in a window of size $N$ and the wave packets are truncated when $n > N/2$ ($n' <  N/2$). At this point, the Gaussian coefficients are less than $10^{-6}$ and the error is negligible. This construction allows us to represent a state of the form \eqref{eq:in_state} as an MPS with bond dimension $2 D_\text{vac}$ \cite{Milsted:2020jmf}.

\subsection{Time Evolution}
To keep track of the evolution we define a window on the spin chain encompassing $N$ spins. For our simulations, $N$ is between $1000$ and $2000$. Within this window, we use the non-uniform ansatz \eqref{eq:mps} and make the tensor coefficients time-dependent. Outside this window, the tensors are fixed to be vacuum tensors $A$. The tensors within the window change to approximate the evolved state, with their bond dimensions increasing up to a maximum allowed value $D_\text{max}$; a typical choice for our simulations is $D_\text{max} = 64$. We require equations of motion $ \dot A(t)$ for the MPS tensors so that
\begin{align}
    \label{eq:schrodinger}
    \begin{split}
    i \partial_t \ket{\psi(\mathbf A(t))} &= i \dot A(t)^I \ket{\partial_I \psi(\mathbf A(t))}\\\
    &\simeq H \ket{\psi(\mathbf A(t))},
    \end{split}
\end{align}
where $I$ is a multi-index that runs over the entries of the tensors that are allowed to change, and where we have abbreviated $\frac{\partial}{\partial A^I} \equiv \partial_I$.

To solve this problem, we evolve the scattering states in time using the time-dependent variational principle (TDVP) algorithm for MPS developed in \cite{PhysRevLett.107.070601,Milsted_2013,PhysRevB.94.165116} and implemented in \cite{evomps}. This approximates the time evolution by projecting $ H \ket{\psi}$ onto the tangent space spanned by $\ket{\partial_I \psi}$,
\begin{align}
\label{eq:tdvp1}
    \braket{\partial_J \psi|\partial_I \psi} \dot A(t)^I = -i \braket{\partial_J \psi | H | \psi}.
\end{align}
Inverting the Gram matrix $\braket{\partial_J \psi|\partial_I \psi}$ then allows us to solve for $\dot A(t)^I$, and use the finite element method to determine the change in $A(t)^I$. In practice, the gauge freedom for our ansatz can be fixed in such a way that the Gram matrix becomes the identity matrix.

At time $t=0$ we start with a two-particle in-state of the form Eq.~\eqref{eq:in_state}. Initially, a split-step integrator \cite{PhysRevB.94.165116} is used which increases the bond dimension from $2D_\text{vac}$ to some $D_\text{max}$. During the subsequent time evolution at fixed bond dimension, we use the Runge-Kutta (RK) 4/5 integrator \cite{Milsted_2013}. The step-size is denoted by $dt_\text{latt}$ such that after $N_\text{t}$ time steps the lattice time $t_\text{latt}$ is given by $t_\text{latt} = N_\text{t} \, dt_\text{latt}$ and can be converted to physical time using the method of App.~\ref{app:conversion}. In our simulations $dt_\text{latt} = 0.05$ or smaller.

\subsection{Phases and Probabilities}
\label{sec:probabilities}
In an ideal scattering scenario, the relation between the state at late and early times is described by the S-matrix
\begin{align}
\label{eq:s_matrix}
\begin{split}
     \ket {\text{out}} &= S \ket {\text{in}}
     \\ &= \lim_{t_{f/i} \to \pm \infty} e^{i H_0 t_f}e^{-iH(t_f - t_i)}e^{i H_0 t_i} \ket {\text{in}},
     \end{split}
\end{align}
where $H = H_0 + H_\text{int}$ is the full Hamiltonian of the system, $H_0$ is the free Hamiltonian that propagates well-separated stable particles and $H_\text{int}$ captures interactions.
Below the inelastic threshold the S-matrix is simply a momentum-dependent phase, $e^{i \phi}$. Our methods do not preserve global phases during time evolution and thus we cannot extract the phase shift. Nonetheless, we can extract the derivative of the phase shift with respect to the center of mass energy of the scattered particles, also known as the time delay. Above the threshold, we can determine probabilities for the scattering of a pair of the lightest particles to produce other stable excitations.

\paragraph{Time Delay.} \label{p:time_delay} A non-trivial momentum dependence of the S-matrix introduces a relative shift in the in- and out-going trajectories of particles which could in principle be extracted from particle tracks such as Fig.~\ref{fig:two_tracks} (a). This strategy has been implemented in the context of time evolution of MPS in \cite{Van_Damme:2021}. However, this approach is not feasible for the present work, partially due to the difficulty of even defining trajectories 
for very broad wave packets. 

We instead follow a different strategy and extract the time delay by projecting the scattering state at different times onto reference states of various momenta. The reference states we choose take a planewave-form,
\begin{align}
    \label{eq:reference_state}
    \ket{\kappa,\kappa'} = \sum_{|n - n'| \geq \Delta n_\text{min}} e^{i (\kappa n + \kappa' n')}\ket{(n,B_1(\kappa)),(n',B_1(\kappa'))}.
\end{align}
The sum runs over all insertions of the excitation tensors which are farther apart than some gap $\Delta n_\text{min}$. This gap ensures that there is no contribution coming from interactions amongst the excitation tensors. For the analyses of this paper we chose $\Delta n_\text{min} = 100$. The fact that for fixed $n$, the sum over $n'$ is cut off at some finite distance generally introduces high momentum contributions into the reference state. However, since we are only considering overlaps with low-energy states with localized excitations, those do not contribute to overlaps. In practice, we compute an overlap matrix $O_{n,n'}$ by taking simulated states and projecting onto localized states with two excitation tensors $\ket{(n,B_1(\kappa)),(n',B_1(\kappa'))}$. As for the one-particle case, the gauge freedom of Eq.~\eqref{eq:gauge_freedom_B} can be fixed for each excitation tensor so that these states form an orthonormal 2-particle ``position basis'': $\langle (n,B_1(\kappa)),(n',B_1(\kappa')) | (m,B_1(\kappa)),(m',B_1(\kappa')) \rangle = \delta_{nm}\delta_{n'm'}$ for $n<n'$, $m < m'$. We then Fourier transform the overlap matrix to obtain the overlaps of the outgoing state with states of the form Eq.~\eqref{eq:reference_state}.

To extract the time delay from overlaps at momentum $\kappa_0$ we consider nearby momenta $\kappa_\pm = \kappa_0 \pm \delta\kappa$ and compute
\begin{align}
    \label{eq:phases_eq}
    \begin{split}
    \phi(t) &= \arg\left(\frac{\braket{\kappa_+,-\kappa_+|\psi(t)}}{\braket{\kappa_+,-\kappa_+|\psi(0)}}\frac{\braket{\kappa_-,-\kappa_-|\psi(0)}}{\braket{\kappa_-,-\kappa_-|\psi(t)}}\right)\\
    &= 2(E_{\kappa_+} - E_{\kappa_-})t + \Delta(\kappa_+) - \Delta(\kappa_-),
    \end{split}
\end{align}
where $t$ is chosen large enough such that the particles are well separated after scattering.
The quantity $\phi(t)$ is independent of the global phase and captures the relative phase-shift between components of $\ket \psi$ with momenta $\kappa_\pm$. The first contribution to $\phi(t)$, which consists of the expected phase shift for a free theory which is known from the one particle spectrum $E_{\kappa_\pm}$, has to be subtracted. The time-independent contribution, $\Delta(\kappa_+) - \Delta( \kappa_-)$, is the phase-shift due to interactions. 
Assuming it to be slowly varying on scales $\delta \kappa$, we can expand $\Delta(\kappa_\pm)$ to first order around $\kappa_0$,
\begin{align}
    \label{eq:time_delay_eq}
    \frac{\Delta(\kappa_+) - \Delta(\kappa_-)}{2(E_{\kappa_+} - E_{\kappa_-})} \simeq \partial_E\Delta(E) = -i \partial_{E}\log S(E),
\end{align}
where we introduced the center-of-mass energy $E = 2 E_{\kappa_0}$ in the last step.
This result is then converted from lattice to physical units as explained in App.~\ref{app:conversion}. In this paper, we compare the time delay obtained this way with theoretical predictions from form-factor perturbation theory. The dependence of our results on the choice of the gap size $\Delta n_\text{min}$ and $\delta \kappa$ is discussed in App.~\ref{app:hyperparameter_timedelay}.

Narrow resonances, corresponding to S-matrix poles in the complex energy plane close to the real axis, can produce a sizable enhancement of the time delay. In Sec.~\ref{sec:resonances} we describe a method for estimating the mass and width of such a resonance by performing temporal and spatial fits to our simulated wavefunctions.

\paragraph{Probabilities.} \label{p:probabilities}
At energies above the inelastic threshold we consider the probability of scattering into a two-particle sector. Similarly to the phases, these scattering probabilities are obtained by projecting the post-collision state onto a basis of localized states, $\ket{(n,B_i(\kappa_0)),(n',B_j(-\kappa_0))}$ with excitation tensors obtained from single particle-localized states with momentum $\pm \kappa_0$. By Fourier transforming $n$ ($n'$) with respect to the lattice momentum $\kappa$ ($\kappa'$) one obtains a good approximation to the overlap with a two-particle state with momenta $\kappa$ and $\kappa'$. 

To improve this approximation, we take into account the so-far neglected $\kappa$-dependence of $B(\kappa)$. This is done by selecting a larger reference set of excitation tensors $B_i(\kappa_a)$, where the momenta $\kappa_a$ are chosen close to $\pm \kappa_0$. We use these to generate an orthonormal basis $\tilde B(\kappa_a)$ onto which a general $B(\kappa)$ can then be projected: $B(\kappa) \approx \sum_a c_a \tilde B(\kappa_a)$. This allows us to use the projection of the state onto the small reference basis to approximate the overlap of the state onto $\ket{(n,B_i(\kappa)),(n',B_j(\kappa'))}$ for general $\kappa$, $\kappa'$.

Squaring the Fourier transformed overlaps yields a ``probability matrix'' $P_{\kappa, i; \kappa', j}(t)$. The quotation marks indicate that our implementation of the Fourier transformation does not keep track of the normalization of the probability matrix. Moreover, as in the procedure for extracting time delays, the sums over lattice sites in the Fourier transformations are subject to a cut-off on the distance between excitation tensors. Therefore, to obtain a proper probability, we have to normalize the ``probability matrix'' by the ``probability'' of being in the 11 sector before scattering. The true scattering probability for an initial state with two particles of type $1$ into two particles of type $i$ and $j$ is then obtained from
\begin{align}
    \label{eq:prob_method}
    P(11 \to ij) = \frac{\sum_{\kappa, \kappa'} P_{\kappa, i; \kappa', j}(t)}{\sum_{\kappa, \kappa'} P_{\kappa, 1; \kappa', 1}(0)},
\end{align}
i.e., by computing the total probability of finding particles of type $i,j$ in the out-state of the simulated scattering process. As before, $t$ must be chosen such that the particles have separated sufficiently after the scattering.


\section{Wavefunctions and Multiverses} \label{sec:multiverses}
Several of the figures shown in this paper, such as Fig.~\ref{fig:two_tracks}, are obtained by computing the energy density expectation value at different spatial positions and time steps. The interpretation of these images is subtle. They capture superpositions of possible scattering outcomes rather than one particular outcome; hence there are combinations of outgoing tracks appearing in the image that do not correspond to any valid inelastic scattering channel \cite{Coleman:2020put}. 

Take for example Fig.~\ref{fig:two_tracks}c which shows particle production. Although the figure shows four outgoing tracks, the state is a superposition of three different outgoing particle configurations with two particles each: The outer two tracks mostly correspond to scattering of $11 \to 11$, i.e., elastic scattering of two particles, each with mass $m_1$. The outer track on the left together with the inner track on the right display scattering of $11 \to 12$, i.e., two particles of mass $m_1$ scattering into one particle with mass $m_1$ moving to the left and one particle with mass $m_2 > m_1$ moving to the right. Likewise, the inner track on the left together with the outer track on the right display scattering of $11 \to 21$. Although not obvious from the image, the two inner tracks, or any combination of more than two tracks, never appear as a scattering outcome.

To confirm this interpretation of Fig.~\ref{fig:two_tracks}c, we can project the outgoing state onto the $11$, $12$, and $21$ sectors as explained above and check that the probabilities to scatter into the respective sectors add up to 1. While this is fairly easy to achieve using numerical methods for our spin chain model, we expect that in the future real-time simulations will run efficiently on quantum computers. Indeed, with quantum computers, we can investigate the regime where the post-collision state is so highly entangled that classical MPS simulations fail \cite{Milsted:2020jmf}.
Since projecting the post-collision state onto various asymptotic outgoing particle states might be rather inefficient in a quantum computation, it is useful to consider other methods for extracting information about possible scattering outcomes from the post-collision quantum state.

\begin{figure}
    \centering
    \includegraphics[scale=0.67]{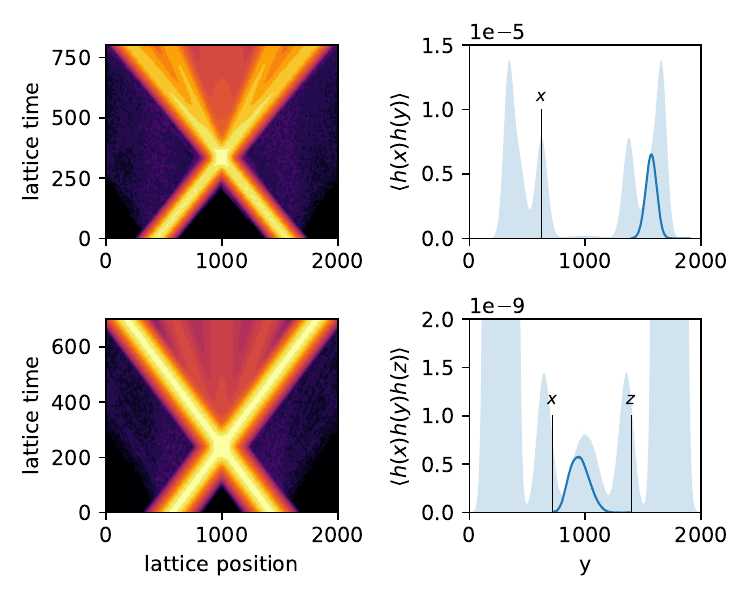}
    \caption{Inelastic particle scattering is visible as the presence of many tracks in the post-collision region (log-scale plot of excess energy on the left). Not every combination of tracks is a possible scattering outcome. Instead, the energy density is strongly correlated between subregions of the post-collision region (right). A comparison with the rescaled energy expectation values, shown as solid blue backgrounds, indicates that only subsets of the outgoing tracks are correlated. The simulation are run with $D = 64$ width a dynamical window of $2000$ sites. Above: $\eta_\text{latt} = 
    .700$ $(g_x = 1.06, g_z = 0.01)$, $E = 3.9$. Below: $\eta_\text{latt} = 1.97$ $(g_x = 1.4, g_z =0.05)$, $E = 3.15$. 
    }
    \label{fig:energy_two_pt_corr}
\end{figure}

One feasible task for quantum computers is estimating correlators of
local operators; for example, we may consider correlation functions for the difference between the local energy density in the post-collision state and the energy density of the vacuum. For a $k$-point correlator, we can fix $k{-}1$ of the operator insertions at the locations of well-separated outgoing particles, and scan over the location of the remaining operator insertion to find the position of the last particle. An example
is shown in the upper row of Fig.~\ref{fig:energy_two_pt_corr}. 
Here the two-point energy-density correlator is plotted as a function of $y$, where $x$ is fixed at a local maximum of the energy density corresponding to the position of a particle with mass $m_2$ moving slowly to the left. 
As can be seen from the figure, the correlator is small for $y$ near the trajectory of a particle of mass $m_2$ moving slowly to the right. As already expected from energy conservation, this indicates that $11 \to 22$ is not a possible final state of the simulated scattering process. Instead, the correlator as a function of $y$ is peaked where $y$ is the position of a particle with mass $m_1$ moving more rapidly to the right. Notably, however, the peak of the correlator is to the left of the peak of the right-most trajectory. This is because the right-most trajectory consists of contributions from the processes $11 \to 11$ and $11 \to 12$. The momentum of the right-moving $m_1$ particle in the former is larger than in the latter, because in the latter process, part of the energy has been converted into the rest mass for $m_2$.

Even more interesting is the scattering process $11 \to 111$, depicted in the lower row of Fig.~\ref{fig:energy_two_pt_corr}. Close to the three-particle threshold one can identify three tracks, each for an outgoing particle with mass $m_1$. However, while for two-particle final states the kinematics uniquely determine the outgoing momenta, for three particles there is a one-parameter family of kinematically admissible outgoing momentum configurations. As a result, the central track appears smeared. Energy-density correlation functions attest that the position of the central wave packet is more sharply defined when we condition on information about the positions of the other two outgoing wave packets. 
Fig.~\ref{fig:energy_two_pt_corr} shows the three-point energy-density correlator as a function of $y$, where $x$ and $z$ are at fixed positions displaced slightly to the right of the local maxima of the energy density associated with the outer tracks moving left and right. As the kinematics requires, this function is maximized when $y$ is displaced slightly to the left of the local maximum of the energy density associated with the central outgoing track.

\begin{figure}
    \centering
    \includegraphics[scale=0.3]{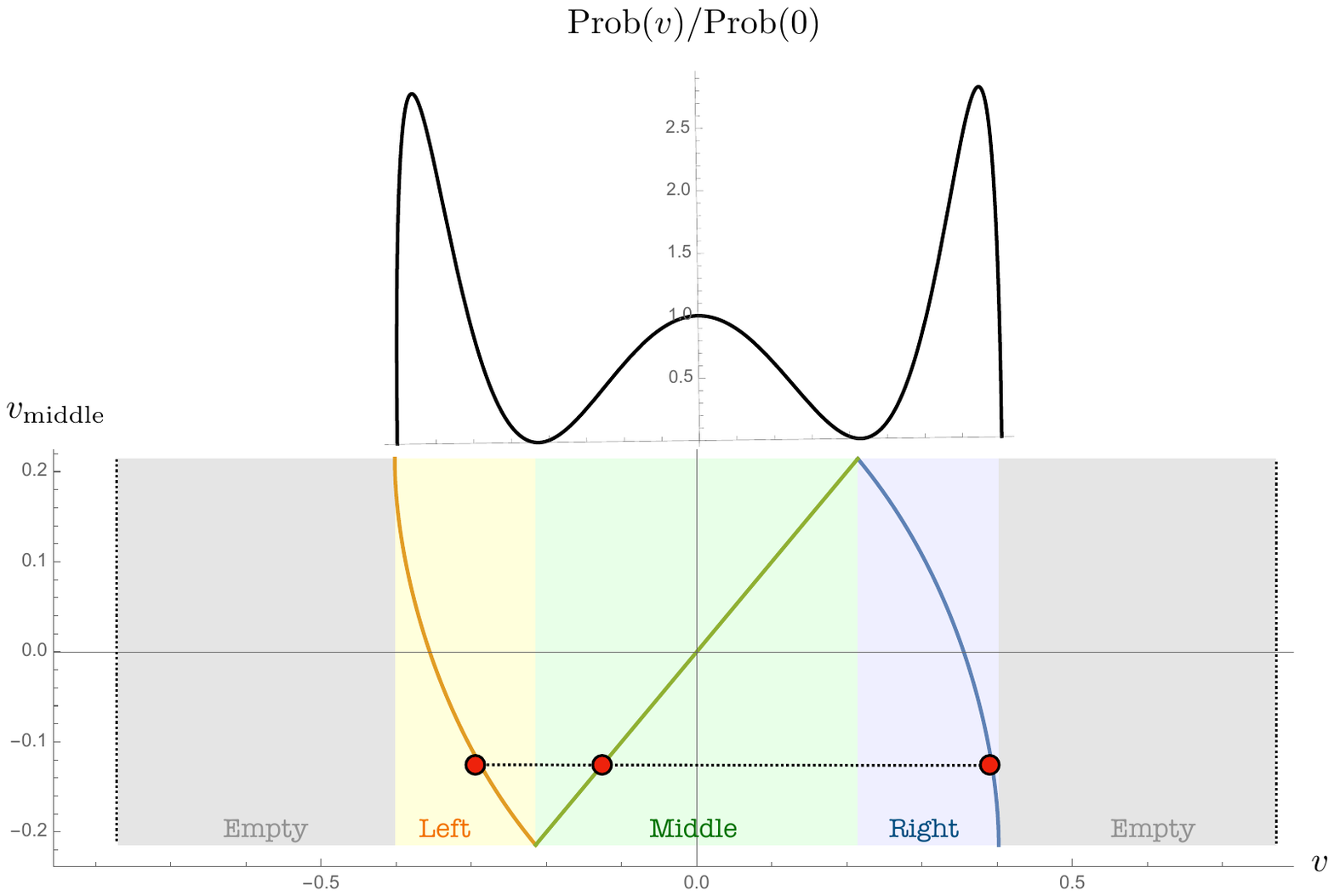}
    \caption{If we detect the position of one of the three particles (one of the red circles), then for a specified total energy we automatically know where the other two particles are. The relative probability of encountering particles in these various configurations -- which can be parameterized by the velocity of the middle particle -- is plotted on the top. For this figure the total incoming energy is $3.33 m_1$.  }
    \label{fig:3particlesKin}
\end{figure}

Our numerical results for the energy-density distribution of the outgoing state, indicated by the blue shading in Fig.~\ref{fig:energy_two_pt_corr}, can be compared with the kinematic constraints on the velocities of the three particles portrayed in Fig.~\ref{fig:3particlesKin}. The lower panel of Fig.~\ref{fig:3particlesKin} shows how, for a state of three particles of type 1 with zero total momentum and specified total energy, fixing the velocity of the middle particles determines the velocities of the other two. Taking into account phase space factors and the momentum dependence of the scattering amplitude, we can obtain the probability distribution for the velocity of the central particle as shown in the upper panel;
see App.~\ref{app:3body} for further discussion. The probability density vanishes at both ends of the kinematically allowed velocity range; at these points the velocity of the middle particle coincides with the velocity of one of the outer particles, which is disallowed because the particles are fermions. The simulated distribution in the lower panel of Fig.~\ref{fig:energy_two_pt_corr} is somewhat smeared compared to the theoretical prediction in the upper panel of Fig.~\ref{fig:3particlesKin} because we scatter wave packets rather than particles with precisely defined momenta. The occurrence of parallel tracks in Fig.~\ref{fig:two_tracks}b is discussed in Sec.~\ref{sec:resonances} and App.~\ref{app:resonance_gap}.

\section{Near Free Fermion Theory} 
\label{sec:nff}

\begin{figure*}
    \centering
    \subfloat[\label{fig:prob}]{
    \includegraphics[scale=0.65]{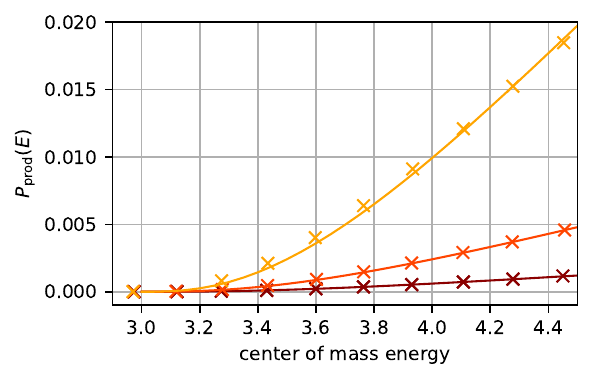}
    }\hspace{0.01\textwidth}
    \subfloat[\label{fig:phase}]{
    \includegraphics[scale=0.65]{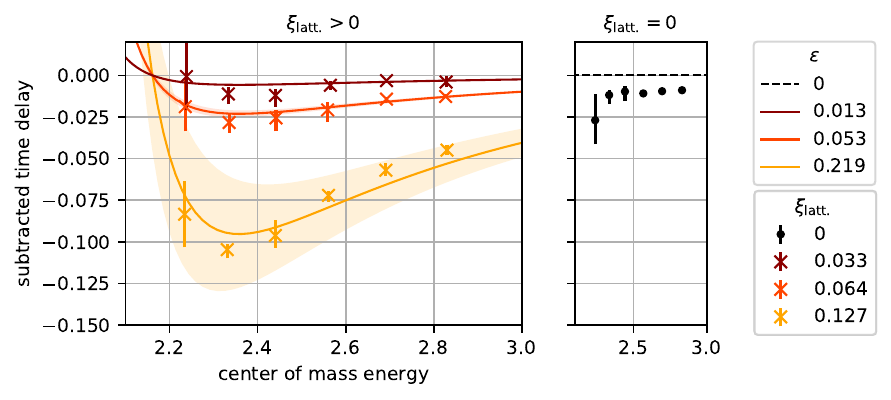}
    }
    \caption{(a) The simulated production probability at three different values of the lattice coupling parameter $\xi_\text{latt}$. The solid lines are single parameter fits of the theoretical production probability at first order in FF perturbation theory, Eqs.~\eqref{NearFFS} and \eqref{eq:SnFF_branchcut_correction}. (b)~The simulated time delay at the same values of $\xi_\text{latt}$. The theoretical predictions from FF perturbation theory are completely fixed by the probability fits. The shaded regions show the expected magnitude of the second-order correction. }
    \label{fig:nFF}
\end{figure*}
We can now demonstrate our techniques by comparing our simulations to theoretical predictions for IFT from perturbation theory.
Zamolodchikov and Ziyatdinov~\cite{ZZ} studied scattering in IFT close to the free fermion limit. At the free fermion point ($\eta=\infty$) the S-matrix for the $11 \to 11$ process, i.e., for the process of scattering two fermions to two fermions, is 
\begin{equation}
S^\texttt{FF}_{11 \to 11}(\theta) =  -1   ;  \label{FFS}
\end{equation}
scattering is trivial, but the phase shift $-1$ is induced by Fermi-Dirac statistics. The argument $\theta$ is the relativistic rapidity,
related to the total scattering energy through $E=2 \cosh(\theta/2)$, where we have set the fermion mass to $1$. 
 
It is useful to introduce
\begin{align}
    \xi = \eta^{-15/8} = \frac{|h|}{\tau^{15/8}}
    \label{eq:newFFparam}
\end{align}
to parametrize deformations away from the free fermion theory
at $\xi = 0$.
The spin chain description of the free fermion theory, Eq.~\eqref{spinChain} with $g_z = 0$, has a  $\mathbb Z_2$ global symmetry in which $\sigma^x$ is applied to every site.
As a result, physical quantities close to free fermion theory may be expanded in powers of $\xi^2$ around $\xi = 0$.
As we deform slightly away from the free fermion point at $\xi = 0$ towards larger $\xi$, two interesting things happen. 

First, we break the $\mathbb{Z}_2$ symmetry by turning on the magnetic field which introduces cubic coupling and correspondingly a self-coupling pole in the S-matrix at $E=1$. By crossing symmetry this new pole is accompanied by a $t$-channel exchange at $E=\sqrt{3}$. The pole must come together with a nearby zero such that the limit $\xi \to 0$ reproduces Eq.~\eqref{FFS}. This leads to a natural guess for an approximate form of the S-matrix near the free fermion point (nFF) at energies well below the inelastic threshold at $E = 3$, which can be written using two Castillejo-Dalitz-Dyson (CDD) \cite{PhysRev.101.453} factors,
\begin{equation}
S^\texttt{nFF}_{11 \to 11}(\theta) \simeq - \frac{\sinh\theta+i\sin\tfrac{\pi}{3}}{\sinh\theta-i\sin\tfrac{\pi}{3}}\times
\frac{\sinh\theta-i\sin(\tfrac{\pi}{3}-\varepsilon)}{\sinh\theta+i\sin(\tfrac{\pi}{3}-\varepsilon)},
\label{NearFFS}
\end{equation}
where $\varepsilon$ is a small, positive parameter related to the strength of the magnetic deformation. Eq.~\eqref{NearFFS} can be expanded at the pole at $E=1$ and takes the form
\begin{equation}
\label{eq:SnFF_pole}
S^\texttt{nFF}_{11 \to 11}(\theta) \simeq \frac{i\varepsilon}{\theta-2\pi i /3} + \dots
\end{equation}
to leading order in $\varepsilon$. For $\varepsilon > 0$ this is the right behavior such that the massive fermion appears as a bound state in the $s$-channel. This shows that $\varepsilon$ is approximately the square of the cubic coupling at the self-coupling point. Using the results of \cite{ZZ} we find that $\varepsilon \propto \xi^2$ at leading order, as expected.

Second, as $\xi$ is tuned away from zero, inelastic channels open up such that particles can be produced in scattering events. The probability of particle production can be perturbatively estimated from form factors of the $\sigma$ operator in the massive free fermion theory. More precisely, Zamolodchikov and Ziyatdinov analytically computed the effect on particle production of the $2\to 3$ processes arising from $5$-point form factors which, as they explain, dominate at most energies and can thus serve as a good proxy for total particle production. This leads to a rich prediction for the probability of particle production $P_\text{prod}(E)=1-|S^\texttt{nFF}_{11 \to 11}(E)|^2$ above $E=3$ given by \cite{ZZ}
\begin{eqnarray}
   P_\text{prod}(E)&=&\varepsilon\times ({E}-3)^3 \times \label{ProdZZ} \\
   &\times&\tfrac{ ({E}+2)^{{5}/{2}} (2 {E}-1)^4}{9 \pi  ({E}-2)^{{3}/{2}} ({E}-1)^{{5}/{2}} {E}^3 ({E}+1) ({E}+3)^{{3}/{2}}} \times \nonumber\\
   &\times& \int\limits_{-1}^{+1} \frac{dt\,\sqrt{1-t^2} }{\Big(1-\tfrac{(E-3)^3 (E+1)}{(E+3)^3(E-1) } \,t^2\Big)^{{3}/{2}}}\nonumber \times \\
      &&\qquad \times  \left(\frac{1-\tfrac{(E-3)^3 (E-2) (E+1) (2 E+1)^2}{(E+3)^3(E-1) (E+2)  (2 E-1)^2} t^2}{1-\tfrac{(E-3)^3 (E+1) (E+2)}{(E+3)^3(E-1)(E-2)  } t^2}\right)^{\!2} \,. \nonumber 
\end{eqnarray}
The $(E-3)^3$ factor in the first line of Eq.~\eqref{ProdZZ} indicates that particle production turns on slowly as $E$ crosses the threshold for inelastic scattering. Importantly, the factor $\varepsilon$ in the first line is the same parameter that appeared in the expression \eqref{NearFFS} for the elastic phase shift. 

By the optical theorem, particle production results in a branch cut at $E>3$ that contributes to the elastic scattering amplitude.
This contribution, computed in \cite{ZZ}, modifies Eq.~\eqref{NearFFS} according to $S^\texttt{nFF}_{11 \to 11}(\theta) \to S^\texttt{nFF}_{11 \to 11}(\theta) + \Delta S^\texttt{nFF}_{11 \to 11}(\theta)$, where to leading order in $\varepsilon$
\begin{align}
    \label{eq:SnFF_branchcut_correction}
    \Delta S^\texttt{nFF}_{11 \to 11}(\theta) =  - \frac{i \sinh \theta}{72 \pi} \int^\infty_{11.25} \frac{P_\text{prod}(v)}{(\sinh^2 \theta - v) \sqrt{v}} dv,
\end{align}
with $v = \frac{E^2(E^2 - 4m_1^2)}{4m_1^2}$. (Recall that we have chosen units such that $m_1=1$.) This expression is $\mathcal O(\varepsilon)$, since $P_\text{prod}(v)$ in Eq.~\eqref{ProdZZ} is proportional to $\varepsilon$.

In Eqs.~\eqref{NearFFS} and \eqref{ProdZZ}, the phase shift below the threshold as well as the particle production above the threshold depend on the single parameter $\varepsilon$. However, the exact relation between $\varepsilon$ (or alternatively $\xi$ in IFT) and the coupling on the lattice, $\xi_\text{latt} = |g_z|/(1-g_x)^{15/8}$, is not known \emph{a priori} and needs to be extracted from the simulation.

Our results for scattering in IFT near the free fermion point are summarized in Fig.~\ref{fig:nFF}. 
As is evident from Fig.~\ref{fig:prob}, the simulations reproduce the expected $(E-3)^3$ increase in particle production probability near the threshold. We obtain the parameter $\varepsilon$ by fitting $1-P_\text{prod}(E)$ from Eq.~\eqref{ProdZZ} to the probability of $11 \to 11$ scattering obtained from MPS simulations for different values of the lattice coupling via Eq.~\eqref{eq:prob_method}. The relation between $\xi^2_\text{latt}$ and $\varepsilon$ is linear at small $\varepsilon$. Since the values of $\xi^2_\text{latt}$ in our simulations are chosen to differ by factors of four, this should also be true for the obtained values of $\varepsilon$, at least within some error which comes from corrections at the next order in $\varepsilon$. And in fact, the ratio between two subsequent values for $\varepsilon$ is $3.97$ and $4.12$, respectively, in good agreement with expectations.

The values for $\varepsilon$ obtained from the probability fits can then be used to make sharp predictions for the time delay at low energies.
The solid curves in
Fig.~\ref{fig:phase} show the time delay, computed using Eqs.~\eqref{NearFFS} and \eqref{eq:SnFF_branchcut_correction} together with the fitted values for $\varepsilon$ and truncated to first order in $\varepsilon$. To indicate the expected magnitude of higher order corrections we also show a shaded region whose total width is twice the magnitude of the $\mathcal O(\varepsilon^2)$ terms in Eq.~\eqref{NearFFS}. 

To compare these theoretical predictions to our simulation we first extract raw values of the time delay from the simulated MPS wavefunctions by using Eq.~\eqref{eq:time_delay_eq} (those values need to be corrected, see below). We choose the state $\ket {\psi(0)}$ in \eqref{eq:phases_eq} to be the initial state at $520$ steps (the precise step number does not matter in practice). This is well before the scattering event, but much after the integrator has acted on the state several times. The final state $\ket{\psi(t)}$ in \eqref{eq:phases_eq} is chosen from the scattering out-region, defined to consist of the last $1000$ simulation steps before the energy $100$ sites from the boundary of the dynamic window becomes larger than $10^{-6}$ in lattice units. This ensures that boundary effects are small. We compute three values for the time delay, using the states at steps $0$, $500$, and $1000$ in the asymptotic out-region. In theory, the three values should agree; however, in practice, we find that the number fluctuates. The data used to produce Fig.~\ref{fig:phase} is the average of the three results. The error bars span the range between the largest and smallest of the three values.

Given time and computing constraints, we were not able to go to a parameter regime that would allow us to reproduce the trivial S-matrix of the FF with high accuracy. Instead, our simulations of particle scattering at $\xi_\text{latt} = 0$ produce a (momentum-dependent) time delay of $\mathcal O(0.01)$, see the right panel of Fig.~\ref{fig:phase}. The simulation data is consistently biased in favor of a negative time delay. We expect that this mismatch comes at least in part from finite step size $dt_\text{latt}$ in the RK4/5 integrator and present some evidence for this claim in App.~\ref{app:subtraction}. To correct for this effect we subtract the time delay at $\xi_\text{latt} = 0$, which should ideally be zero, from the time delays at $\xi_\text{latt} \neq 0$. This corrected data is displayed in Fig.~\ref{fig:phase} and agrees well with the theoretical predictions.


\section{Near \texorpdfstring{$E_{8}$}{E8} Theory}
\label{sec:ne8}
We can also consider small perturbations away from the other integrable point, the $E_{8}$ theory at $\eta=0$. At the $E_{8}$ point, the S-matrix for elastic scattering for the two lightest particles consists of three CDD factors,
\begin{align}
    \label{eq:SmatrixE8}
    S^\text{\texttt{E$_{8}$}}_{11\to 11}(\theta) = \prod^3_{j=1} \frac{\sinh \theta + i \sin \alpha_j}{\sinh \theta - i \sin \alpha_j}.
\end{align}
The poles at $\alpha_1 = \frac{2 \pi}{3}$, $\alpha_2 = \frac{2 \pi}{5}$, $\alpha_3 = \frac{2 \pi}{30}$ correspond to the three lightest particles with known masses $m_1 = 2 \cos(\alpha_1/2) = 1$, $m_2 = 2 \cos(\alpha_2/2) \approx 1.618$, and $m_3 = 2 \cos(\alpha_3/2) \approx 1.989$, c.f.~Table~\ref{fig:E8_spec_det}. Upon deforming the theory by perturbing the transverse field away from $g_x=1$ the locations of the poles change \cite{Zamolodchikov:2013ama}. The first few corrections
in perturbation theory are \cite{Delfino:2005bh}: 
\begin{align}
    \label{eq:alpha2pert}
    \alpha_2 &= \frac{2 \pi}{5} - 2.377 \eta - 4.55 \eta^2 + \mathcal O(\eta^3),\\
    \label{eq:alpha3pert}
    \alpha_3 &= \frac{2 \pi}{30} - 8.497 \eta - 43.42 \eta^2 + \mathcal O(\eta^3),
\end{align}
where $m_1$ is fixed at 1 (or equivalently $\alpha_1$ is fixed at $\frac{2\pi}{3}$). The parameter $\eta$, defined in Eq.~\eqref{eq:cont_eta}, determines the deformation away from the Ising CFT and is related to the lattice deformation $\eta_\text{latt}$, given in Eq.~\eqref{eq:lattice_eta}, by $\eta = \beta \eta_\text{latt} + \mathcal O(\eta_\text{latt}^2)$, 
for some proportionality constant $\beta$ that can be extracted from our simulations. In contrast to the situation near the free fermion theory, we can use the fact that we have more than one stable particle to compute $\beta$ by matching Eqs.~\eqref{eq:alpha2pert} or \eqref{eq:alpha3pert} to the spectrum obtained from an MPS ansatz; we find  $\beta \approx 0.28$. See App.~\ref{app:Ising-chain_and_conversion} for details. 
As a result, and unlike near the free fermion point, we obtain predictions for scattering probabilities and the time delay without any free parameters.

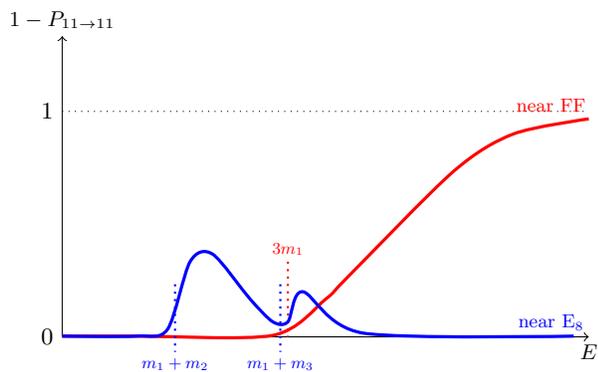
\begin{figure}[t]
    \centering
    \begin{tikzpicture}
        \draw[->] (0,0) node [left] {$0$} -- (7,0) node [below, scale=0.9] {$E$};
        \draw[->] (0,0) -- (0,4) node [above, scale=0.9] {$1- P_{11 \to 11}$};
        \draw[dotted] (0,3) node[left] {$1$} -- (7,3);
        \draw[very thick, red, smooth] plot [smooth, tension=0.5] coordinates {(0.,0.01) (0.99,0.01) (2.8,0.02) (3.5,0.5) (3.8,0.8) (5.2,2.2) (6,2.7) (7,2.9)};
        \draw[thick, dotted, red] (3.0,1.0)  node [above, scale=0.7] {$3 m_1$} -- (3.0,-0.);
        \draw[red] (6.5,2.9) node[above, scale=0.8] {near FF};
        \draw[very thick, blue, smooth] plot [smooth, tension=0.5] coordinates {(0.,0.01) (0.99,0.01) (1.4,0.1) (1.7,1.0) (2.0,1.1) (2.5,0.5) (2.8,0.2) (3,0.2) (3.2,0.6) (4,0.05) (6.8,0.01)};
        \draw[thick, dotted, blue] (1.5,0.7) -- (1.5,-0.2) node [below, scale=0.7] {$m_1 + m_2$};
        \draw[thick, dotted, blue] (2.9,0.7) -- (2.9,-0.2) node [below, scale=0.7] {$m_1 + m_3$};
        \draw[blue] (6.5,0) node[above, scale=0.8] {near E$_8$} ;
    \end{tikzpicture}
    \caption{Blue curve: Close to $E_8$ we expect particle production to increase each time a new inelastic channel opens up, but to fade away eventually so that $P_{11\to 11}$ approaches $1$ at high energy. Red curve: Close to the free-fermion limit we expect inelastic scattering to become dominant so that $P_{11\to 11}$ approaches $0$ at high energy.
    }
    \label{fig:cartoon}
\end{figure}

At small but non-zero $\eta$ the particles can scatter inelastically. We show the qualitative behavior of particle production close to $E_{8}$ in blue in Fig.~\ref{fig:cartoon}. The shape is dominated by sudden increases in production probability every time a new inelastic channel opens up, followed each time by a subsequent drop in the production probability as the energy increases.
In particular, sizable jumps in the production probability occur at threshold for the channels $11 \to 12$ and $11 \to 13$. 
Other two-particle channels such as $11 \to 22$ occur at higher order 
in $\eta$ and are thus practically invisible when $\eta$ is small. The multi-particle production threshold for $11 \to 111$ starts out very smooth, because three-particle production is kinematically suppressed near the threshold. 
Our expectations at high energies come from a conjecture by Zamolodchikov \cite{Zam_2013}, which we will discuss in Sec.~\ref{sec:high}.

When $m_3$ crosses the threshold it becomes an anti-bound state and the $11 \to 13$ channel closes.\footnote{See \cite{Zamolodchikov:2013ama} for a discussion of the fate of the $m_2$ and $m_3$ pole as we increase the coupling.} However, it is possible that new anti-bound states or resonances appear which increase the probability of scattering into stable final states, such as $12$ or $111$. We do not have a clear picture of what happens, but as we will see below, our simulations suggest enhancement of inelastic scattering probability around $E = m_1 + m_3$ even after $m_3$ has left the spectrum of stable particles.

\begin{figure}[t]
    \centering
    \includegraphics[scale=0.5]{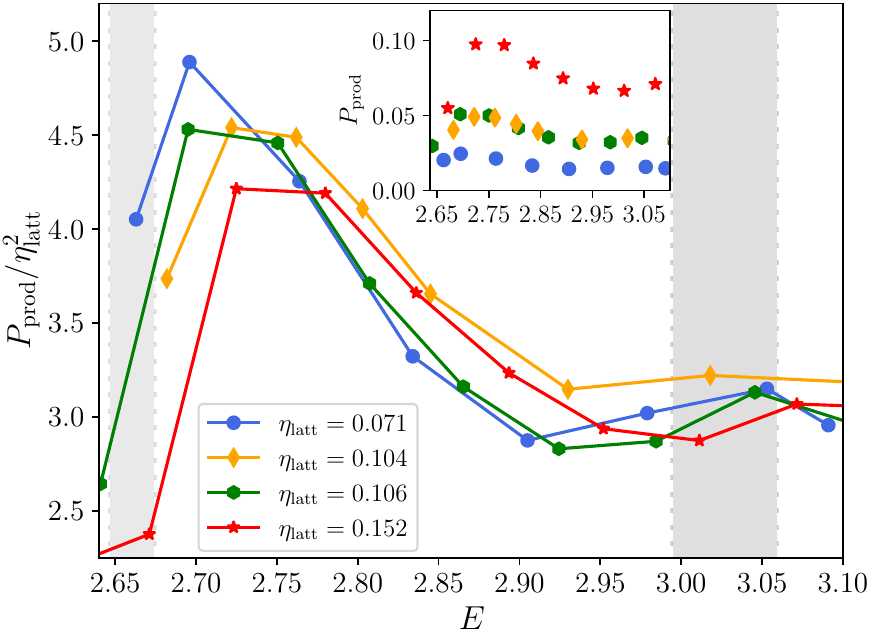}
    \caption{The rescaled production probability for values of lattice $\eta_\text{latt}$ close to $E_{8}$ and the unscaled probability (inset). The results are mostly obtained for $D=64$ but the dependence on bond dimension is negligible, see App.~\ref{app:hyperparameter_probability}. There appears to be a jump in production close to $m_{1}+m_{2}$ and $m_{1} + m_{3}$. We shade the threshold regions $m_{1} + m_{2}$ and $m_{1} + m_{3}$ for the range of $\eta$ used in the numerical results. }
    \label{fig:probabilities_simulationE8}
\end{figure}

Fig.~\ref{fig:probabilities_simulationE8} shows the inelastic scattering probabilities obtained from our simulations.
Ignoring higher-order corrections, the inelastic scattering probabilities are proportional to $\eta^2$;  we have therefore divided the probabilities by $\eta^2$, and we find as expected that plots of the rescaled production probability versus energy for different (small) values of $\eta$ nearly coincide. 

\begin{figure*}[t!]
    \centering
    \includegraphics[scale=0.7]{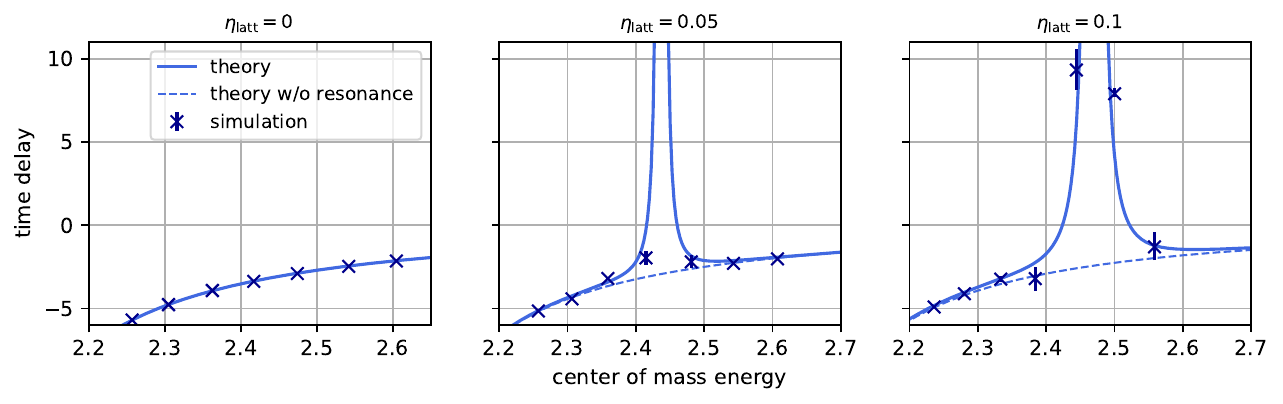}
    \caption{Time delay for elastic $11 \to 11$ scattering near $E_{8}$. The solid lines show the predictions from form factor perturbation theory. The dashed lines show the expected time delay in the absence of resonances. Neither prediction contains contributions from the $11\to 12$ threshold or any higher threshold. }
    \label{fig:phase_nE8}
\end{figure*}

Because the inelastic production probability as a function of energy has an intricate structure in the regime near $E_{8}$, and the wave packets used in our simulations are relatively broad in momentum space, it is difficult to perform a detailed comparison between our simulation results and theoretical expectations (such a comparison is easier in the regime near free fermions, because there the production probability is a reasonably smooth function of energy). 
We can, however, verify that the simulation results in Fig.~\ref{fig:probabilities_simulationE8} capture the basic qualitative features shown in Fig.~\ref{fig:cartoon}. We perceive, for example, a bump in production probability near the thresholds for the $12$ and $13$ channels.

A more quantitative comparison of theoretical predictions with our numerical results can be achieved by computing the time delay in the $11\to 11$ elastic scattering channel. In particular, elastic scattering near $E_8$ exhibits narrow resonances, a feature which is absent near the free fermion limit, and these resonances can be characterized in detail. To analyze these resonances, one observes that the $E_{8}$ theory has five stable particles with masses above the two-particle threshold. Out of these, the two lightest have masses $m_4 \approx 2.405 \,m_1$ and $m_5 \approx 2.956\, m_1$ \cite{zamolodchikov1989integrals}. Since they are not protected by any symmetry, these stable particles of the $E_8$ theory become resonances when $\eta$ is nonzero and small; the associated poles in the S-matrix move off the real axis into the complex energy plane.
Hermitian analyticity requires that any generic resonance corresponds to a product of two CDD factors;\footnote{The exception is $m_3$ which turns from a virtual state into a resonance when $\operatorname{Re}(\alpha_3) = - \frac \pi 2$. In this case, one CDD factor is sufficient to satisfy analyticity.} for each particle type $j = 4,\dots, 8$,
the S-matrix acquires a factor
\begin{align}
    \label{eq:resonance}
    S^\text{\texttt{E$_{8}$}}_{11\to 11}(\theta) = \dots \times \frac{\sinh \theta + i \sin \alpha_j}{\sinh \theta - i \sin \alpha_j} \, \frac{\sinh \theta + i \sin \alpha^*_j}{\sinh \theta - i \sin \alpha^*_j}.
\end{align}
Each pole is accompanied by a nearby zero and thus the effect on the scattering phase shift is well localized around the resonant energy. Therefore, when the energy is close to $m_4$, it is a good approximation to include only the contribution from the $m_4$ resonance and ignore heavier resonances. We have checked that the $m_5$ resonance is negligible in the following discussion.

As a result, below threshold and near $E_{8}$ the S-matrix can be well approximated using Eq.~\eqref{eq:SmatrixE8} with $\alpha_2$ and $\alpha_3$ given by Eqs.~\eqref{eq:alpha2pert} and \eqref{eq:alpha3pert}, multiplied by Eq.~\eqref{eq:resonance} with~\cite{Delfino:2005bh}
\begin{align}
\label{eq:m4}
\begin{split}
    \operatorname{Re}(m_4) &= 2.404 + 2.33 \eta + \mathcal O(\eta^2),\\
    \operatorname{Im}(m_4) &= - 3.71 \eta^2 + \mathcal O(\eta^3),
    \end{split}
\end{align}
from which $\alpha_4$ is obtained through $m_4 = 2 \cos(\alpha_4/2)$. Since we do not have data on the $\mathcal O(\eta^2)$ correction to $\operatorname{Re}(m_4)$, we set it to zero and effectively used $\operatorname{Re}(m_4) = 2.404 + 2.33 \eta + \mathcal O(\eta^3)$.
At $\eta = 0$, $\alpha_4$ is purely imaginary and Eq.~\eqref{eq:resonance} does not contribute to the S-matrix since all poles and zeros in this expression cancel. For non-vanishing $\eta$ the zeros in Eq.~\eqref{eq:resonance} are moved to the physical sheet ($0 < \operatorname{Im} \theta < \pi$)  of the S-matrix, while the poles are shifted onto the unphysical sheet ($\pi < \operatorname{Im} \theta < 2\pi$). This creates a narrow resonance.

The resulting time delay and simulated data are shown in Fig.~\ref{fig:phase_nE8}. As one can see from these plots, the simulation agrees well with theory in the vicinity of the $E_8$ point, including the presence of a resonance at the expected energy. Note that the error bars, which are obtained in the same way as for the free fermion simulations by measuring the time delay at different times in the asymptotic out-region, are much smaller than in the near free fermion case. The reason is that the time delay is much larger in the vicinity of $E_8$ limit. Thus, the constant magnitude simulation error results in smaller relative error.


\section{Resonances}
\label{sec:resonances}
In interacting theories, particles that participate in scattering processes and have masses above the two-particle threshold generically decay into lighter particles. Such unstable particles are characterized by their mass $m$ and their decay width $\Gamma$. In scattering theory, unstable particles appear as resonances, i.e.~features in phase shifts localized in a region of size $\mathcal O(\Gamma)$ around the total scattering energy $E=m$. In the S-matrix, unstable particles are poles at complex values of the energy $E_\text{pole} = m - i \frac \Gamma 2$. 

As mentioned in the previous section, the integrable $E_8$ theory at $\eta = 0$ contains particles with masses above the two-particle threshold $E=2m_1$ which become unstable when we perturb away from the integrable point. Their effect on the time delay can clearly be seen in Fig.~\ref{fig:phase_nE8}. 
These resonances have been studied both analytically and numerically \cite{Zamolodchikov:2013ama, Gabai:2019ryw}, but their properties in IFT far from the integrable theory have not been well established. 
 
The preceding section suggests that properties of the resonances can be obtained at least in principle by fitting the parameters $\alpha_j$ that appear in Eq.~\eqref{eq:resonance} to the time delay shown in Fig.~\ref{fig:phase_nE8}. Here we describe an alternative method of extracting the resonance mass $m_4$ and decay width $\Gamma_4$ from our simulations, by performing temporal and spatial fits to simulated wavefunctions.

To understand how a resonance is manifested in the wavefunction, consider a simple model S-matrix for scattering in the presence of a resonance.
Close to a resonance of mass $m$ and decay-width $\Gamma$, the $11 \to 11$ S-matrix, which describes scattering of two particles of mass $m_1$, takes the form
\begin{equation}
\label{eq:reso_smatrix}
    S^\text{\texttt{Toy}}_{11\to 11}(E) = \frac{E - m - i\frac{\Gamma}{2}}{E - m + i\frac{\Gamma}{2}},
\end{equation}
with a pole (zero) at $E = m \mp i\Gamma/2$. $S^\text{\texttt{Toy}}_{11\to 11}(E)$ is a pure phase $e^{i\phi(E)}$, with $S^\text{\texttt{Toy}}_{11\to 11}(E=m)=-1$ at resonance  and $S^\text{\texttt{Toy}}_{11\to 11}\rightarrow 1$ at high energies, where $E$ dominates both numerator and denominator.

The associated time delay of the outgoing particles near the resonance compared to the free trajectory, Eq.~\eqref{eq:def_time_delay}, is given by a Breit-Wigner distribution with width $\Gamma$ peaked at $E = m$ and approaching zero as $E$ goes further from $m$,
\begin{align}
    \Delta t = \frac{\Gamma}{(E - m)^2 + \frac{\Gamma^2}{4}}.
\end{align}
Hence, the resonance has only a small effect on scattering when $|E-m|\gg \Gamma$.
However, as can be seen from Fig.~\ref{fig:phase_nE8}, the time delay varies sharply with energy when the total energy $E$ is close to the resonance mass $m$. 

\begin{figure}
    \centering
    \includegraphics[width=\linewidth]{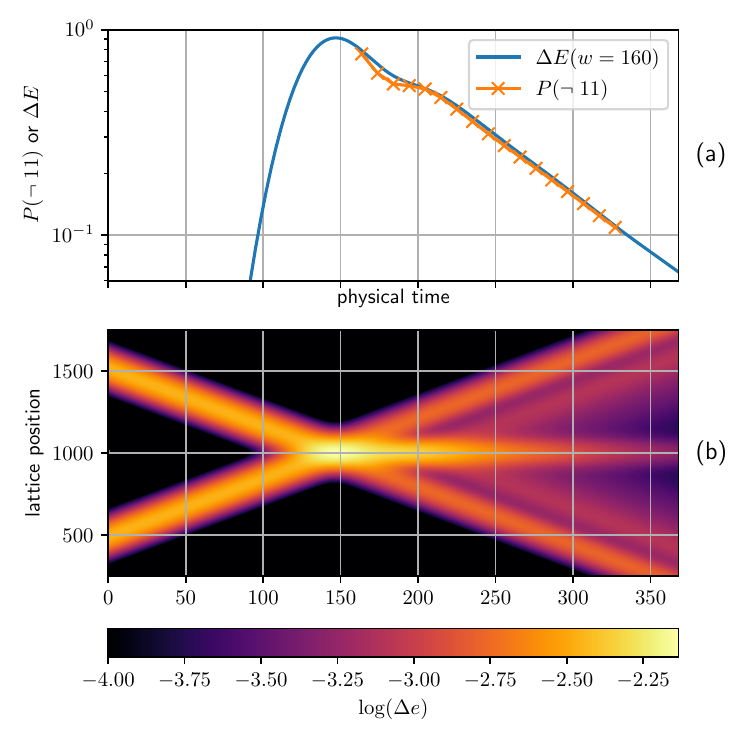}
    \caption{Upper plot (a): Exponential decay in time near the form-factor predicted $m_4$ resonance energy, showing the the excess energy $\Delta E$ in a spatial window of width 160 at the scattering center and the portion $P(\lnot \; 11)$ of the wavefunction outside the 11 sector. Here, $\eta_{\text{latt}}\approx 0.7052$ ($g_x = 1.07, g_z = 0.01315$), $E/m_1 \approx 2.81$, $\sigma/2\ell\approx 13.3$, and $D \leq 64$. 
    Lower plot (b): the local excess energy $\Delta e(n,t) = \langle h^{\textrm{symm}}_{n,n+1} \rangle_t - \langle h^{\textrm{symm}} \rangle_{\textrm{vac}}$, as a function of time and position.
    }
    \label{fig:decay_ex}
\end{figure}

In the simplified case where the state of one particle is a Gaussian wave packet centered at momentum $p_0$ and the other is a plane wave of energy $E_2$, the outgoing wavefunction takes the form
\begin{equation}
  \psi(t,x) = \frac{\sigma}{\sqrt{2 \pi}}  \int_{-\infty}^\infty dp \; e^{-\frac {\sigma^2}{2} (p - p_0)^2} S(E(p) + E_2) e^{-i(Et - px)}.
\end{equation}
As discussed in detail in App.~\ref{app:resonance_toy}, this integral can be approximately evaluated by deforming the contour to a path along which the phase is stationary. At sufficiently late times (or large distances), the integration contour is deformed across the pole in the S-matrix at $E(p) = m - E_2 - i\frac{\Gamma}{2}$, which by the residue theorem gives an additional contribution centered at $E=m$. If the incoming particles are wave packets with momentum distributions much broader than the decay width $\Gamma$, the contribution from the deformed contour and the contribution from the pole can be of the same magnitude and interfere \emph{destructively}, owing to an approximate phase difference of $\Delta\phi \approx \pi$. As explained in App.~\ref{app:resonance_gap}, this gives rise to the appearance of two parallel tracks in the out-state, which are clearly visible in Fig.~\ref{fig:two_tracks}b as well as Fig.~\ref{fig:decay_ex}. The spacing of these two tracks is mostly determined by the width of the wave packet, not by the lifetime of the resonance.

An explicit computation (App.~\ref{app:resonance_toy}) shows that the pole contribution to the wavefunction is proportional to a temporal decay factor $\exp({-\frac{\Gamma}{2}t})$ and thus we expect the probability of finding a particle at any fixed lattice site $n$ to decay exponentially with rate $\Gamma$. Moreover, since the wave packets in our scattering simulations are narrow in momentum space, one can compute energy density expectation values using the same saddle point approximation that is used in App.~\ref{app:resonance_toy} to compute the wavefunction. One thus expects to find the same exponential decay in energy density expectation values. Therefore the above discussion suggests that one can extract the decay width $\Gamma$ of resonances by fitting an exponential to energy densities or probabilities at late times.

However, in practice, the energy density and wavefunction at late times show oscillations and possibly other spurious effects coming from the presence of thresholds and poles in the S-matrix. Such oscillations can for example arise if the scattered wavefunction contains several resonance-like contributions which, when computing the probability, lead to interference effects. Nonetheless, if there is a dominant contribution coming from a single resonance, we should expect that the decay rate of the energy or probability averaged over a very large window shows an exponential decay with rate $\Gamma$.

In fact, as demonstrated in Fig.~\ref{fig:decay_ex}, the exponential decay can indeed be observed in our simulations by estimating two quantities which have approximately the same decay rate: The first quantity is the excess energy (relative to the vacuum) within a spatial window
\begin{equation}
    \label{eq:deltaE}
    \Delta E(n_0, w,\; t) := \sum_{n=n_0-w/2}^{n_0+w/2-1} \left( \langle h^{\textrm{symm}}_{n,n+1} \rangle_t - \langle h^{\textrm{symm}} \rangle_{\textrm{vac}} \right),
\end{equation}
where $h^{\textrm{symm}}$ is the unique spatially symmetric energy-density operator on the lattice, $w$ is the window width, and we choose the center $n_0$ to be the spatial center of the scattering event. The second quantity is the probability
\begin{equation}
    \label{eq:Pnot11}
    P(\lnot\; 11, t) := 1-P(11, t)
\end{equation} 
of finding the system outside of the numerically-defined asymptotic $11$ sector at time $t$.\footnote{Note that $P(\lnot\; 11, t)$ is not equivalent to $P_\text{prod}$ of Sec.~\ref{sec:nff}. The latter is the probability of not being in the asymptotic $11$ sector when all particles are well-separated. Thus, $P_\text{prod}$ is the limit of $P(\lnot\; 11, t)$ for sufficiently late times.} This probability is easily accessible, since $P(11, t)$ can be obtained using the methods described in Sec.~\ref{sec:probabilities}. In App.~\ref{app:reso_toy_asymp} we argue that the time-dependence of $P(\lnot\; 11, t)$ directly follows from the behavior of the wavefunction at late times and that the rate of decay of $P(\lnot\; 11, t)$ is indeed expected to be given by $\Gamma$.

The intuitive reason for the decay of $P(\lnot\; 11, t)$ is that the projection onto the $11$-sector approximately excludes everything from the state that is \emph{not} a combination of two freely propagating $m_1$ particles, hence $P(\lnot\; 11)$ captures any portion of the wavefunction which is ``still resonating.'' In addition, it also captures any part of the wavefunction that escapes $11$ via inelastic channels -- something we must account for when scattering near the inelastic threshold due to the finite energy width of our wave packets. Indeed, such contributions due to inelastic channels can also show up in the energy expectation value, which does not discriminate between particle sectors.

We find fitting the decay of $P(\lnot\; 11)$ to be a more robust means of determining the resonance width $\Gamma_4$ than fitting the decay of $\Delta E(w, t)$ (see App.~\ref{app:resonance}).
To obtain $\Gamma_4$ from our simulation, we scatter two particles of type $1$ at energies close to the resonance energy. We then compute the temporal evolution of $P(\lnot\; 11,t)$ by projecting the wavefunction at time $t$ onto an MPS basis for $11$, as defined in Eq.~\eqref{eq:reference_state}, where the excitation tensors are separated in space by a specified minimum distance to avoid interaction effects, defining an approximate asymptotic regime. For this projection, we account for momentum dependence of the excitation tensors as described in \cite{Milsted:2020jmf}, although we observe the correction to the probabilities to be small, presumably owing to a weak dependence of the tensors on momentum and the spatial broadness of our wave packets.
We fit $P(\lnot\; 11, t)$ using the ansatz (c.f.\ App.~\ref{app:reso_toy_asymp})
\begin{equation}
    \label{eq:ansatz}
    A e^{-\Gamma_4 t} + P_{\textrm{leak}},
\end{equation}
where the factor $A$ represents a time offset (the start of the decay) as well as the amplitude of the decaying portion of the state, and $P_{\textrm{leak}}$ represents a constant offset from zero, accounting for accidental leakage into inelastic channels due to proximity to the $m_1 + m_2$ threshold. In our simulations, leakage is especially important far from $E_8$, where the resonance moves closer to $m_1 + m_2$ and our wave packets have non-negligible support at and beyond this threshold; see App.~\ref{app:resonance} for details. The best-fit value for $\Gamma_4$ in \eqref{eq:ansatz} is then reported as the decay width of the resonance.

\begin{figure}[t]
    \centering
    \includegraphics[width=\linewidth]{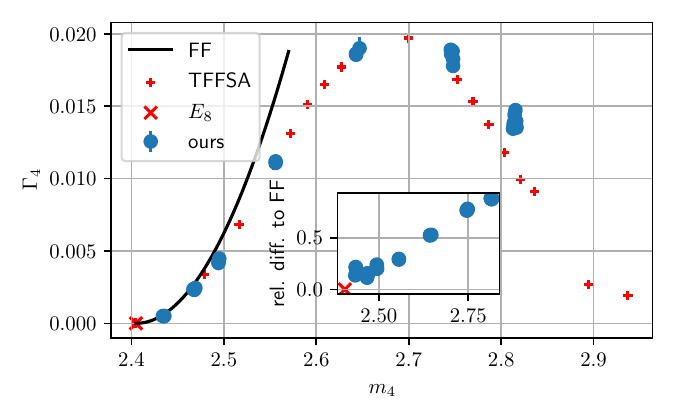}
    \caption{Comparison of $\Gamma_4$ and $m_4$ obtained from fitting to our simulations (blue dots) with the form-factor prediction (black line) and TFFSA data (red crosses) taken from Ref.~\cite{Gabai:2019ryw}. The maximum MPS bond dimension in our simulations is 64 and multiple simulations at different lattice spacings and energies are shown at each $m_4$ value (see Table~\ref{tab:reso_params}). Fitting $\Gamma_4$ is more difficult further from $E_8$, at larger $m_4$, as evidenced by the greater vertical spread. \emph{Inset}: Relative difference between simulated and form-factor values, showing convergence as we approach the $E_{8}$ point. }
    \label{fig:delta_t_m4}
\end{figure}

We can also estimate the mass $m_4$ of the resonance, for example by simulating scattering events at different energies and finding the energy for which the contribution of the second set of tracks to the scattering process is largest. However, a more efficient and accurate method is to exploit the fact that any simulation close enough to the resonance to witness its decay already contains information about its energy: In App.~\ref{app:resonance}, we show that a Fourier analysis of the spatial secondary tracks in the outgoing wavefunction projected into the $11$ sector can deliver a precise estimate of $m_4$. We also show in App.~\ref{app:resonance_toy} that a \emph{spatial} section of the wavefunction, and thus also the excess energy density, has a decay rate related to a certain combination of the mass of the resonance and its decay width, in principle providing a further means of estimating $m_4$. However, in practice we find the method just described more useful.

We compute $\Gamma_4$ and $m_4$ using simulations across a range of $\eta_{\textrm{latt}}$, both close to and further from $E_8$, at a few different values of the correlation length in lattice units. We also check the robustness with respect to the MPS bond dimension and integration time-step size, as well as the wave-packet width. For details, see App.~\ref{app:resonance}. In Fig.~\ref{fig:delta_t_m4}, we plot our fitted $\Gamma_4$ values from $P(\lnot\; 11,t)$ against $m_4$ estimated via Fourier analysis of the outgoing wavefunctions. We compare the data with the form-factor perturbation theory prediction Eq.~\eqref{eq:m4} as well as with data computed using the truncated free-fermion space approach (TFFSA) of Ref.~\cite{Gabai:2019ryw}. We note good agreement with the latter, especially near $E_8$. Further from $E_8$ the agreement is less clear and our data gives us a broader spread of decay rates. Our analysis suggests the spread is mainly due to the difficulty of fitting accurately in a regime where leakage into the 12 and 21 sectors becomes significant, limiting the window of time in which resonance decay dominates $P(\lnot\; 11,t)$. Increasing spatial wave-packet widths (and hence the MPS window size $N$) further would likely improve these results.


\section{High Energy Scattering}
\label{sec:high}

\begin{figure}[t]
    \centering
    \begin{tikzpicture}[font=\sffamily]
    
    \def\xscale{2};
    \def\yscale{5};
    
    \fill [left color=gradientblue, right color=gradienteta2] (0,0) rectangle (\xscale*0.93,\yscale*1);
    \fill [left color=gradienteta2, right color=gradientFF] (\xscale*0.92,0) rectangle (\xscale*3.5,\yscale*1);
    \draw [-latex] (0,0) -- (\xscale*3.6,0) node[right] {$\eta_\text{latt}$};
    \foreach \x in {0, 0.5, 1.0, 1.5, 2.0, 2.5, 3.0, 3.5}
        \draw (\xscale*\x,0) node [below] {\x}-- (\xscale*\x,0.1);
    
    \draw [-latex] (0,0) -- (0,\yscale*1.1) node[above] {P$_{11\to11}$};
    \foreach \y in {0, 0.2, 0.4, 0.6, 0.8, 1.0}
        \draw (0, \yscale*\y) node [left] {\y}-- (0.1, \yscale*\y);

    \node[align=left] at (\xscale*0.3,\yscale*1.05) {near E$_8$};
    \node[align=right] at (\xscale*3.2,\yscale*1.05) {near FF};
    \foreach \pos in {{{3.062, 0.9583}}, {{2.4, 0.90151}}, {{2., 0.81915}}, {{1.531, 0.60882}}, {{1.225, 0.39138}}, {{1.08, 0.30014}}, {{1., 0.2685}}, {{0.9186, 0.25723}}, {{0.6995, 0.38189}}, {{0.6124, 0.47889}}, {{0.3062, 0.87659}}}
        \fill[white!75!black] (\xscale*\pos[0], \yscale*\pos[1]) circle (1mm);
  
    \foreach \pos in {{{3.062, 0.9464}}, {{2.4, 0.8743}}, {{2., 0.77198}}, {{1.531, 0.533}}, {{1.225, 0.31482}}, {{1.08, 0.2414}}, {{1., 0.22528}}, {{0.9186, 0.2333}}, {{0.6995, 0.4062}}, {{0.6124, 0.51407}}, {{0.3062, 0.8979}}}
        \fill[white!50!black] (\xscale*\pos[0], \yscale*\pos[1]) circle (1mm);     
        
    \foreach \pos in {{{3.062, 0.938}}, {{2.4, 0.8532}}, {{2., 0.73645}}, {{1.531, 0.476}}, {{1.225, 0.264573}}, {{1.08, 0.2068}}, {{1., 0.2037}}, {{0.9186, 0.2261}}, {{0.6995,0.4274}}, {{0.6124, 0.5454}}, {{0.3062, 0.9116}}}
        \fill[black] (\xscale*\pos[0], \yscale*\pos[1]) circle (1mm);

    \draw [very thick, yellow, -latex] (\xscale*0.6124, \yscale*0.47889) -- (\xscale*0.6124, \yscale*0.5454);
    \draw [very thick, yellow, -latex] (\xscale*1.531, \yscale*0.60882) -- (\xscale*1.531, \yscale*0.476);

    \draw[black] (\xscale*0.4,\yscale*0.2) node[align=center] { \scriptsize P$_{11\to11}$\\\scriptsize increasing\\ \scriptsize with $E$};
    \draw[black] (\xscale*3,\yscale*0.2) node[align=center] { \scriptsize P$_{11\to11}$\\\scriptsize decreasing\\ \scriptsize with $E$};
        
    
    \draw[fill=white!75!black] (\xscale*0.5, -0.8) circle (1mm);
    \draw (\xscale*0.55, -0.8) node[right] {$E = 6 m_1$};
    \draw[fill=white!50!black] (\xscale*1.5, -0.8) circle (1mm);
    \draw (\xscale*1.55, -0.8) node[right] {$E = 7 m_1$};
    \draw[fill=white!0!black] (\xscale*2.5, -0.8) circle (1mm);
    \draw (\xscale*2.55, -0.8) node[right] {$E = 8 m_1$};

    \end{tikzpicture}
    \caption{The high energy behavior of scattering in IFT for $E=6,7,8$ in units of the lightest particle mass. The results indicate that
    the large-$E$ behavior changes around $0.7 \lesssim \eta_{\text{c,latt}} \lesssim 0.9$.
    } 
    \label{fig:highE}
\end{figure}
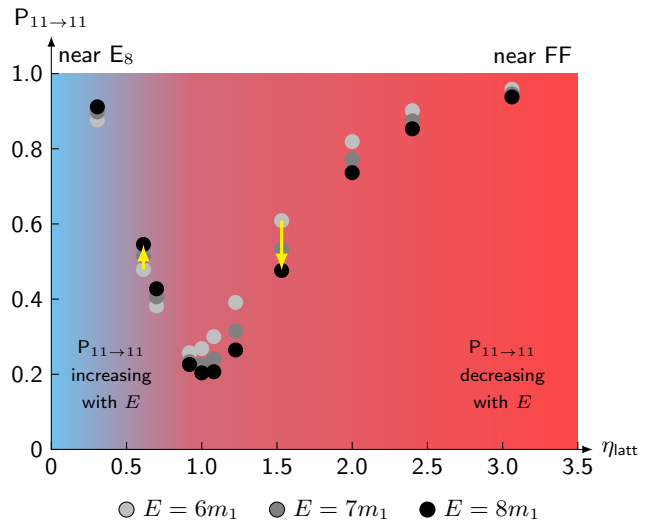

In the previous sections, we considered particle production close to the two integrable limits of the IFT.
We now focus on another interesting limit: scattering at high energies for values of $\eta$ that interpolate between the two integrable points. In general, the high energy behavior of the continuum theory is difficult to study on the lattice because we are limited to energies such that lattice artifacts are under control. Hence to accurately probe very large energies, we need a very small ratio of lattice spacing to correlation length, which is numerically expensive. Nevertheless, our numerical results arguably capture some instructive features of the continuum IFT at relatively high energy.

The main question of interest is whether elastic or inelastic scattering becomes dominant in the high energy limit. We find evidence suggesting a change in behavior as $\eta_{\text{latt}}$ varies, with elastic scattering dominating at high energies for small $\eta_{\text{latt}}$, and inelastic scattering dominating at high energies for large $\eta_{\text{latt}}$. Our studies of inelastic particle production in Sec.~\ref{sec:nff} and \ref{sec:ne8} already point in this direction. We found that the production probability increases with increasing center of mass energy in Fig.~\ref{fig:prob} (close to free fermion theory), and decreases with center of mass energy in Fig.~\ref{fig:probabilities_simulationE8} (close to $E_8$), apart from transient spikes near thresholds.

This behavior supports Zamolodchikov's proposal~\cite{Zam_2013} that a crossover/transition occurs at some $\eta_{\text{c}}$ such that $S_{11\to 11}(\infty) = +1$ for $\eta<\eta_{\text{c}}$ and $S_{11\to 11}(\infty) = 0$ for $\eta> \eta_{\text{c}}$. A toy S-matrix illustrating how this change of behavior might arise is 
\begin{equation}
\label{eq:toy_model_zam}
S(E) \simeq f(E)+\frac{E}{E+i/\epsilon} g(E)  \,,
\end{equation}
where $\epsilon$ parametrizes a pole in the complex energy plane.
For any finite $\epsilon$, $S(\infty) \simeq f(\infty) + g(\infty)$. But if $\epsilon = 0$, we have $S(\infty) \simeq f(\infty)$ instead. A possible scenario is that as $\eta$ increases $1/\epsilon$ also increases, reaching infinity at a finite value $\eta_c$, and thus triggering a transition between two different values of $S(\infty)$. In IFT, this behavior could occur because as $\eta$ increases the particle with mass $m_3$ becomes an unstable resonance, and the corresponding pole in the S-matrix contributes as in Eq.~\eqref{eq:toy_model_zam} and reaches infinity at some $\eta=\eta_c$; see Sec. 3.2 in \cite{Zamolodchikov:2013ama}. 

We sought evidence for such a crossover by estimating the probability of elastic/inelastic scattering for energies where our simulations provide a reasonable approximation to the continuum IFT. 
The results for three different energies up to $E=8m_1$ are shown in Fig.~\ref{fig:highE}. We find that close to the free fermion theory (large $\eta_{\text{latt}}$) the probability $P_{11\to 11}(E) \approx |S_{11\to 11}(E)|^2$ of elastic scattering decreases as energy increases, while close to the $E_8$ theory (small $\eta_{\text{latt}}$) it increases with $E$. The crossover between inelastic-dominated scattering 
and elastic-dominated scattering occurs at $0.7 \lesssim \eta_{\text{c,latt}} \lesssim 0.9$; it would be fascinating to investigate whether the third particle resonance $m_3$ indeed reaches infinity in the second sheet around this value.
Extrapolated to high energy, these findings are at least qualitatively compatible with the particle production probability sketched in Fig.~\ref{fig:cartoon}.
The sensitivity of our results to the simulation parameters is presented in App.~\ref{app:hyperparameter_highE}.

\section{Summary and Discussion}
Using real-time MPS simulations, we have studied the time delays and particle production probabilities as a function of scattering energy close to the free fermion and $E_8$ integrable points of Ising Field Theory. In all cases we found good agreement with predictions from form-factor perturbation theory. Additionally, we studied the $m_4$ resonance near the $E_8$ point and found reasonable agreement with results obtained using TFFSA~\cite{Gabai:2019ryw}.  

Further, we have used our method to explore the high energy behavior of scattering in IFT. Zamolodchikov conjectured that near the $E_8$ point the scattering should be purely elastic at high energies, while near the free fermion point, scattering should be completely inelastic. By extrapolating scattering probabilities at moderate energies we found support for this conjecture and were able to locate a possible region far inside the non-perturbative region of the parameter space where a crossover from trivial to completely inelastic scattering at high energies might occur.

The S-matrix is an analytic function of energy $E$ on a multi-sheeted Riemann surface which encodes the spectrum and couplings of the theory. Features like resonances and anti-bound states are encoded in poles that are a finite distance away from the physical scattering region $4m_1^2 \leq E \in \mathbb R$. We have demonstrated that we can extract masses and decay rates of resonance poles from our real-time simulations, and hence infer features of the S-matrix's analytic structure. Our work demonstrates how real-time simulations of elastic and inelastic scattering phenomena provide an instructive non-perturbative tool
for exploring the space of two-dimensional quantum field theories that arise as scaling limits of quantum spin chains. 

However, to extract more precise data further work is needed to improve our methods. For example, when extracting the time delays near the free fermion theory, we leveraged our knowledge that time delays vanish in the free theory to subtract a systematic error and hence obtain accurate results.
Far from the integrable points, however, we know little about IFT and alternative schemes will be needed;
hence we have not yet attempted to
predict the time delay far away from integrability. 
One might attain more accurate estimates of scattering phase shifts in this regime
by increasing the computational resources and/or modifying the algorithms used for time evolution or extraction of the phase shift.

Already in IFT there are many more open questions worth pursuing such as numerical studies of IFT in Euclidean time near the Yang-Lee point (see, e.g.~\cite{Zamolodchikov:2013ama}).
Moreover, our method based on real-time simulation could be extended to hunt for resonances in the higher-energy region where inelastic scattering can occur. 
By extending our analysis far into the inelastic regime one might track the location of resonances as a function of the deformation parameter $\eta$, which could provide, e.g., further insight into the role of $m_3$ for the high energy behavior of the scattering amplitude. Our methods can also be applied to other theories beyond IFT, such as scalar $\phi^{4}$ theory~\cite{Milsted:2013rxa}, the $O(3)$ sigma model~\cite{Bruckmann:2018usp}, the three-state Potts model~\cite{Zamolodchikov:1987zf}, and the Schwinger model~\cite{Papaefstathiou:2024zsu}. 

Once precision data is obtained one might use numerical results from real-time simulations as an input in the S-matrix bootstrap \cite{Paulos:2016but,Homrich:2019cbt,Correia:2022dyp}, to further narrow down the space of allowed S-matrices of non-integrable, ($1{+}1$)-dimensional quantum field theories. (Attempts to bootstrap the IFT include \cite{Paulos:2016but,Guerrieri:2020kcs,Karateev:2019ymz,Correia:2022dyp}.) Injecting approximate inelastic scattering probabilities into the bootstrap program is expected to yield $O(1)$ improvements in bounds constraining the S-matrix \cite{Paulos:2016but,Correia:2020xtr,Antunes:2023irg,Guerrieri:2024ckc,Tourkine:2021fqh,Tourkine:2023xtu}. Even qualitative estimates indicating whether high energy amplitudes are mostly elastic or mostly inelastic could be very useful. For example, the scattering of the excitations of the one-dimensional QCD flux tube was introduced in \cite{Dubovsky:2012sh} and bootstrapped in  \cite{EliasMiro:2019kyf,EliasMiro:2021nul,Gaikwad:2023hof}. We know little about how this S-matrix behaves at high energy. Simulations akin to our studies of IFT might help, starting with two-dimensional adjoint QCD recently revisited in~\cite{Dempsey:2021xpf}.

However, MPS methods like the ones discussed in the present paper are best suited for studies of one-dimensional systems, and even there are limited to simulations of processes that do not produce profoundly entangled final states. Eventually, simulations performed on quantum computers will open the door to compelling visualizations as well as quantitative studies of higher-energy scattering events producing many particles; see \cite{ciavarella2024quantum,farrell2024scalable,farrell2024quantum} for steps towards this goal. Until then, as we have demonstrated, real-time MPS simulations of scattering in ($1{+}1$)-dimensional quantum field theories provide a powerful tool to gain both intuition and quantitative results addressing many open questions. 

\begin{acknowledgments}
\label{sec:acknowledgments}
We thank Roland Farrell, Liam Fitzpatrick, Barak Gabai, Andrea Guerrieri, Alexandre Homrich, Joao Penedones, Federica Surace, Hao-Lan Xu, and  Sasha Zamolodchikov for discussions and valuable comments on our draft, and Daniel Ranard for initial collaboration.
This material is based upon work supported by the U.S.\ Department of Energy, Office of Science, Contract No. DE-AC05-06OR23177, under which Jefferson Science Associates, LLC operates Jefferson Lab. The research was also supported by the U.S.\ Department of Energy, Office of Science, National Quantum Information Science Research Centers, Co-design Center for Quantum Advantage under contract number DE-SC0012704.
DN acknowledges support from the Heising-Simons Foundation ``Observational Signatures of Quantum Gravity" collaboration grant 2021-2817, as well as the Würzburg-Dresden Cluster of Excellence ``Complexity and Topology in Quantum Matter" (ct.qmat).
JP acknowledges funding provided by the Institute for Quantum Information and Matter, an NSF Physics Frontiers Center (PHY-1733907, PHY-2317110), the DOE QuantISED program through the theory consortium ``Intersections of QIS and Theoretical Particle Physics'' at Fermilab, the DOE Office of High Energy Physics (DE-SC0018407), the DOE Office of Advanced Scientific Computing Research, Accelerated Research in Quantum Computing (DE-SC0020290), and the Air Force Office of Scientific Research (FA9550-19-1-0360).
Research at the Perimeter Institute is supported in part by the Government of Canada 
through NSERC and by the Province of Ontario through MRI. This work was additionally 
supported by a grant from the Simons Foundation (Simons Collaboration on the Nonperturbative Bootstrap \#488661) and ICTP-SAIFR FAPESP grant 2016/01343-7 and FAPESP grant 2017/03303-1. 

\end{acknowledgments}

\appendix

\section{Ising Spin Chain to Ising Field Theory}
\label{app:Ising-chain_and_conversion}

The two-dimensional Ising CFT and its deformations can be obtained as a scaling limit of the Ising spin chain which has the Hamiltonian 
\begin{equation}
H = -\sum_{j=0}^{N-1} \Big(\sigma^{z}_{j} \sigma^{z}_{j+1}  + g_x \sigma^{x}_{j} + g_z \sigma^{z}_{j}\Big),  
\end{equation}
where $\sigma_z$ and $\sigma_x$ are Pauli matrices.
Since the Hamiltonian is geometrically local, one can approximate its ground state 
using an MPS with a modest bond dimension. When $g_z=0$, the model can be mapped to the free Majorana fermion
with dispersion relation
\begin{equation}
\epsilon(k) = 2 \sqrt{1 + g_x^2 -2g_x \cos(k)},
\end{equation}
where $k$ denotes the wave number in lattice units.
For $g_x=1$ and in the limit $N \to \infty$ we obtain the free fermion Ising CFT with central charge equal to $1/2$, where the associated correlation length $\ell$ of the lattice model diverges.

When $g_z=0$ and $g_x = 1+\delta$, we obtain a free fermion QFT with mass proportional to $\delta$. More generally, taking the $N \to \infty$ limit for finite couplings, we can access a family of gapped continuum QFTs by scaling $g_z \to 0$ and $g_x \to 1$ while keeping a suitable ratio of those two coupling constants fixed. 
Specifically, we may choose the couplings to scale with a parameter $\mu \to 0$ as
\begin{eqnarray}
    g_x -1 \sim \mu, \quad g_z\sim \mu^{15/8},
\end{eqnarray}
hence fixing the ratio $\eta_\text{latt}$ defined in Eq.~\eqref{eq:lattice_eta}.
For generic values of $\eta_{\text{latt}}$ between $0$ and $\infty$, we obtain a non-integrable IFT with a varying number of stable particles. Choosing special values of $\eta_{\text{latt}}$ as $\mu \to 0$ yields integrable theories: the massive free fermion theory is obtained in the limit $\eta_{\text{latt}}\to \infty$, and the integrable $E_8$ theory is obtained for $\eta_{\text{latt}} = 0$.

This one-parameter family of continuum QFTs can also be obtained by deforming the continuum theory by relevant operators, as explained in the introduction. The resulting theories can be parametrized using $\eta$, a dimensionless ratio constructed from the dimensionful couplings which control the deformation, defined in Eq.~\eqref{eq:eta-definition}. The relationship between the parameters $\eta$ describing IFT and $\eta_{\text{latt}}$ defined in terms of lattice Hamiltonian is not straight-forward but can be determined by comparing physical data between the lattice and CFT definition of the non-integrable theories. 

\begin{figure}[t]
    \centering
    \includegraphics[scale=0.4]{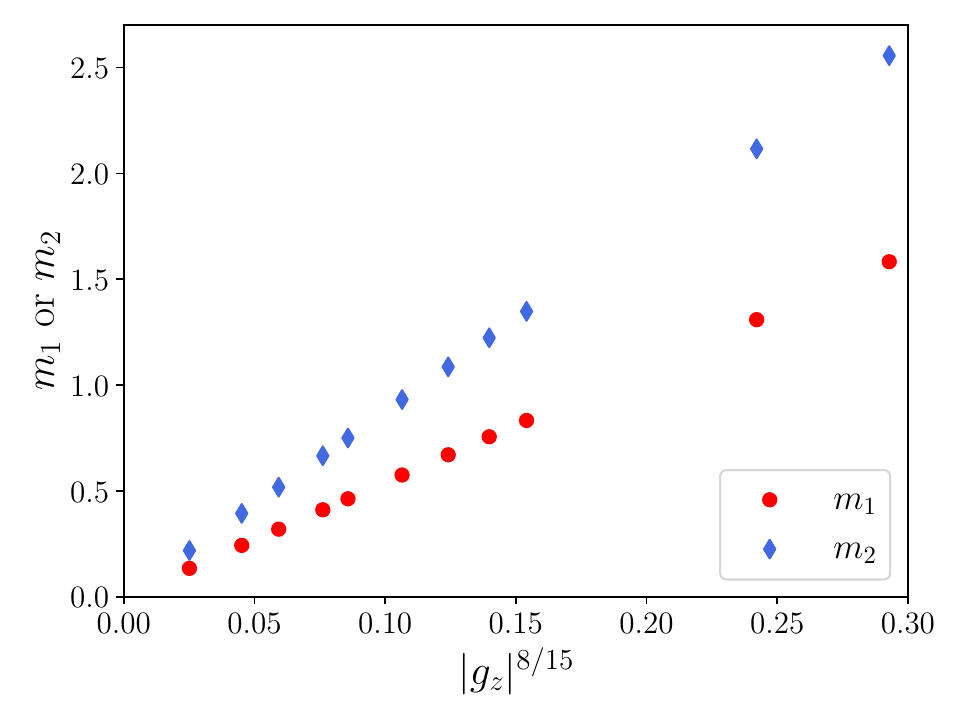}
    \caption{The mass of two lightest particles in IFT as a function of magnetic deformation at $g_{x} = 1$. 
    } 
    \label{fig:m1m2_h}
\end{figure}

For example, at the $E_8$ point $\eta=0$, there are three stable particles with masses $m_1$, $m_2$, $m_3$ (Table \ref{fig:E8_spec_det}). As $\eta$ increases from $0$ towards the free-fermion integrable point ($\eta \to \infty$), the masses $m_2$ and $m_3$ increase, eventually crossing the stability threshold and becoming virtual states. This phenomenon has been explored in previous work, and the values and $\eta_{2}$, $\eta_{3}$ where the second and third stable particles of $E_8$ theory become unstable have been calculated using continuum form-factor perturbation theory. In particular, Ref.~\cite{Delfino:2005bh} found $\eta_2 \approx 0.333(7)$ and $\eta_3 \approx 0.022$, in agreement with earlier results of Ref.~\cite{Fonseca:2001dc}. 

Thus, one can compute $m_{2}/m_1$ as a function of $\eta$ and obtain Eq.~\eqref{eq:alpha2pert} where $m_2 = 2 \cos(\alpha_2/2)$. On the lattice side, the spectrum of low-lying excitations can be computed using a variational ansatz \cite{Haegeman_2013}. This again produces $m_{2}/m_1$, but now for a range of $\eta_{\text{latt}}$, which can then be fitted to the functional dependence of $m_2/m_1$ on $\eta$. 
Surprisingly, we find that for a large range of values, namely all the way to the point where $m_2$ becomes unstable, the relationship between the lattice and continuum couplings is almost linear $\eta \approx \beta \eta_{\text{latt}}$ with $\beta \approx 0.28$.

This method is limited by the reliability of form-factor perturbation theory, as well as limits of the variational method we use to extract the particle spectrum from the lattice theory. Notably, the variational method is limited to particles below the two-particle threshold $E < 2m_{1}$, since it relies on there being a gap between the targeted particle-like excitation and other nearby excitations~\cite{Haegeman_2013}. Indeed, convergence of the method slows down as this gap closes. Beyond the two-particle threshold there exists a continuum of scattering states in addition to any further stable particles, which violates the gap assumption. Indeed, convergence problems mean that we cannot reliably compute $m_2$ or $m_{3}$ when they are very close to $2m_1$. This makes it difficult to estimate $\eta_{\text{3,latt}}$ in particular, because $m_3$ is already very close to $2m_1$ at $E_8$. This also implies that near and above $\eta_2$ we need to rely on matching other physical predictions such as phases or probabilities, as is done in Sec.~\ref{sec:nff}.

As another check that our lattice computations match continuum results, we consider how masses of stable particles depend on $g_z$ when $g_x=1$. The renormalization-group analysis predicts that masses scale as $|g_z|^{8/15}$. As shown in Fig.~\ref{fig:m1m2_h}, we find good agreement with this prediction for both $m_1$ and $m_2$, and the ratio of the slopes is consistent with $m_2 = 1.618\, m_1$ as expected from the continuum theory. The masses of particles at $E_{8}$ were also computed in Ref.~\cite{Banuls:2019qrq} using tensor network methods.

\section{Converting Lattice to Continuum Units}
\label{app:conversion}
\begin{figure}[t]
    {\centering
    \includegraphics[scale=0.65]{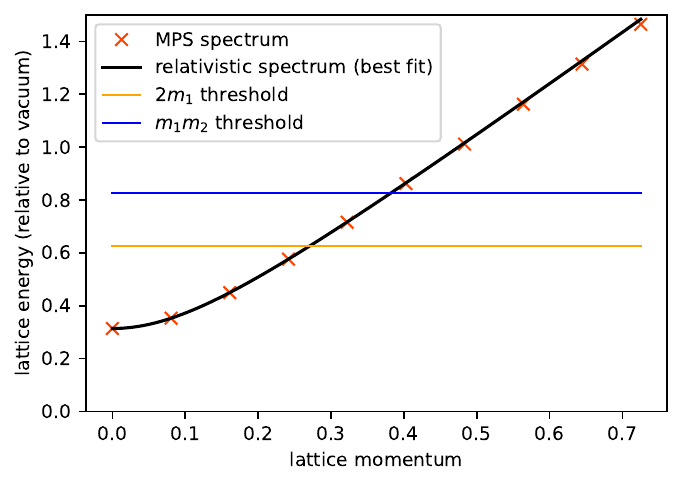}}
    \caption{The spectrum obtained from an excitation tensor ansatz can be fitted with a relativistic spectrum to obtain the conversion factors between lattice and continuum. In this example, we have chosen $g_x = 1.003$ and $g_z = 0.005$ with $\eta_{\text{latt}} \approx 0.0506$. The fit parameters are given by $\mathfrak c \approx 2.000$, $\mathfrak m \approx 0.079$.}
    \label{fig:spectrum}
\end{figure}

The parameters used in the simulations are given in dimensionless lattice units where $m_1 = c = 1$. We obtain the conversion factors from the lattice by computing the spectrum for the lowest mass and fitting a relativistic spectrum of the form
\begin{align}
    E_\text{latt}(p_\text{latt}) = \sqrt{\mathfrak{m}^2 \mathfrak{c}^4 +  \mathfrak{c}^2 p_\text{latt}^2}
\end{align}
to the first two data points. Conversion between lattice and continuum units is then given by $E_\text{phys} = E_\text{latt} / (\mathfrak{mc}^2)$, $p_\text{phys} = p_\text{latt} / (\mathfrak{mc})$, $t_\text{phys} = \mathfrak{mc}^2 \, t_\text{latt} $, and $x_\text{phys} = \mathfrak{mc} \, x_\text{latt}  $. An example is shown in Fig.~\ref{fig:spectrum}. 

\section{Time Delay}
\label{app:time_delay}
To understand why the quantity given in Eq.~\eqref{eq:def_time_delay} is called \emph{time delay} consider a scattering experiment. For simplicity we take the case where a wave packet is scattered against a target in $1{+}1$ dimensions. At early times the wave packet is given by some asymptotic state $\ket{\psi_\text{in}(t)}$. At late time, after the scattering, the quantum state is given by $\ket{\psi_\text{out}(t)}$ which is related to the in-state via
\begin{align}
    \ket{\psi_\text{out}(t)} =  S \ket{\psi_\text{in}(t)},
\end{align}
where $S$ is the S-matrix and both $\ket{\psi_\text{in}(t)}$ and $\ket{\psi_\text{out}(t)}$ time-evolve with the free Hamiltonian.
In a position space representation this results in a wavefunction of the form
\begin{align}
    \label{eq:wf_out}
    \psi_\text{out}(t,x) \sim \int dp e^{- i E t + i p x} e^{- \frac{(p - p_0)^2}{2 \sigma^2}} S(p),
\end{align}
and $\psi_\text{in}(t,x)$ is given by the same expression with $S(p) = 1$. To get a spatial profile of the wavefunction after scattering, we can approximate the integral in Eq.~\eqref{eq:wf_out} by the leading saddle-point contribution, which is valid if the momentum space width $\sigma$ of the wave packets is sufficiently small. 
Up to a phase and constant factors, this yields a Gaussian wave packet 
\begin{align}
    \psi_\text{out}(t,x) \sim \exp{\left(- \sigma^2 (t \partial_p E_0 - x + i \partial_p \log S(p_0))^2\right)},
\end{align}
where $E_0 = E(p_0)$. Thus, while before the scattering the wave packet was centered around the trajectory $x(t) = v t$, with $v = \partial_p E_0$, after the scattering it follows the trajectory
\begin{align}
    x(t) = v (t - \Delta t)
\end{align}
with time delay
\begin{align}
    \Delta t = - i \partial_E \log S(p_0).
\end{align}
Depending on the momentum dependence of $S(p)$, the wave packet thus arrives later ($\Delta t >0$) or earlier ($\Delta t <0$) at a given location, which justifies the name time delay. If instead we scatter two wave packets, at leading order each wave packet suffers a time delay given by the derivative of the S-matrix with respect to the wave packet's momentum evaluated at the scattering energy.

\section{Hyperparameter Dependence of Time Delay}
\label{app:hyperparameter_timedelay}
\begin{figure*}[ht!]
    \centering
    \subfloat[]{
    \includegraphics[scale=0.65]{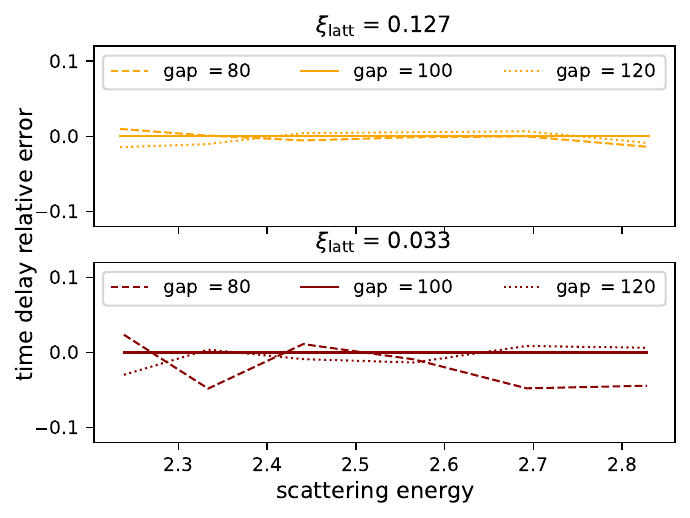}
    }\hspace{0.01\textwidth}
    \subfloat[]{
    \includegraphics[scale=0.65]{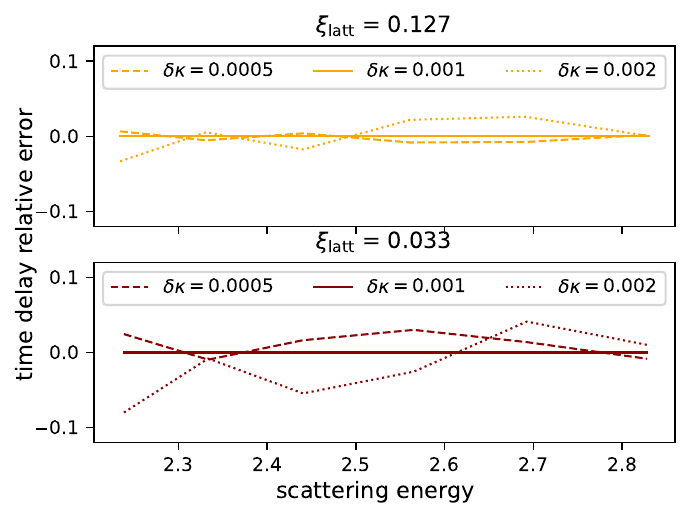}
    }
    \caption{(a) The relative error of the time delay as a function of energy for different choices of minimal distance between excitation tensors in the reference state. (b) The relative error of the time delay as a function of energy for different $\delta \kappa$. The gap is $\Delta n_{\text{min}}$ as defined in Eq.~\eqref{eq:reference_state}.}
    \label{fig:gap_dependence}
\end{figure*} 
\paragraph{Minimum Distance of Excitation Tensors $\Delta n_\text{min}$} 
In this work we are, amongst other things, interested in the time delay of two-particle scattering below the inelastic threshold. To obtain the time delay, we project outgoing states onto a certain set of plane-wave states, as explained in Sec.~\ref{p:time_delay}.
To construct those plane-wave states we insert excitation tensors into a vacuum MPS and Fourier transform with respect to their location. The Fourier transform of a single excitation tensor creates, to good approximation, a state in the one-particle subspace of the Hilbert space. However, bringing two excitation tensors close to each other creates a state whose overlap with other $n$-particle states is not controlled. We therefore implemented a constraint that the insertions of two excitation tensors in the Fourier transform have to be at least $\Delta n_\text{min}=100$ sites apart.

To check that our results do not significantly depend on this choice we have repeated the computation for different choices of this gap size and for different datasets. Examples of such data from near the free fermion point are shown on the left in Fig.~\ref{fig:gap_dependence}, which indicates that the relative error is typically below $5\%$.

\paragraph{Finite Momentum Difference.}
Similarly, in the computation of the time delay we have to take a difference between phases corresponding to different momenta. Due to numerical artifacts, the result will depend on a choice of $\delta \kappa$, the difference of the lattice momenta. In the right panel of Fig.~\ref{fig:gap_dependence} we show the relative error in the time delay for different choices of $\delta \kappa$. As can be seen from the plots, the results for different choices agree to within $10\%$ with the value used in the main text ($\delta \kappa = 0.001$).

\paragraph{Bond Dimension}
In Table \ref{tab:timedelay_hyper1} we show the dependence of some our simulations used to determine the time delay on bond dimension. All simulations are close to the free fermion theory, since here the time delay is smallest and thus the relative error would be biggest. As we get closer to the free fermion theory, there still is some dependence on the bond dimension $D$, even for $D \geq 64$. However, as we discuss in App.~\ref{app:subtraction}, the main source of uncertainty comes from the choice of time-step size. Unfortunately computational constraints prevent us from increasing both the bond dimension and decreasing the step size. Therefore, we focus on improving our simulations by going to step-size $dt_\text{latt} = 0.04$ but stay at a bond dimension of $D=64$.  
\begin{table}[h!]
\centering
\renewcommand{\arraystretch}{1.05}
\setlength{\tabcolsep}{8pt}
\begin{tabular}{|c|c|c|c|c|c|}
\hline
$D$ & $g_x$ & $g_z$ & $\eta_\text{latt}$ & $E$ & $- i \partial_E \log S$ \\
\hline
64 & 1.2 & 0.0062 & 3.0 & 2.33 & $-0.116$ \\ 
72 &  1.2 & 0.0062 &3.0 & 2.33 & $-0.116$ \\ 
\hline
64 & 1.2 & 0.0062 &3.0 & 2.44 & $-0.100$ \\ 
72 & 1.2 & 0.0062 &3.0 & 2.44 & $-0.106$ \\
\hline
64 & 1.2 & 0.0031 &4.3 & 2.34 & $-0.040$ \\ 
72 & 1.2 & 0.0031 &4.3 & 2.34 & $-0.037$ \\
\hline
64 & 1.2 & 0.0031 &4.3 & 2.44 & $-0.035$ \\ 
72 & 1.2 & 0.0031 &4.3 & 2.44 & $-0.029$ \\
\hline
64 & 1.2 & 0.0016 &6.2 & 2.33 & $-0.023$ \\ 
72 & 1.2 & 0.0016 &6.2 & 2.33 & $-0.022$ \\
\hline
64 & 1.2 & 0.0016 &6.2 & 2.44 & $-0.022$ \\ 
72 & 1.2 & 0.0016 &6.2 & 2.44 & $-0.017$ \\
\hline
\end{tabular}
\caption{The dependence of the simulations for time delay on the bond dimension with $dt_\text{latt} = 0.05$, $\sigma = 100$ and $N=2000$.}
\label{tab:timedelay_hyper1}
\end{table}

\section{Near Free Fermion Theory}

\subsection{Relative Probabilities of \texorpdfstring{$11 \to 111$}{11 to 111} Scattering}
\label{app:3body}
The relative probability plotted in Fig.~\ref{fig:3particlesKin} is given by $(A/B)/(A/B)_{y=1}$ where 
\begin{eqnarray*}
       A&=& 4 (x+y)^2 (x y+1)^2\\
      &\times& \left(x^4 y-x^3 y^2-x^3-x^2 y-x y^2-x+y\right) \\
    &\times&\left(x^4 y-2 x^3 y^2-2 x^3+5 x^2 y-2 x y^2-2 x+y\right)^2 \\
    &\times&\left(2 x^4 y-2 x^3 y^2-x^3+5 x^2 y-x y^2-2 x+2 y\right)^2 \\
    &\times&\left(2 x^4 y-x^3 y^2-2 x^3+5 x^2 y-2 x y^2-x+2 y\right)^2 \\
   B&=&y (x-y)^2 \left(x^2-x y+1\right) (x y-1)^2 \left(x^2 y-x+y\right)\\ &\times&\left(x^2-x y+y^2\right)^2 \left(x^2 y^2-x y+1\right)^2 \\
   &\times&\sqrt{\left(x^2-x y+1\right) \left(x^2 y-x+y\right)} \\
   &\times& \sqrt{x^4 y-x^3 y^2-x^3-x^2 y-x y^2-x+y}
\end{eqnarray*}
with $x=\sqrt{\frac{1+v}{1-v}}$ and $y=\sqrt{\frac{1+v_\text{middle}}{1-v_\text{middle}}}$ where $v>0$ is the velocity of the right moving incoming particle (we assume two particles colliding head-on in the center of mass frame) and $v_\text{middle}$ is the velocity of the middle particle.  The ratio $(A/B)$ is the five particle form factor of the $\sigma$ operator times the relevant phase space kinematical factor to convert it into an appropriate probability. 

Kinematically we have 
\begin{equation}
   x \in [2.618,3.732] \label{rangeX}
\end{equation}
since by assumption $3<E<4$ so we are in the regime where no more than three particles can be produced. We also have
\begin{equation}
    |y|<\frac{x^4+5 x^2+\sqrt{x^8-6 x^6-5 x^4-6 x^2+1}+1}{4 \left(x^3+x\right)}
\end{equation}
    defining the green region in Fig.~\ref{fig:3particlesKin}. In that figure we took $x=3$ inside the range (\ref{rangeX}). The outer boundaries of the blue and yellow regions in that figure are given by the maxima and minima velocities 
    \begin{equation}
        \pm \frac{\sqrt{x^8-6 x^6-5 x^4-6 x^2+1}}{x^4-x^2+1}
    \end{equation}
    which three particles can kinematically access. The total probability Eq.~(\ref{ProdZZ}) is the integral over all such allowed three particles states. Indeed, the remaining integral over $t$ in that expression is nothing but this phase space integral after a simple rescaling \cite{ZZ}.
    
\subsection{Correcting Bias in Time Delay}
\label{app:subtraction}

\begin{figure*}[ht!]
    \centering
    \subfloat[\label{fig:correction}]{
    \includegraphics[scale=0.65]{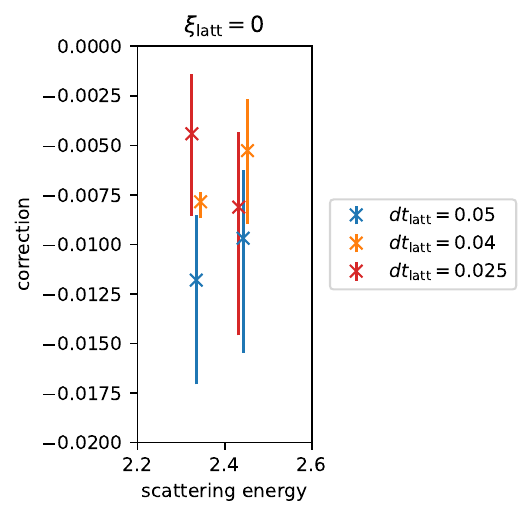}
    }\hspace{0.01\textwidth}
    \subfloat[\label{fig:subtraction}]{
    \includegraphics[scale=0.65]{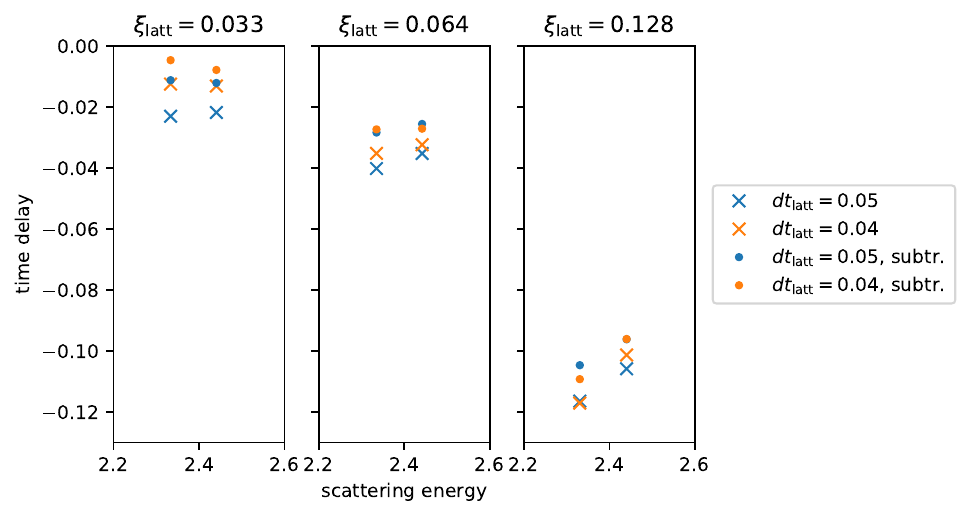}
    }
    \caption{
    (a) Time delays extracted from scattering simulations in free fermion theory ($\xi_\text{latt}=0$), for three different values of the time step size $dt_\text{latt}$. The points are slightly displaced to make the error bars more visible. (b) Time delays with and without subtracting the value found for free fermions.
    }
\end{figure*}

Simulating IFT at $\eta = \infty$, i.e.\ at the free fermion point, we find an approximate time delay between $-0.01$ and $-0.03$, depending on the scattering energy, while the expected value is $0$; see Fig.~\ref{fig:phase}. This one-sided bias towards negative values in the simulated data motivated us to hypothesize that the error can be attributed to a systematic bias which is approximately given by the negative of the time delay we find at the free fermion point. After adding the bias to the simulated data taken near the free fermion value of $\eta$ we found good agreement between the simulation and the theoretical expectations.

We checked the dependence of the time delay against various bond dimensions and while there is some change when going from $D=64$ to $D=72$ as shown in Table \ref{tab:timedelay_hyper1}. We find an even stronger dependence on the step size $dt_\text{latt}$ of the RK4/5 integrator. 
In this appendix we discuss the dependence of the time delay on the step size and give supporting evidence that the employed subtraction indeed improves the quality of the data.

It seems plausible that errors in the MPS time evolution due to the finite step size, while very small per step, accumulate over the $20\,000$ time-steps of each simulation in a way that biases relative phases between two different states.
To obtain a qualitative picture of the dependence, we simulated scattering in the free fermion theory at two energies for $dt_\text{latt} = 0.04$ and $dt_\text{latt} = 0.025$ in addition to the usual $dt_\text{latt} = 0.05$. As is shown in Fig.~\ref{fig:correction} we do find a significant dependence on $dt_\text{latt}$ although the dependence does not monotonically become smaller as we decrease $dt_\text{latt}$. However, note that the errors we associate with those data points are relatively large and so we can merely conclude that while the value of $dt_\text{latt}$ seems to play an important role in our analysis, there are additional factors affecting the time delay obtained from simulations. 

To test our hypothesis that the data suffers from a systematic bias that can at least in part be removed by our proposed subtraction, we simulated scattering at two energies for all values of $\xi$ shown in Fig.~\ref{fig:phase}. We show the time delay with and without subtraction in Fig.~\ref{fig:subtraction} for a step size of $dt_\text{latt} = 0.04$ and $dt_\text{latt} = 0.05$. While we are not able to make a very precise quantitative statement, we point out that qualitatively the subtracted data points for different values of $dt_\text{latt}$ are closer together than the uncorrected ones. This indicates that at least some part of the error comes from a $dt_\text{latt}$-dependent bias which is shared between all simulations at the same value of $dt_\text{latt}$.

\section{Near \texorpdfstring{$E_{8}$}{E8} Theory}
\subsection{Hyperparameter Dependence of Probabilities}
\label{app:hyperparameter_probability}
 
In Tables.~\ref{tab:nearE8_prob1} and ~\ref{tab:nearE8_prob2}, we show the negligible dependence of the bond dimension $D$ on $P_{11 \to 11}$ for two different $\eta_{\text{latt}}$ close to $E_8$ limit in Fig.~\ref{fig:probabilities_simulationE8}.

\begin{table}
\centering
\renewcommand{\arraystretch}{1.05}
\setlength{\tabcolsep}{8pt}
\begin{tabular}{|c|c|c|}
\hline
$D$ & $P_{11 \to 11}$ & $E$ \\
\hline
32 & 0.9332 & 3.195 \\
64 & 0.9353 & 3.195 \\ \hline 
32 & 0.9299 & 3.072 \\
64 & 0.9290 & 3.072 \\ \hline 
32 & 0.9337 & 3.011 \\
64 & 0.9335 & 3.011 \\ \hline 
32 & 0.9332 & 2.952 \\
64 & 0.9320 & 2.952 \\ \hline 
32 & 0.9258 & 2.894 \\
64 & 0.9252 & 2.894 \\ \hline 
32 & 0.9154 & 2.836 \\
64 & 0.9153 & 2.836 \\ \hline 
32 & 0.9023 & 2.780 \\
64 & 0.9031 & 2.780 \\ \hline 
32 & 0.9011 & 2.725 \\
64 & 0.9026 & 2.725 \\ \hline 
32 & 0.9459 & 2.671 \\
64 & 0.9452 & 2.671 \\
\hline
\end{tabular}
\caption{Table of $D$, $P_{11 \to 11}$, and $E$ with $dt_\text{latt}=0.05$, $\sigma=80$, $N=1400$, $\eta_\text{latt}=0.152~(g_x = 1.015, g_z = 0.013)$ for results shown in Fig.~\ref{fig:probabilities_simulationE8}.}
\label{tab:nearE8_prob1}
\end{table}

\begin{table}
\centering
\renewcommand{\arraystretch}{1.05}
\setlength{\tabcolsep}{8pt}
\begin{tabular}{|c|c|c|}
\hline
$D$ & $P_{11 \to 11}$ & $E$ \\
\hline
32 & 0.9656 & 3.108 \\
64 & 0.9667 & 3.108 \\ \hline 
32 & 0.9653 & 3.046 \\
64 & 0.9648 & 3.046 \\ \hline 
32 & 0.9681 & 2.985 \\
64 & 0.9677 & 2.985 \\ \hline 
32 & 0.9649 & 2.865 \\
64 & 0.9645 & 2.865 \\ \hline 
32 & 0.9588 & 2.807 \\
64 & 0.9584 & 2.807 \\ \hline 
32 & 0.9490 & 2.751 \\
64 & 0.9501 & 2.751 \\ \hline 
32 & 0.9482 & 2.695 \\
64 & 0.9499 & 2.695 \\ \hline 
\end{tabular}
\caption{Table of $D$, $P_{11 \to 11}$, and $E$ with $dt_\text{latt}=0.05$, $\sigma=80$, $N=1400$, $\eta_\text{latt}=0.106~(g_x = 1.01, g_z = 0.012)$ for results shown in Fig.~\ref{fig:probabilities_simulationE8}.}
\label{tab:nearE8_prob2}
\end{table}

\section{Resonances}
\label{app:resonance_section}

\subsection{Toy Model}
\label{app:resonance_toy}

As discussed in Sec.~\ref{sec:resonances}, when the total energy $E$ in an elastic scattering process is close to a resonance, we expect to observe exponential temporal decay in outgoing scattering wavefunctions. The decay rate is closely related to the resonance width~$\Gamma$, as we will now explain. 

\begin{figure}[ht]
    \centering
\begin{tikzpicture}[scale = 1]
\draw[->] (-3,0) -- (3,0);
\draw[->] (0,-2) -- (0,2);
\draw (2-0.05, -0.5-0.05) -- ++(0.1,0.1);
\draw (2+0.05, -0.5-0.05) -- ++(-0.1,0.1);
\node[below] at (2.0,-0.5) (at) {$E(p) = m - i\frac{\Gamma}{2}$};
\draw (-2-0.05, 0.5-0.05) -- ++(0.1,0.1);
\draw (-2+0.05, 0.5-0.05) -- ++(-0.1,0.1);
\draw (-2,-0.5) circle (0.05);
\draw (2,0.5) circle (0.05);
\draw[very thick, dotted] (0,1) -- (0,2);
\node[right] at (0,1) (bct) {$p = i m_1$};
\draw[very thick, dotted] (0,-1) -- (0,-2);
\end{tikzpicture}
    \caption{Analytic structure of $S^{\texttt{Toy}}_{11\to11}(E)$ in the complex momentum plane. The circles denote zeros and crosses denote the poles. The dotted line along the imaginary axis shows the location of branch cuts.}
    \label{fig:complex_plane}
\end{figure}
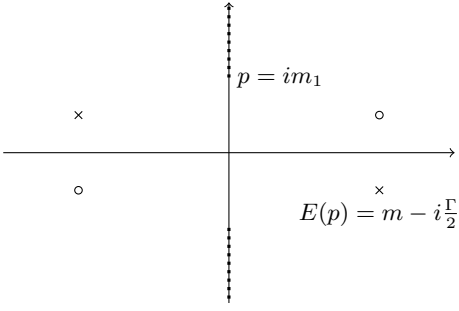

Consider the toy S-matrix 
\begin{equation}
    \label{eq:reso_smatrix_app}
    S^{\texttt{Toy}}_{11\to11}(E) = \frac{E - m - i\frac{\Gamma}{2}}{E - m + i\frac{\Gamma}{2}},
\end{equation}
of Eq.~\eqref{eq:reso_smatrix}. It describes $11 \rightarrow 11$ scattering of two particles of mass $m_1$ near a resonance at mass $m$. To simplify our analysis, we suppose that the wavefunction of the first particle is a Gaussian wave packet in momentum space with mean $p_0$ and variance $\sigma^{-2}$, while the wavefunction of the other particle is a plane wave with energy $E_2 = \frac{m}{2}$. The (unnormalized) outgoing wavefunction takes the form
\begin{equation}
    \label{eq:reso_psi}
  \psi(t,x) = \frac \sigma {\sqrt{2 \pi}} \int_{-\infty}^\infty dp \; e^{-\frac {\sigma^2}{2} (p - p_0)^2} S\left(E(p)\right) e^{-i(E(p)t - px)},
\end{equation}
where $E(p) = E_1(p) + \frac{m}{2}$ is the total energy, and $E_1(p) = \sqrt{p^2+m_1^2}$.
In the complex plane, the integrand has poles (zeros) where $E(p) = m \mp i\frac{\Gamma}{2}$, as well as branch cuts extending from $p=\pm i m_1$ along the imaginary axes, as shown in Fig.~\ref{fig:complex_plane}.

Using $E(p) = E_1(p) + \frac {m} 2$ we can also express the locations of the poles in terms of $p$ and find that the poles occur at
\begin{equation}
  \label{eq:psq_parts}
  p^2 = \frac{1}{4}(m^2 - \Gamma^2) - m_1^2 - i\frac{\Gamma m}{2}.
\end{equation}
If we let $p^2 = A + iB$, with $A > 0$ and $B < 0$ implicitly defined in Eq.~\eqref{eq:psq_parts}, we can write the corresponding momentum as $p = \pm (a + ib)$ where
\begin{equation}
  a = \sqrt{\frac{A}{2}} X, \;
  b = - \sqrt{\frac{B^2}{2A}} X^{-1}, \; X := \sqrt{1 + \sqrt{1 + \frac{B^2}{A^2}}}.
\end{equation}
As a further simplification, we assume that $2m_1 \ll m$, so the resonance is far from the threshold and the scattered particles are highly relativistic. If additionally $\Gamma$ is small compared to $m$, we have $B \ll A$, $X \approx \sqrt{2}$ and thus $a \approx \sqrt A \approx \frac{m}{2}$ and $b \approx - \frac{\Gamma} 2$. 
If instead we write $E(p)=E_1(p) + E_2$, and expand to leading order in $\Gamma$, we obtain another useful approximation:
\begin{align}\label{eq:b-formula}
    b  \approx - \frac \Gamma 2\sqrt{\frac{(E_2-m)^2}{(E_2-m)^2 - m_1^2}} .
\end{align}

As the resonance is at very high energy, we can approximate the energy $E(p)$ in the S-matrix by its ultra-relativistic limit and hence see that the momentum integral has an isolated pole. If we suppose that the wave packet is peaked at the resonant energy $p_0 = a$, then by changing the integration variable to $\tilde p := p - p_0$, we have
\begin{equation}
  \label{eq:reso_psi_approx}
  \psi(t,x) \approx \frac \sigma {\sqrt{2 \pi}}  \int_{-\infty}^\infty d\tilde p \; e^{-\frac{\sigma^2}{2}\tilde p^2} \frac{\tilde p - i\frac{\Gamma}{2}}{\tilde p + i\frac{\Gamma}{2}} e^{-i(E(\tilde p)t - \tilde px)} e^{i\frac{m}{2}x},
\end{equation}
again assuming $m\gg 2m_1$ and $\Gamma \ll m$; we have also assumed that $\sigma$ is large enough so that the contributions far from the resonance are negligible. While we used the approximation $E_1(p)\approx |p|$ to derive Eq.~\eqref{eq:reso_psi_approx}, there will be a contribution from the pole located at $\tilde p = i b$ even if the energy is not highly relativistic. 

To capture the most important terms in the wavefunction, we may evaluate \eqref{eq:reso_psi} or \eqref{eq:reso_psi_approx} using the stationary phase approximation, which is justified if $\sigma^2$ is sufficiently large. Thus 
we deform the $\tilde p$ integration contour in \eqref{eq:reso_psi} or \eqref{eq:reso_psi_approx}, such that it passes through a point of stationary phase. The phase factor is stationary for $\tilde p =\tilde p_\text{stat}= \frac{1}{\sigma^2} i(x - v t)$, where $v = \partial E/\partial \tilde p$. For large negative $x-vt$, the deformed contour that passes through $\tilde p_\text{stat}$ sits well below the resonance pole; therefore, by deforming the contour from the real axis into the lower half plane, we pick up a contribution from the residue of the pole. Thus, replacing $E(\tilde p) = m - \frac{i \Gamma}{2}$ and $\tilde p = ib$ (their values at the pole) we find
\begin{equation}
  \label{eq:reso_pole}
  \psi_\text{pole}(t,x) \approx - \sqrt{2\pi} \Gamma \sigma e^{\frac{\sigma^2 b^2}{2}}  e^{-\frac \Gamma 2 t} e^{-b x} e^{i (\frac{m}{2} x - m t)}.
\end{equation}
Using the approximation $b\approx -\Gamma/2$, we find that this pole contribution decays exponentially with rate $\Gamma/2$ when $t-x$ is large and positive. 
If, on the other hand, $t-x$ were large and negative, we would deform the $\tilde p$ integration contour into the upper half plane, picking up a contribution from the pole at $\tilde p = -2p_0 - i b$.  
The pole contribution would then be suppressed by the factor. $\exp(-2 \sigma^2 (p_0^2 - b^2/4))$.
Under the assumption $\Gamma \ll m$, this contribution is negligible if the wave packet is centered at the resonant energy $p_0 \approx m$.

Away from the ultra-relativistic limit, $b$ deviates from $-\Gamma/2$ as in Eq.~\eqref{eq:b-formula}, and therefore the temporal and spatial decay rates are not precisely the same. Instead we have
\begin{align}\label{eq:b-formula-again}
    b  \approx - \frac \Gamma 2\sqrt{\frac{(E_2-m)^2}{(E_2-m)^2 - m_1^2}} 
    \approx-\frac \Gamma 2 \sqrt{\frac{m^2}{m^2 - 4 m_1^2}} .
\end{align}
Both $\Gamma$ and $b$ can be determined from our simulations as explained below. From these values we can, at least in principle, determine the mass $m$ of the resonance by inverting Eq.\eqref{eq:b-formula} to obtain
\begin{align}
\label{eq:m_from_gamma_and_b}
    m = E_2 + \frac{m_1}{\sqrt{1- \frac{\Gamma^2}{4b^2}}}.
\end{align}
As can be seen from this expression, $\Gamma/2|b|$ is the velocity of particle 1 such that the total energy of the scattering process is exactly the resonant mass $m$. In practice, we use a different method to determine the mass $m$ of a resonance, explained in App.~\ref{app:resonance}.

\subsection{Decay and Multiple Tracks}
\label{app:resonance_gap}
This model helps us understand the origin of the gap between tracks visible in Fig.~\ref{fig:two_tracks}b. As we will now explain, the two tracks arise when the momentum-space width $\sigma^{-1}$ of the scattered wave packet is large compared to the width of the resonance (but still small enough to justify our stationary phase approximation).  After scattering there are two contributions to the wavefunction, one due to trivial scattering and one coming from the interaction mediated by the resonance. Interference between those two contributions produces the two parallel tracks seen in Fig.~\ref{fig:two_tracks}b. Interestingly, and perhaps contrary to naive expectations, the scale of the distance between the two tracks is not set by the lifetime of the resonance, but instead by the width of the wave packet. 

For simplicity, we again consider the relativistic limit $E \approx p$, which implies that $b \approx - \frac \Gamma 2$. 
As explained in the previous section, the full wavefunction after scattering is given by $\psi = \psi_\text{pole} + \psi_\text{contour}$, where $\psi_\text{pole}$ is given in Eq.~\eqref{eq:reso_pole} and $\psi_\text{contour}$ is Eq.~\eqref{eq:reso_psi_approx} evaluated along the contour of stationary phase,
\begin{align}
\label{eq:cont}
\psi_\text{contour}(t,x) \approx e^{- \frac{(t-x)^2}{2 \sigma^2}} e^{i (\frac{m}{2} x - m t)}.
\end{align}
Here we have assumed that $\frac{t-x}{\sigma^2} \gg \Gamma$, such that the pole is picked up when we deform the contour from the real axis, and the phase factor coming from the S-matrix is approximately $1$. 

The total wavefunction $\psi_\text{pole} + \psi_\text{contour}$ is only determined up to a global phase. As can be seen from the explicit expressions Eqs.~\eqref{eq:reso_pole} and \eqref{eq:cont}, the phase-factors $e^{i (\frac{m}{2} x - m t)}$ of both contributions agree and we can thus choose the global phase such that both contributions are real. Crucially, the two contributions have opposite sign, which leads to the aforementioned destructive interference and the appearance of two tracks. Due to corrections to the stationary phase approximation that we have ignored, this destructive interference is not complete.
Therefore, in practice the energy expectation value between the two tracks does not reach its vacuum value. 

The relative importance of the contribution to scattering due to the resonance is controlled by the dimensionless combination $\Gamma \sigma$. We are interested in the case of a narrow resonance and therefore $\Gamma \sigma \ll 1$. Since the contour contribution decays exponentially with $\exp(-(t-x)^2/(2\sigma^2))$ while the pole contribution decays as $\exp(- \Gamma (t-x)/2)$, we expect that the pole contribution becomes important when $t-x$ is sufficiently large. This is exactly when the approximation of the total wavefunction as a contour contribution plus a pole contribution is reliable.

Given our choice of phase, the total wavefunction (made real by an appropriate choice of overall phase) is positive near the unscattered trajectory at $t-x = 0$ and negative when the pole contribution becomes important at some $t - x > 0$. At very large $t-x$ it vanishes. Thus, there is a local maximum of the probability $|\psi(t,x)|^2$ at some $u :=t-x = u_\text{track}$, which becomes the location of the second track.
This local maximum of the probability $|\psi(t,x)|^2$ is a local minimum of the wavefunction $\psi(t,x)$; by differentiating $\psi_\text{pole}+\psi_\text{contour}$ with respect to $u$ and approximating $e^{b^2\sigma^2/2}\approx 1$ we find that that the minimum occurs at $u_\text{track}$ satisfying
\begin{align}
    \label{eq:track_equation}
    \sqrt{\frac{\pi}{2}}
    \Gamma ^2 \sigma 
    e^{-\Gamma  u_\text{track}/2 }= \frac{u_\text{track} e^{-u_\text{track}^2/2 \sigma ^2}}{\sigma ^2}.
\end{align}

Though we cannot give an explicit solution to this equation, we can estimate the behavior of solutions in the regime of interest. Defining
\begin{align}
\Delta = u_\text{track}/\sigma, \quad \gamma= \Gamma \sigma,
\end{align}
Eq.~\eqref{eq:track_equation} becomes
\begin{align}
\log\left(\frac{1}{\gamma^2\sqrt{\pi/2} }\right)+\frac{\gamma\Delta}{2} = -\log \Delta + \frac{\Delta^2}{2}.
\end{align}
which for $\gamma \ll 1$ has the approximate solution
\begin{align}
  u_\text{track}/\sigma  =\Delta\approx 2 \sqrt{\log\frac{1}{\gamma}} = 2 \sqrt{\log\left(\frac{1}{\Gamma\sigma}\right)}
\end{align}
We conclude that the scale $u_\text{track}$ of the distance between the two tracks is set primarily by the wave-packet width $\sigma$ and depends only weakly on the resonance lifetime $\Gamma^{-1}$. 

We should check that our approximate solution for $u_\text{track}$ is within the domain of validity of our approximate evaluation of the wavefunction.  In our computation of $\psi_\text{contour}$ we replaced the phase factor in the S-matrix by 1, which is a reasonable approximation for $u/\sigma^2 \gg \Gamma$. Furthermore, in this regime the stationary phase condition is satisfied at $\tilde p = -i u/\sigma^2$, far below the pole in the lower half plane, so the contribution $\psi_\text{pole}$ arises when we deform the contour.

\subsection{Decay into the Asymptotic 11 Sector}
\label{app:reso_toy_asymp}
The resonating component of the outgoing wavefunction -- a decaying, unstable particle -- is, by definition, outside of the asymptotic sectors of the stable particles. We see in our simulations, for example in Fig.~\ref{fig:two_tracks}b, that resonant scattering leaves an excitation which remains in the scattering region for a long time. We observe that, as expected, this excitation has little to no overlap with the asymptotic 11 sector. This can be clearly seen for example in Fig.~\ref{fig:app_reso_spatial_decay}, where we plot ``position-basis'' probabilities for the left particle
\begin{equation}
    P(11,\; n) := \sum_{n'} P(11,\; n,n'),
\end{equation}
where 
\begin{equation}
P(11,\; n,n') := |\langle (n,B_1),(n',B_1) | \psi \rangle|^2,    
\end{equation}
is defined using the orthonormal excitation position states of Eq.~\eqref{eq:reference_state}. The probability of observing an excitation at position $n$ decreases near the central peak in the energy expectation value at $n \approx 1000$. We should therefore expect the probability $P(\lnot\; 11, t)$ of being outside of the asymptotic 11 sector to decay in time with the resonance excitation.

In this section we argue that, in our resonance model Eq.~\eqref{eq:reso_smatrix_app} describing elastic scattering of two particles of type 1,  $P(\lnot\; 11, t)$ indeed decays exponentially in $t$ at the same rate $\Gamma$ as the pole contribution \eqref{eq:reso_pole}.

If an on-resonance scattering event leaves a resonance excitation in the scattering region, we can conclude that the probability of the state 
not being in the 11 sector behaves as 
\begin{align}
    \label{eq:probabilities_exp_decay_eq}
    P(\lnot\; 11, t) \sim \alpha_0 \left(1- P_{|x|>x_0}(11,t)\right),
\end{align}
where $1 - P_{|x|>x_0}(11,t)$ is the probability to find two particles in the $1$ sector outside a band of width $2 x_0$ around the scattering region. Here $x_0$ must be chosen such that excitations at $|x|>x_0$ can be treated as asymptotic states. The proportionality constant $\alpha_0$ takes into account that 
our choice of $x_0$ may exclude some 11 contributions within this central band.
In this case the time-dependence of the function $P_{|x|>x_0}(11,t)$ can be computed from the results of App.~\ref{app:resonance_toy}.

To obtain the time dependence of $P_{|x|>x_0}(11,t)$ (we will assume that $x > 0$, the case of $x < 0$ follows analogously) we proceed as follows. First, we split the region $x > x_0$ into two parts. The first region $x_0 < x < x_1(t)$ is the region which is far enough from the trajectory such that the exponential decay of the wavefunction in $t$ and $x$ as discussed in App.~\ref{app:resonance_toy} is a good approximation. The second region $x_1(t) < x$ contains the tracks. We choose $x_1(t) = \bar x_1 + v t$, where $v$ is the velocity with which the outgoing trajectories move outwards. With this choice $x_1(t)$ is co-moving with the trajectories and this guarantees that $P_{x>x_1(t)}(11,t)$ is constant. The remaining contribution $P_{x_1(t)>x>x_0}(11,t)$ is then proportional to a spatial integral over the absolute value squared of the wavefunction $\sim \exp(- \Gamma t - 2 b x)$ as discussed in App.~\ref{app:resonance_toy}, i.e.\ the integral 
\begin{align}
\begin{split}
    e^{-\Gamma t} \int_{x_0}^{x_1(t)} dx e^{-2 b x} &\sim e^{-\Gamma t} \left(e^{-2 b (\bar x_1 + v t)} -  e^{-2 b x_0} \right) \\
    &\sim  e^{-2 b \bar x_1} -  e^{-2 b x_0} e^{-\Gamma t} ,
    \end{split}
\end{align}
where we have used that $\Gamma/2b = -v$ -- see the discussion around Eq.~\eqref{eq:m_from_gamma_and_b}.
In total, we therefore find the functional dependence
\begin{align}
P_{|x|>x_0}(11,t)\approx \alpha_1 + \alpha_2 e^{-\Gamma t},
\end{align}
with some time-independent quantities $\alpha_i$. Using the relation \eqref{eq:probabilities_exp_decay_eq} it immediately follows that the functional dependence of $P(\lnot\; 11, t)$ on time has the same structure,
\begin{align}
P(\lnot\; 11, t) \approx \beta_1 + \beta_2 e^{-\Gamma t},
\end{align}
where $\beta_1 = \alpha_0(1 - \alpha_1)$ and $\beta_2 = -\alpha_0\alpha_2$. We can fix the value of $\beta_1$ by requiring that as $t \to \infty$ the probability $P(\lnot\; 11, t)$ approaches $P_\text{leak}$, which is the probability of ending up in another asymptotic state, and thus we can conclude that 
\begin{align}
P(\lnot\; 11, t) =A e^{-\Gamma t} + P_\text{leak},
\end{align}
for some constant $A$.

\subsection{Extracting Resonance Data}
\label{app:resonance}

\begin{figure}[t]
    {\centering
    \includegraphics[width=\linewidth]{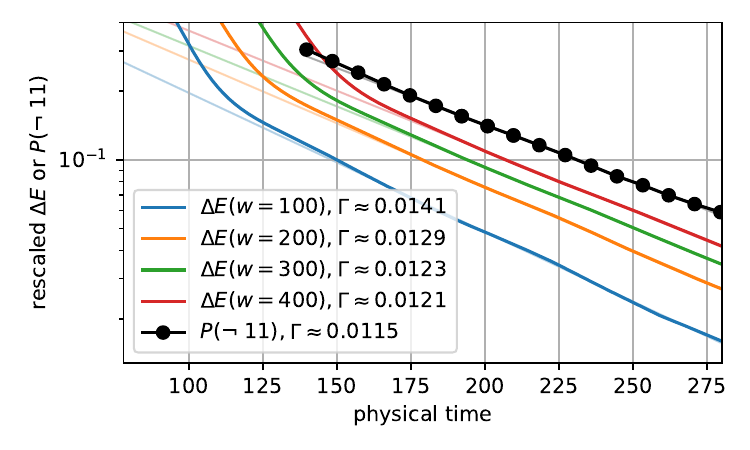}}
    \caption{Decay of $P(\lnot\; 11, t)$ and the excess energy expectation values $\Delta E(w)$ for a subregion of spatial width $w$, centered at the scattering location, for simulations at $E/m_1 \approx 2.746$, near the $m_4$ resonance, at $\eta_{\text{latt}}\approx 0.7052$ ($g_x = 1.06, g_z = 0.00985$), with $\sigma = 80$, bond dimension $D \le 64$ and time-step size $0.05$. Relative position on the $y$ axis is not meaningful and probabilities are scaled for ease of comparison with energies. We plot the excess energy for several sizes of the subregion window over which we sum the energy contributions. We observe fairly strong oscillations in the energy data, which become smaller with larger window sizes. The light-colored lines indicate fits, used to estimate $\Gamma$, from data between $t=180$ and $t=260$.
    }
    \label{fig:app_reso_en_prob}
\end{figure}

\begin{figure}[t]
    {\centering
    \includegraphics[width=\linewidth]{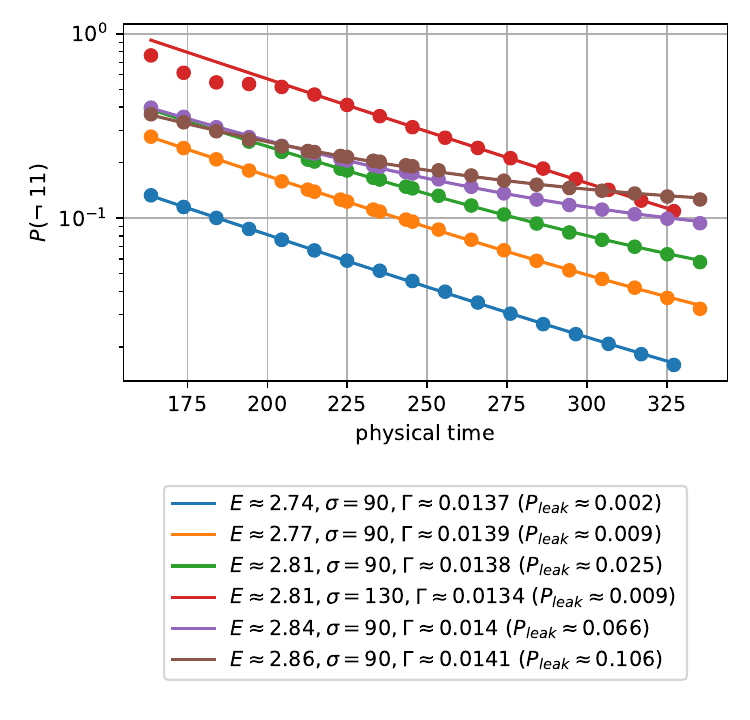}}
    \caption{Decay of $P(\lnot\; 11, t)$ for $\eta_{\text{latt}}\approx 0.7052$ ($g_x = 1.07, g_z = 0.01315$), for simulations at different energies. The correlation length is $\ell \approx 4.88$ and all of the simulations use initial wave-packet width $\sigma = 90$ except the red curve, which has $\sigma = 130$. We see that there is flattening at late times that is more pronounced at higher energies. This indicates some scattering into additional sectors beyond 11. We also see that increasing the wave-packet width reduces the flattening, as we focus the energy more on the resonance and less on other nearby channels. The resonance energy is $m_4 \approx 2.814$. Each curve is fitted (lines) to estimate $\Gamma$ and a constant probability offset $P_{\textrm{leak}}$. For the red curve, the fitted data begins at $t=220$, while for others all data is used. } 
    \label{fig:app_reso_offsets}
\end{figure}

In simulations of scattering of two type-1 particles with total energy close to the resonance energy $m_4$ predicted by form-factor perturbation theory, we observe that the probability of not being in the asymptotic 11 sector, denoted $P(\lnot\; 11, t)$ and defined in Eq.~\eqref{eq:Pnot11}, decays exponentially in time after the collision.
We also observe exponential temporal decay of the excess energy $\Delta E(w, t)$, as noted in the main text.

Because the energy of the $m_4$ resonance is below the threshold $m_1 + m_2$ for inelastic scattering into other asymptotic particle sectors, we expect the outgoing state to be dominated by the 11 sector, and thus for any exponential decay of $P(\lnot\; 11, t)$ and $\Delta E(w, t)$ to be associated with the resonance. If there is just one type of stable particle and a single resonance, the toy model of App.~\ref{app:resonance_toy} is applicable. For that model we found that the contribution to the scattering amplitude arising from the pole decays in time at a rate $\Gamma/2$ at any point $x$ in space. In App.~\ref{app:reso_toy_asymp} we argued that, as a consequence, the portion $P(\lnot\; 11, t)$ of the outgoing state outside the asymptotic 11 sector, decays at rate $\Gamma$.

If $P(\lnot\; 11, t)$ decays with rate $\Gamma$, and if the resonating part of the wavefunction is spatially separated from the off-resonant part -- an assumption we also used in App.~\ref{app:reso_toy_asymp} -- we should expect $\Delta E (w, t)$ to also decay with rate $\Gamma$ insofar as the window of width $w$ contains the resonating excitation. Indeed, it should not matter if the window also captures some particles in the 11-sector resulting from the decay of the resonance; the probability of finding such particles in the window is roughly proportional to the probability $\sim e^{-\Gamma t}$ that the resonance has not yet decayed. Therefore, we can generously set $w$ to include much of the scattering region as long as it does not contain the off-resonant tracks.

\begin{figure*}[ht!]
    \centering
    \subfloat[\label{fig:app_reso_bdim_steps}] {\centering
    \includegraphics[width=0.45\linewidth]{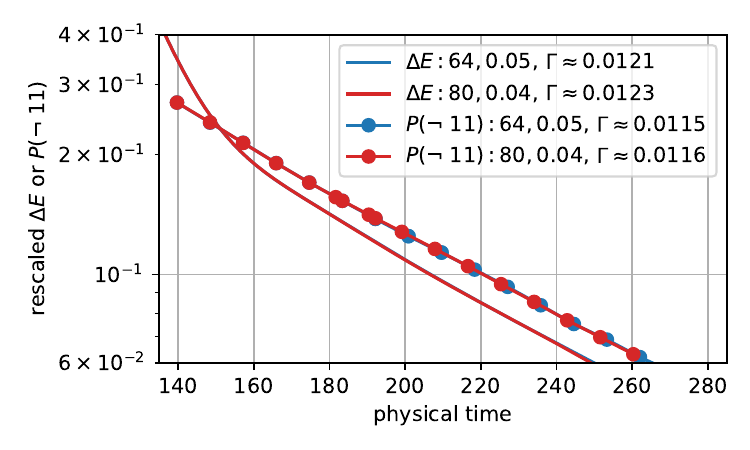}}\hspace{0.01\textwidth}
    \subfloat[\label{fig:app_reso_lattice}]{
    \includegraphics[width=0.45\linewidth]{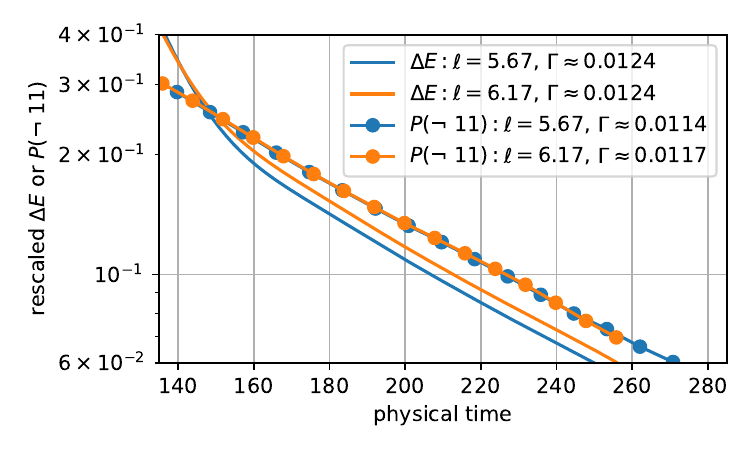}
    }
    \caption{
    ($a$) Decay of $P(\lnot\; 11, t)$ and the excess energy expectation values $\Delta E(w=400)$ with a 400-site window, for the same lattice parameters shown in Fig.~\ref{fig:app_reso_en_prob}, but with two different maximum bond dimensions $D \le 64, 80$ and integration step sizes $0.05, 0.04$. We see some small deviations for late times in the energy, but the results are generally robust. Each estimated $\Gamma$ is from a fit using data with $t \in [180, 260]$. 
    \\    
    ($b$) Decay of $P(\lnot\; 11, t)$ and the excess energy expectation values $\Delta E(w=400)$ with a 400-site window, with bond dimension $D \le 64$ and time-step size $0.05$, at two different lattice spacings, shown here via the correlation length $\ell$, in lattice sites. The lattice parameters are $g_x = 1.06, g_z = 0.00985, \sigma = 80$ ($\ell = 5.67$, $\sigma/2\ell \approx 7.05$) and $g_x = 1.055, g_z = 0.008368, \sigma = 90$ ($\ell = 6.17$, $\sigma/2\ell \approx 7.29$) with energies $E/m_1 \approx 2.746$ and $E/m_1 \approx 2.758$ respectively. In both cases $\eta_{\text{latt}} \approx 0.7052$. Each estimated $\Gamma$ is from a fit using data with $t \in [180, 250]$. 
    } 
\end{figure*}

\subsubsection{Fitting \texorpdfstring{$\Gamma_4$}{Gamma4}}
We fit $P(\lnot\; 11, t)$ using the ansatz Eq.~\eqref{eq:ansatz}. We illustrate these fits, and the necessity of the offset, in Fig.~\ref{fig:app_reso_offsets}. We prefer fitting $P(\lnot\; 11, t)$ to fitting $\Delta E(w, t)$ due to two phenomena in our data that hinder fitting the latter: First, we observe temporal oscillations in $\Delta E(w, t)$, most pronounced at smaller $w$, particularly for larger $\eta_{\textrm{latt}}$. This is possibly due to interference from small 12 and 21 contributions, since these contributions would be near threshold so that outgoing particle velocities are small, these contributions could linger in the window region $w$ for extended times. Second, we find a bias toward larger fitted decay rates for smaller $w$, with the rate approaching that of $P(\lnot\; 11, t)$ at large $w$. Both phenomena are illustrated in Fig.~\ref{fig:app_reso_en_prob}.

We speculate that the bias at small $w$ may be due to contributions to $\Delta E(w, t)$ from portions of the resonating excitation, or small 12 and 21 contributions, leaving the window region. In this case, large $w$ would guarantee that, except at late times, we only capture decay into the $11$ sector and no inelastic scattering whose asymptotic states remain closer to the center. One could then attempt to fit the energy density also allowing for an offset as in Eq.~\eqref{eq:ansatz}. We did not do this because larger fixed-size windows start to pick up contributions from the primary (off-resonant) tracks at early times. To get good data to fit the resonance decay, this must be avoided by allowing the primary tracks to leave the window, necessitating longer simulation times. It may be possible to get good data from time-dependent window sizes $w(t)$, or by attempting to extrapolate in $w$. We did not pursue this as both of these procedures are more complicated than simply fitting $P(\lnot\; 11)$ and introduce extra variables to the analysis.

To test robustness, we fit $\Gamma_4$ at different energies, lattice spacings, and with different settings for the maximum MPS bond dimension $D_{\max}$ and the integrator step size $dt_\text{latt}$. We show in Figs.~\ref{fig:app_reso_bdim_steps} and~\ref{fig:app_reso_lattice} that our results are robust in these parameters for selected simulations at larger $\eta_{\text{latt}}$, where we observe the most spread in Fig.~\ref{fig:delta_t_m4}.

\subsubsection{Estimating \texorpdfstring{$m_4$}{m4}}
\begin{figure}[t]
    {\centering
    \includegraphics[width=\linewidth]{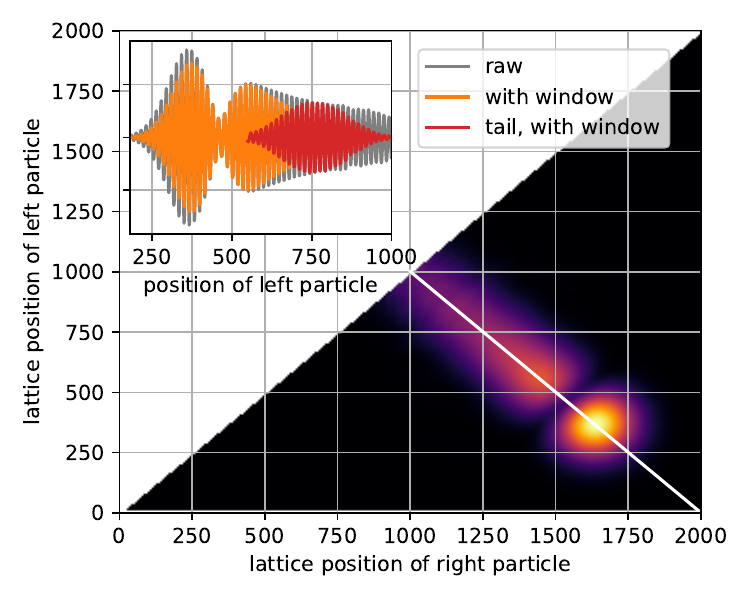}}
    \caption{Wavefunction for $\eta_{\text{latt}}\approx 0.7052$ ($g_x = 1.07, g_z = 0.01315$), $E/m_1 \approx 2.81$, near the $m_4$ resonance, at time $t_\text{phys} \approx 285$ ($t_\text{latt} = 800$), projected into the 11 particle lattice position basis. Here we neglect momentum dependence of the basis excitation tensors and fix them according to the initial wave-packet momenta. We observe a ``blob'' near the edge of the spatial window followed by a resonance-decay ``tail'' extending to the middle, with an interference gap in between. The inset shows a linescan, indicated by a white line in the main plot, and shows how the linescan data may be modulated via multiplication by a window function (here a Hamming function) for FFT frequency analysis.} 
    \label{fig:app_reso_projection}
\end{figure}

To estimate $m_4$, we first project the outgoing state on a timeslice onto the 11 position basis formed by the double-excitation states defined in Eq.~\eqref{eq:reference_state}. This results in a 2-particle wavefunction, an example of which is shown in Fig.~\ref{fig:app_reso_projection}, excluding any resonating component of the state, as well as any components that have leaked via inelastic channels. What remains is the off-resonance 11 component, visible as a ``blob'' near the edge of the system, and the 11 decay products from the on-resonance component, visible as a ``tail'' leading toward the scattering center. These features correspond respectively to the primary and secondary tracks visible in the energy expectation-value plots such as Fig.~\ref{fig:decay_ex}. We do not account for momentum-dependence of the excitation tensors in the projection used for this analysis -- we find the corrections obtained from doing this are very small.

Because the resonance is narrowly focused in momentum space, the energy of the tail part of the 11 wavefunction will be focused close to the resonance energy $m_4$, even when the total scattering energy deviates significantly. The resonance acts as an energy filter on the incoming state that spatially separates the components of the state close to the resonance energy from the remainder. This is convenient because we can then hope to cleanly extract the resonance energy from a Fourier analysis of the spatial tail, whose dominant spatial frequency should be given by the momentum of type-1 particles at energy $m_4$.

\begin{figure}[t]
    {\centering
    \includegraphics[width=\linewidth]{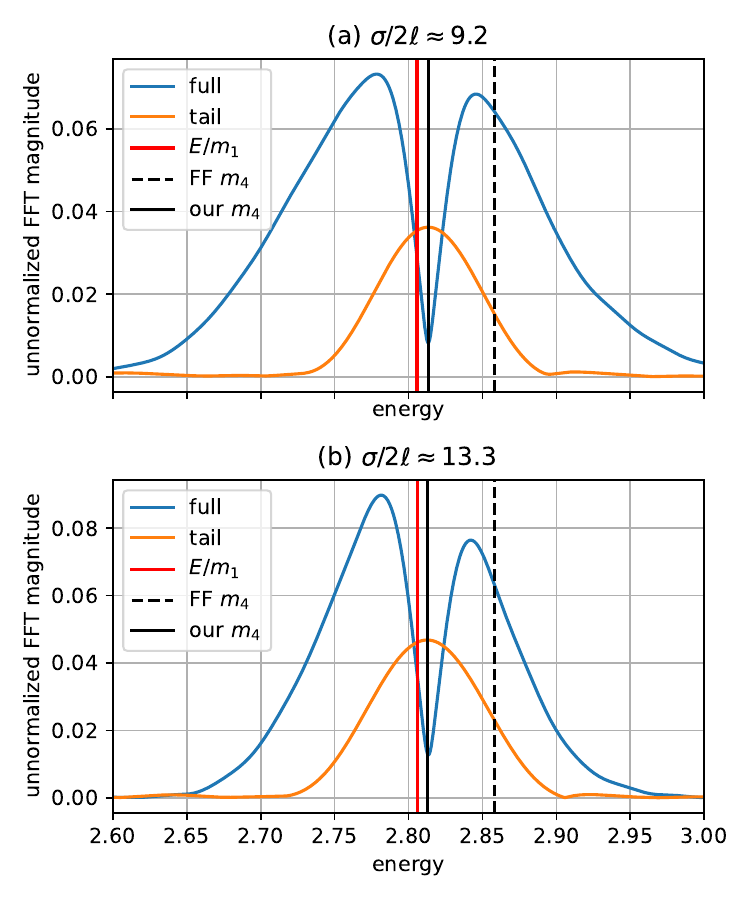}}
    \caption{Fast-Fourier transform (FFT) of a linescan through the projected 11 position-basis wavefunction at chosen times after scattering near the $m_4$ resonance. We show the FFT of the (Hamming-window modulated, zero-padded) full linescan as well as the FFT of the ``tail'' feature (also windowed and padded). The peak frequency of the ``tail'' FFT gives us our $m_4$ estimate and is indicated by a solid black vertical line. It is different than the form-factor (FF) predicted $m_4$, indicated by a black dashed line. The parameters are $\eta_{\text{latt}}\approx 0.7052$ ($g_x = 1.07, g_z = 0.01315$), $E/m_1 \approx 2.81$ with two different wave-packet widths. The spatial data for plot (b) is shown in the inset of Fig.~\ref{fig:app_reso_projection}. } 
    \label{fig:app_reso_m4_fourier}
\end{figure}

To extract this frequency, we consider a one-dimensional ``linescan'' through the two-particle position-basis wavefunction, defined by setting the left and right particle positions to be an equal distance from the scattering center. Such a linescan captures the ``blob'' and ``tail'' features, as illustrated in Fig.~\ref{fig:app_reso_projection}. We then apply an FFT to the amplitudes inside a spatial subregion chosen within the tail feature, using a Hamming window function to reduce ringing effects due to the finite size of the subregion. 
This FFT of the tail feature yields a function of momentum peaked at a value $k_\text{tail}$, which is related to the mass of the resonance via
\begin{align}
    \label{eq:freq_vs_resonant_mass}
    m = \sqrt{4 m_1^2 + k_\text{tail}^2}.
\end{align}

To understand Eq.~\eqref{eq:freq_vs_resonant_mass}, consider the total wavefunction of the left- and right-moving particles after scattering (ignoring overall factors which are irrelevant for the present discussion),
\begin{align}
\label{eq:two_particle_wavefunction}
\begin{split}
    \psi(x,y,t) \sim \int  dp dq \, S(E)&  e^{-\frac {\sigma^2}{2} (p - p_0)^2-\frac {\sigma^2}{2} (q + p_0)^2}  \\ \times &e^{-i(E(p)t + E(q)t - px - qy)},
    \end{split}
\end{align}
where $p$ and $q$ are the momenta of the right- and left-moving particle, respectively. We are interested in finding the dominant frequency mode of the tail feature of a diagonal slice $\psi(x,-x,t)$ of this wavefunction (corresponding to the linescan in our numerical analysis) for a scattering process in the vicinity of the resonance. Around the resonance, the S-matrix takes the form~\eqref{eq:reso_smatrix_app}. As discussed in Sec.~\ref{app:resonance_toy}, the contribution relevant for the tail feature comes from the S-matrix pole which is picked up when we deform one of the integration contours to a saddle point. To evaluate the residue of this pole it is useful to write the momenta as
\begin{align}
    \label{eq:expansion_p_q}
    p = \sqrt{\frac{m^2}{4} - m_1^2} + \delta p, && q = - \sqrt{\frac{m^2}{4} - m_1^2} + \delta q,
\end{align}
such that for $\delta p = \delta q = 0$ the center of mass energy sits exactly at the resonance, $E_\text{tot} = m^2$. Expanding the denominator of the S-matrix to linear order in $\delta p$ and $\delta q$ gives
\begin{align}
S(E) \sim \frac{1}{\delta p - \delta q - i b},
\end{align}
with $b$ defined in Eq.~\eqref{eq:b-formula-again}. This is a good approximation at least as long as $\delta p^2, \delta q^2 \ll m^2$, which is valid in our simulations and demonstrates that the integrand has a pole at a purely imaginary value of $\delta p - \delta q = i b$. Integrating over one of the momenta picks up the residue of this pole. The $x$-dependence of the remaining momentum integral in $\psi(x,-x,t)$ thus reduces to
\begin{align}
\begin{split}
    \label{eq:phases_mass_detector}
    e^{-i(px - qx)} \Large|_\text{pole} &= e^{-i(\sqrt{m^2 - 4 m_1^4} + \delta p - \delta q) x} \Large|_\text{pole}\\
    &= e^{-i\sqrt{m^2 - 4 m_1^2} x} \, e^{b x} ,
\end{split}
\end{align}
which does not depend on momenta and thus can be factored out of the integral. We thus see that if we scatter in the vicinity of the resonance, i.e., if $\delta p^2, \delta q^2 \ll m^2$, the tail part of the wavefunction oscillates in the $x = -y$ direction with momentum
\begin{align}
    \label{eq:dominiant_freq}
    k_\text{tail} = \sqrt{m^2 - 4 m_1^2}.
\end{align}
Therefore, Fourier transforming the diagonal wavefunction $\psi(x,-x,t)$ multiplied with a window function in $x$ with support on the tail feature produces a function in momentum space which is peaked at $\sqrt{m^2 - 4 m_1^2}$. This can be verified by Fourier transforming Eq.~\eqref{eq:phases_mass_detector} after multiplying it by a Gaussian. Furthermore, the choice of the window function does not affect the peak of the Fourier transform and we thus use the more conventional Hamming window function instead of a Gaussian. Once the dominant frequency $k_\text{tail}$ is extracted from the Fourier transform the resonance mass $m$ is obtained by inverting the relation \eqref{eq:dominiant_freq} which yields Eq.~\eqref{eq:freq_vs_resonant_mass}.

In Fig.~\ref{fig:app_reso_m4_fourier}, we show the FFT results, plotted against energy, for two different wavepacket widths. Note that, because we preprocess the spatial data by applying window functions, these are \emph{not} equivalent to a momentum-basis projection of the wavefunction. We plot the FFT of both the full linescan and of the tail feature. We see that the peak frequency of the tail (marked with a vertical line) matches a dip feature in the full linescan, illustrating the destructive interference between the on-resonance and off-resonance parts of the outgoing state. We also plot the form-factor prediction for $m_4$, which we see is significantly different than the observed $m_4$ in this case. 

\subsubsection{Spatial Decay}

By examining the spatial wavefunction on a timeslice, we can also test the relation, described in Eqs.~\eqref{eq:b-formula} and~\eqref{eq:m_from_gamma_and_b}, between spatial and temporal decay rates. Although these relations are derived from scattering of a wave packet with a plane wave, we might expect them to be at least approximately valid if we scatter very close to the resonance and let $E_2 = E/2$, which is half the total simulation energy $E$. So, replacing $E_2 = E/2$ in Eq.~\eqref{eq:b-formula}, and using $\Gamma = \Gamma_4$ from our temporal decay fits and $m = m_4$ from Fourier analysis of the spatial wavefunction, we get a prediction for the spatial decay rate $b$. This must be converted into lattice units by dividing by the correlation length $\ell$.

\begin{figure}[t]
    {\centering
    \includegraphics[width=\linewidth]{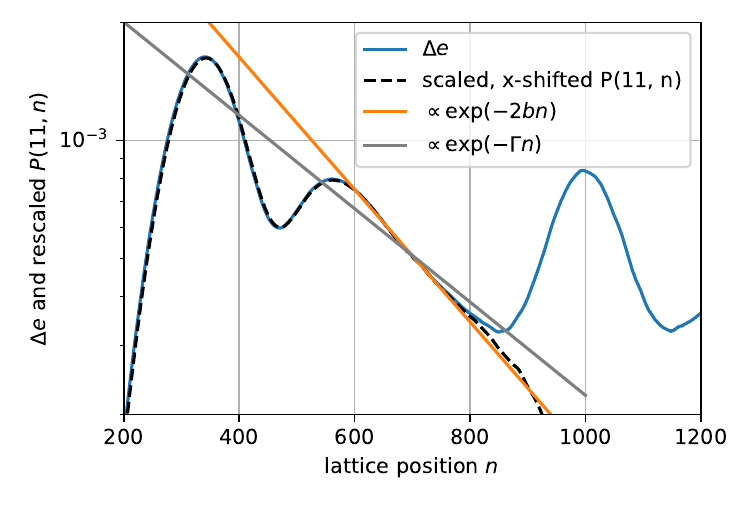}}
    \caption{Spatial behavior of the energy-density expectation value $\Delta e(n) = \langle h^\text{symm}_{n,n+1} \rangle_t- \langle h^\text{symm}\rangle_\text{vac}$ and 11 position-basis probability for the left particle $P(11, n)$ on a timeslice at $t_\text{phys} \approx 285$ ($t_\text{latt} = 800$). The probability data is scaled to match the energy density and spatially shifted to account for the location of the position-basis excitation being offset from the excitation tensor. Physical parameters are $\eta_{\text{latt}}\approx 0.7052$ ($g_x = 1.07, g_z = 0.01315$), $E/m_1 \approx 2.81$. The orange line shows the spatial decay predicted by Eq.~\eqref{eq:b-formula}, using the fitted temporal decay (see Fig.~\ref{fig:app_reso_offsets}) and $m_4$ estimated as in Fig.~\ref{fig:app_reso_m4_fourier}(b). For comparison, we also show in gray the temporal decay at rate $\Gamma$. Here, both $\Gamma$ and $b$ are converted to lattice units.} 
    \label{fig:app_reso_spatial_decay}
\end{figure}

We see for an example simulation in Fig.~\ref{fig:app_reso_spatial_decay} that the predicted rate matches a section of the spatial wavefunction on a timeslice. Note that we do not use Eq.~\eqref{eq:m_from_gamma_and_b} to estimate $m_4$, preferring the Fourier-analysis method described above, for two reasons: First, we find that the spatial decay is quite difficult to fit accurately in our simulated data, requiring long-running simulations to get a good fit range. Second, this estimate would also depend on $\Gamma$ obtained from temporal fits, adding further to the error.

\section{Hyperparameter Dependence of High Energy Scattering}
\label{app:hyperparameter_highE}
In Fig.~\ref{fig:highE}, the dependence of $P_{11 \to 11}$ for different values of scattering energy across wide range of $\eta_\text{latt}$ was shown. We provide the numerical data for the dependence of parameters on the probability data in Table~\ref{tab:highE_hyper1} and Table~\ref{tab:highE_hyper2} where we find almost no dependence on the parameters.

\begin{table}[ht!]
\centering
\renewcommand{\arraystretch}{1.05}
\setlength{\tabcolsep}{5pt}
\begin{tabular}{|c|c|c|c|c|c|c|}
\hline
$g_x$ &    $g_z$ &  $\eta_\text{latt}$ &  $\sigma$ &  $N$ & $E$ & $\ell$ \\
\hline
1.0043 & 0.010000 &               0.050 &        75 & 1400 & 2.433 &    4.28 \\
1.0043 & 0.010000 &               0.050 &        85 & 1400 & 2.433 &    4.28 \\
1.0035 & 0.006800 &               0.050 &        85 & 1400 & 2.435 &    5.23 \\
\hline
1.0100 & 0.012000 &               0.106 &        85 & 1400 & 2.464 &    3.99 \\
1.0080 & 0.007900 &               0.106 &        75 & 1400 & 2.465 &    4.95 \\
1.0070 & 0.006150 &               0.106 &        88 & 1400 & 2.448 &    5.66 \\
\hline
1.0110 & 0.007450 &               0.150 &      80 & 1400 & 2.543 &    5.23 \\
1.0110 & 0.007450 &               0.150 &      90 & 1400 & 2.543 &    5.23 \\
1.0150 & 0.013000 &               0.152 &      85 & 1400 & 2.518 &    3.91 \\
\hline
1.0250 & 0.013000 &               0.253 &      85 & 1400 & 2.565 &    4.09 \\
1.0200 & 0.008555 &               0.253 &      90 & 1400 & 2.520 &    5.09 \\
\hline
1.0500 & 0.020000 &               0.403 &      70 & 1400 & 2.664 &    3.51 \\
1.0500 & 0.020000 &               0.403 &      90 & 1400 & 2.664 &    3.51 \\
1.0400 & 0.013162 &               0.403 &      90 & 1400 & 2.689 &    4.36 \\
\hline
1.0500 & 0.010000 &               0.583 &      90 & 1400 & 2.652 &    5.41 \\
1.0500 & 0.010000 &               0.583 &      90 & 1400 & 2.685 &    5.41 \\
1.0500 & 0.010000 &               0.583 &      80 & 1400 & 2.760 &    5.41 \\
1.0500 & 0.010000 &               0.583 &      90 & 1400 & 2.760 &    5.41 \\
1.0400 & 0.006580 &               0.583 &      90 & 1400 & 2.703 &    6.72 \\
\hline
1.0700 & 0.013150 &               0.705 &      90 & 1400 & 2.771 &    4.89 \\
1.0700 & 0.013150 &               0.705 &      90 & 1400 & 2.806 &    4.89 \\
1.0700 & 0.013150 &               0.705 &     130 & 2000 & 2.806 &    4.89 \\
1.0600 & 0.009850 &               0.705 &      80 & 1400 & 2.747 &    5.68 \\
1.0550 & 0.008368 &               0.705 &      90 & 1400 & 2.758 &    6.18 \\
1.0550 & 0.008368 &               0.705 &     120 & 1800 & 2.758 &    6.18 \\
1.0500 & 0.007000 &               0.705 &      90 & 1400 & 2.732 &    6.78 \\
1.0500 & 0.007000 &               0.705 &     120 & 1900 & 2.733 &    6.78 \\
1.0500 & 0.007000 &               0.705 &     120 & 1900 & 2.810 &    6.78 \\
\hline
\end{tabular}
\caption{Parameters used for resonance data in Fig.~\ref{fig:delta_t_m4}. In all cases, $D_\text{vac}$, $D=64$, $dt_\text{latt}=0.05$.}
\label{tab:reso_params}
\end{table}

\begin{table}[ht!]
\centering
\renewcommand{\arraystretch}{1.05}
\setlength{\tabcolsep}{8pt}
\begin{tabular}{|c|c|c|c|c|c|c|}
\hline
$D$ & $dt_\text{latt}$ & $g_x$ & $g_z$ & $\eta_\text{latt}$ & $E$ & $P_{11 \to 11}$ \\
\hline
64 & 0.025 & 1.1 & 0.006 & 1.53 & 7 & 0.533 \\
64 & 0.05 & 1.1 & 0.006 & 1.53 & 7 & 0.533 \\
48 & 0.05 & 1.1 & 0.006 & 1.53 & 7 & 0.534 \\
\hline
64 & 0.025 & 1.04 & 0.006 & 0.61 & 7 & 0.514 \\
64 & 0.05 &  1.04 & 0.006 & 0.61 & 7 & 0.514 \\
48 & 0.05 & 1.04 & 0.006 & 0.61 & 7 & 0.520 \\
\hline
64 & 0.025 & 1.04 & 0.006 & 0.61 & 8 & 0.544 \\
64 & 0.05 & 1.04 & 0.006 & 0.61 & 8 & 0.545 \\
48 & 0.05 &  1.04 & 0.006 & 0.61 & 8 & 0.558 \\
\hline
64 & 0.025 & 1.1 & 0.006 &1.53 & 8 & 0.476 \\
64 & 0.05 & 1.1 & 0.006 & 1.53 & 8 & 0.476 \\
48 & 0.05 &  1.1 & 0.006 & 1.53 & 8 & 0.478 \\
\hline
\end{tabular}
\caption{The negligible dependence of $P_{11 \to 11}$ on the bond dimension $D$ and the integrator time step $dt_\text{latt}$ for two different $\eta_\text{latt}$ on either side of the transition region and scattering energy $E$ (in units of $m_1$).}
\label{tab:highE_hyper1}
\end{table}

\begin{table}[ht!]
\centering
\renewcommand{\arraystretch}{1.05}
\setlength{\tabcolsep}{5pt}
\begin{tabular}{|c|c|c|c|c|c|c|c|}
\hline
$D$ & $dt_\text{latt}$ & $g_x$ & $g_z$ & $\eta_\text{latt}$ & $\sigma$ & $N$ & $P_{11 \to 11}$ \\
\hline
64 & 0.05 & 1.06 & 0.006 & 0.92 & 70 & 1000 & 0.257 \\
64 & 0.05 & 1.06 & 0.006 & 0.92 & 100 & 1200 & 0.256 \\
\hline
64 & 0.05 & 1.06532 & 0.006 & 1.0 & 70 & 1000 & 0.30 \\
64 & 0.05 & 1.06532 & 0.006 & 1.0 & 100 & 1200 & 0.269 \\
\hline
64 & 0.05 & 1.07054 & 0.006 & 1.08 & 70 & 1000 & 0.30 \\
64 & 0.05 & 1.07054 & 0.006 & 1.08 & 100 & 1200 & 0.30 \\
\hline
\end{tabular}
\caption{The dependence of the probability on number of sites $N$ and the packet width $\sigma$ for three different $\eta_\text{latt}$ at $E = 6m_{1}$. For $E > 6m_{1}$ we do not see any dependence even for this $\eta_\text{latt}$.}
\label{tab:highE_hyper2}
\end{table}

\section{Computational Resources}
The simulations were run on \texttt{c5a.4xlarge}, \texttt{c6i.4xlarge}, and \texttt{c7i.4xlarge} instances provided by AWS, each with 8 physical cores (16 vCPUs, or hardware threads) using the publicly available code \emph{evoMPS}~\cite{evomps}. The simulations with $D = 64$ and $N = 2000$ run at approximately $10000$ RK4 time steps/week on \texttt{c5a.4xlarge} instances. Some numerical computations were also done on Symmetry which is an HPC system hosted by Perimeter Institute.

\vspace{5em}

\bibliographystyle{utphys}
\raggedleft
\bibliography{v1.bib}

\end{document}